\documentclass[a4paper,11pt]{article}
\pdfoutput=1
\usepackage{jheppub}
\usepackage[toc,page]{appendix}
\usepackage{graphicx}
\usepackage{amsmath,amssymb,mathrsfs}
\usepackage{bbm}
\usepackage{color}
\usepackage{xcolor}
\usepackage{dsfont} 
\usepackage{cancel}
\usepackage{dsfont}
\usepackage{epstopdf}
\usepackage{epsfig}
\usepackage{bm}
\usepackage{dcolumn}
\usepackage{hyperref}
\usepackage{enumitem}
\usepackage{multirow}
\usepackage{lineno}
\usepackage{mathtools,slashed}
\usepackage{url}
\usepackage{lineno}
\usepackage{wrapfig}
\usepackage[load-configurations=abbreviations, redefine-symbols=true]{siunitx}

\newcommand{\ben}{\begin{enumerate}}
\newcommand{\een}{\end{enumerate}}
\newcommand{\bit}{\begin{itemize}}
\newcommand{\eit}{\end{itemize}}

\newcommand{\beqa}{\begin{eqnarray}}
\newcommand{\eeqa}{\end{eqnarray}}
\newcommand{\beq}{\begin{equation}}
\newcommand{\eeq}{\end{equation}}
\newcommand{\bay}{\begin{array}}
\newcommand{\eay}{\end{array}}

\newcommand{\nn}{\nonumber}
\def\ifmath#1{\relax\ifmmode #1\else $#1$\fi}

\def\gsim{\ \rlap{\raise 3pt \hbox{$>$}}{\lower 3pt \hbox{$\sim$}}\ }
\def\lsim{\ \rlap{\raise 3pt \hbox{$<$}}{\lower 3pt \hbox{$\sim$}}\ }

\def\ls#1{\ifmath{_{\lower1.5pt\hbox{$\scriptstyle #1$}}}}
\def\lsup#1{^{\lower 6pt\hbox{$\scriptstyle#1$}}}

\def\bracket#1#2 {\mathinner{\langle{#1}|{#2}\rangle}}

\def\bracket#1#2 {\mathinner{\langle{#1}|{#2}\rangle}}

\newcommand{\be}{\begin{equation}}
\newcommand{\ee}{\end{equation}}
\newcommand{\bea}{\begin{eqnarray}}
\newcommand{\eea}{\end{eqnarray}}

\graphicspath{{figs/}}

\begin{document}

\title{Electroweak bubble wall expansion: gravitational waves and baryogenesis  in Standard Model-like thermal plasma}

\abstract{  Computing the properties of the bubble wall of a cosmological first order phase transition at electroweak scale is of paramount importance for the correct prediction of the baryon asymmetry of the universe and the spectrum of gravitational waves. By means of the semi-classical formalism we calculate the velocity and thickness of the wall using as theoretical framework the scalar singlet extension of the SM with a parity symmetry and the SM effective field theory supplemented by a dimension six operator. 
We use these solutions to carefully predict the baryon asymmetry and the gravitational wave signals. The singlet scenario can easily accommodate the observed asymmetry but these solutions do not lead to observable effects at future gravity wave experiments. In contrast the effective field theory fails at explaining the baryon abundance due to the strict constraints from electric dipole moment  experiments, however, the strongest solutions we found fall within the sensitivity of the LISA experiment. We provide a simple analytical approximation for the wall velocity which only requires calculation of the strength and temperature of the transition and works reasonably well in all models tested.
We find that generically the weak transitions where the fluid approximation can be used to calculate the wall velocity and verify baryogenesis produce signals too weak to be observed in future gravitational wave experiments. Thus, we infer that GW signals produced by simple SM extensions visible in future experiments are likely to only be produced in strong transitions described by detonations with highly relativistic wall velocities. 
}

%%%%%%%%%%%%%%%%%%%%%%%%%%%%%%%%%%%%%%%%%%%%%%%
\author[1]{Marek Lewicki}
\author[1]{Marco Merchand}
\author[1]{Mateusz Zych}
\affiliation[1]{Faculty of Physics, University of Warsaw ul.\ Pasteura 5, 02-093 Warsaw, Poland}
\emailAdd{marek.lewicki@fuw.edu.pl}
\emailAdd{mmerchand@fuw.edu.pl }
\emailAdd{mateusz.zych@fuw.edu.pl}
\maketitle
%%%%%%%%%%%%%%%%%%%%%%%%%%%%%%%%%%%%%%%%%%%%%%%
\section{Introduction}

The Standard Model (SM) of particle physics is one of the most successful theories ever devised. Although a plethora of experiments have verified the validity of the SM to high level of precision the SM is incapable to offer an explanation for the existence of dark matter, the baryon asymmetry of the universe (BAU) or the masses of neutrinos.

Cosmological first order phase transitions (FOPT) could be a vital ingredient in understanding the mechanism of dark matter production \cite{Baker:2019ndr, Azatov:2021ifm},  account for the baryon asymmetry of the universe \cite{Cohen:1993nk,Kuzmin:1985mm,Rubakov:1996vz, Morrissey:2012db} and be related to the origin of neutrino masses \cite{DiBari:2020bvn}.  Additionally they could lead to an observable stochastic gravity wave (GW) background \cite{Witten:1984rs}. With recent detection of GWs from black hole mergers ~\cite{TheLIGOScientific:2017qsa,Abbott:2020niy} and hints of a stochastic background~\cite{Arzoumanian:2020vkk,Goncharov:2021oub} that could have been produced in the early Universe \cite{Ellis:2020ena,Blasi:2020mfx,Vaskonen:2020lbd,DeLuca:2020agl,Nakai:2020oit,Ratzinger:2020koh,Kohri:2020qqd,Vagnozzi:2020gtf,Neronov:2020qrl,Middleton:2020asl} this possibility has motivated many studies into probing various beyond the SM (BSM) scenarios predicting a FOPT through their GW signals~\cite{Grojean:2006bp,Espinosa:2008kw,Dorsch:2014qpa,Jaeckel:2016jlh,Jinno:2016knw,Chala:2016ykx,Chala:2018opy,Artymowski:2016tme,Hashino:2016xoj,Vaskonen:2016yiu,Dorsch:2016nrg,Beniwal:2017eik,Baldes:2017rcu,Marzola:2017jzl,Kang:2017mkl,Iso:2017uuu,Chala:2018ari,Bruggisser:2018mrt,Megias:2018sxv,Croon:2018erz,Alves:2018jsw,Baratella:2018pxi,Angelescu:2018dkk,Croon:2018kqn,Brdar:2018num,Beniwal:2018hyi,Breitbach:2018ddu,Marzo:2018nov,Baldes:2018emh,Prokopec:2018tnq,Fairbairn:2019xog,Helmboldt:2019pan,Dev:2019njv,Ellis:2019flb,Jinno:2019bxw,Ellis:2019tjf,Azatov:2019png,vonHarling:2019gme,DelleRose:2019pgi,Mancha:2020fzw,Vanvlasselaer:2020niz,Giese:2020znk,Hoeche:2020rsg,Baldes:2020kam,Croon:2020cgk,Ares:2020lbt,Cai:2020djd,Bigazzi:2020avc,Wang:2020zlf} (for a comprehensive review see~\cite{Caprini:2015zlo,Caprini:2019egz}). 

First order transitions are characterized by the nucleation of bubbles of a symmetry breaking vacuum phase~\cite{Coleman:1977py,Callan:1977pt,Linde:1981zj} which subsequently expand and eventually collide ending the transition. Throughout the growth of the bubbles their spherical symmetry prohibits GW production. As the bubbles collide GWs can be sourced by the collisions themselves~\cite{Kosowsky:1992vn,Cutting:2018tjt,Ellis:2019oqb,Lewicki:2019gmv,Cutting:2020nla,Lewicki:2020jiv}, however, this source is relevant only in strongly supercooled~\cite{Ellis:2019oqb,Ellis:2020nnr} transitions in which plasma effects are suppressed. The source we will focus on comes from the motion of the plasma generated by its interactions with the bubble walls~\cite{Kamionkowski:1993fg,Hindmarsh:2015qta,Hindmarsh:2016lnk,Hindmarsh:2017gnf,Ellis:2018mja,Hindmarsh:2019phv,Ellis:2020awk}.

Between formation and collision, in a thermal plasma environment, the bubbles reach a steady state which is described by the velocity of the front interface and by the shape of its profile. Determination of these properties is crucial for making reliable predictions of the BAU and the GW spectrum. 
The first analytic formula, to the best of our knowledge, for the wall velocity was given by Andrei Linde in Ref. \cite{Linde:1981zj}. Assuming non-relativistic walls $v \ll 1$, the plasma has a large heat conductivity and the bubble expansion is isothermal. There is an outward pressure given by the difference in energy density of the false vacuum  to that in the true vacuum, i.e.,  
\begin{equation}
\epsilon  \equiv V(0,T)  - V (\phi_0,T),
\end{equation} 
this is compensated by the extra pressure from particles that are reflected and accelerated after reaching the phase boundary. All particles that are massless in the false vacuum but obtain a mass $m \gg T$ cannot penetrate inside and gain energy after bouncing back into the space of false vacuum. Thus the extra pressure on the wall is 
\be
\Delta p  =  \left( \sum_i \frac{\pi^2}{30}g_i T^4 \right) v
 \ee
where the sum is over all particles that become massive and $g_i$ the number of degrees of freedom.  From these equations one obtains
\be
v = \frac{30 \epsilon}{\sum_i \pi^2 g_i T^4} \label{Linde_estimate}. 
\ee
This simple picture is very intuitive and gives the right prediction, $v \rightarrow 0  $ as $\epsilon \rightarrow 0$ as well as suggesting that lower temperatures produce faster walls which holds true generically.  
Another analytic estimate appeared in \cite{Dine:1990fj} where the authors used energy and momentum conservation across the wall and obtained instead the quadratic relation $v^2 \sim \Delta p/\Delta \rho$. It was 
assumed that the temperature and velocity across the wall are constant however we will show in the present paper that these changes cannot be brushed aside.

In ref.~\cite{Dine:1992wr} it was shown that the analytic estimate of eqn. \eqref{Linde_estimate} can only be applicable if the phase transition is strongly first order since in this case particles acquire a large mass inside the bubble and are reflected off of the wall  therefore damping its propagation. This case corresponds to the "thin-wall" approximation where the thickness of the wall is much smaller than all the other relevant length scales. Other important length scales to evaluate are the mean free paths for elastic scattering $\lambda_{\text{elastic}}$ and for particle number changing processes $\lambda_{\text{inelastic}}$.  

Three limiting situations were first identified in ref.~\cite{Dine:1992wr} (see also  \cite{ Liu:1992tn}) ; (1) the thin-wall limit  with  $L_w < \lambda_{\text{elastic}}, \lambda_{\text{inelastic}} $, where $L_w$ is the thickness of the wall and which corresponds to maximal departure from thermodynamic equilibrium, (2) the thick-wall case which is the opposite scenario with  $\lambda_{\text{elastic}}, \lambda_{\text{inelastic}} < L_w$, in this case the particles that cross the wall have enough time to interact with the other particles in the plasma and thermal equilibrium is maintained and (3) the intermediate situation $\lambda_{\text{elastic}} < L_w < \lambda_{\text{inelastic}}$ where  some approximate form of thermal equilibrium is expected. 

 The aforementioned results   make clear that an estimation of the relevant hierarchy of length scales is pivotal for making a judicious choice of method for the computation of the wall properties. This, nonetheless, cannot be done a priori since the wall thickness is an unknown to begin with. For a given model the only thing one can do is to assume the propagation of the bubble falls into one of these limiting situations and to check the validity of the assumption a posteriori after applying a certain methodology.

 Investigations of the wall properties assuming case (2), local equilibrium, include \cite{Konstandin:2010dm, BarrosoMancha:2020fay, Balaji:2020yrx, Ai:2021kak} and in this case the only friction on the wall propagation comes from hydrodynamic effects of the plasma. On the other hand, if the wall is thin (but thicker than the particles thermal wavelength so that WKB approximation is valid), case(2), reflection and transmission of particles  is appropriate to quantify the friction. Studies that fall into to this case can be found in  \cite{Arnold:1993wc,Bodeker:2009qy,Bodeker:2017cim,BarrosoMancha:2020fay,Azatov:2020ufh, Hoche:2020ysm}. Recent methods aiming to compute the bubble wall velocity in strongly coupled theories  using holography can be found in Refs.~\cite{Bea:2021zsu, Bigazzi:2021fmq}.

In the impractical scenario of a FOPT in the SM with a light Higgs mass, it has been found \cite{Dine:1992wr, Liu:1992tn} that the most likely limiting case is that of the intermediate region with a small departure from thermodynamic equilibrium, the case (3). One of the reasons being that in the SM we cannot ignore the effect of particles crossing the wall since some of the masses are comparable to the temperature.  In Ref. \cite{ Liu:1992tn} it was found that the thin-wall scenario is unrealistic in the SM as it requires tuning the parameters of the theory. In this reference the authors showed that the typical mean free path of the heavier particles is of the same order as the width of the bubble wall and that a good approximation is to assume a small departure from local thermal equilibrium. In the same paper, as well as in~\cite{Turok:1992jp, Dine:1992wr} it was recognized that in order to calculate the friction forces that stop the bubble wall requires solving the non-equilibrium distributions of the massive particles in the plasma. To do so one must solve a complicated system of Boltzmann transport equations. This was first performed in the electroweak theory in~\cite{ Liu:1992tn} and later in more detail by Moore and Prokopec in \cite{Moore:1995ua}.

 The method appropriate for case (3), henceforth called the $\textit{semi-classical}$ approximation \cite{Moore:1995ua},  utilizes a three parameter $\textit{fluid}$ ansatz, which corresponds to perturbations in the chemical potential, temperature and velocity. The linear transport equations that follow are supplemented by the Higgs equation of motion (EOM) and a dynamical solution to the wall shape and velocity can be obtained.

 The idea that the expansion of nucleated vacuum bubbles at finite temperature could correspond to the motion of detonation waves was first proposed by Steinhardt in Ref.~\cite{Steinhardt:1981ct}. This was subsequently expanded~\cite{Espinosa:2010hh} to include all possible solutions in which the bubble walls reach a constant finite velocity due to interactions with the plasma. The hydrodynamic effects governing friction were taken into account for the electroweak theory in~\cite{Enqvist:1991xw, Liu:1992tn, Ignatius:1993qn}.

 The full inclusion of non-equilibrium particle populations, hydrodynamic effects as well as a leading order treatment ($\textit{leading-log}$ approximation) of all the relevant scattering and decay rates which enter the collision term was first undertaken in~\cite{Moore:1995si}. Here it was discovered that including the jump effect from hydrodynamics, the solutions become subsonic. 
 Furthermore it was pointed out that their results underestimate the friction since the contribution from the Higgs self coupling and from infrared $W$ bosons was ignored.

  The semi-classical approximation has also been used for the singlet extension of the SM in Refs.~\cite{Konstandin:2014zta, Kozaczuk:2015owa,Dorsch:2018pat}. 
  The inclusion of scattering processes that include the Higgs boson was done in Ref.~\cite{Kozaczuk:2015owa} and a calculation of the collision terms beyond $\textit{leading-log}$ was investigated in~\cite{Wang:2020zlf}. The effect of infrared gauge boson (also termed transition radiation) has been studied in~\cite{Bodeker:2009qy, Bodeker:2017cim} and more recently in~\cite{Hoeche:2020rsg}.
  
 A simplified method of estimating the friction term in the Higgs EOM, called phenomenological  approach has also been implemented in  \cite{Ignatius:1993qn, Megevand:2009ut, Megevand:2009gh, Sopena:2010zz, Megevand:2012rt, Huber:2013kj, Megevand:2013hwa}. This  approach consists in adding a term $\propto \eta(v_w, \phi ) u^{\mu}\partial_{\mu}$ to the Higgs EOM, where $u^{\mu}$ is the four velocity of the fluid and $\eta$ is an ad-hoc parameter which can depend non-trivially on the wall velocity and on the Higgs field. It was shown in  Ref.~\cite{Konstandin:2014zta} under which cases the phenomenological approach can reproduce all features of the Boltzmann equations.

In this paper we undertake the study of the wall expansion assuming the limiting situation (3) when there is a sizeable (but still small) departure from equilibrium. We base our calculations for the properties of the wall on the recently improved semi-classical fluid equations by Cline and Laurent \cite{Laurent:2020gpg}. Our primary focus here is to dissect the qualitative properties of the wall and their dependence on the parameters of the theory and how they correlate with the characteristics of the phase transition. 

In order to remain consistent with current phenomenology of the SM while allowing for FOPTs we use as benchmark models the gauge singlet extension with parity symmetric potential and the SM effective field theory with a dimension six operator. While previous studies have focused \cite{Cline:2021iff} on the complementarity of GW signals with the BAU, here we build on these analyses by assessing the range of applicability of the semi-classical treatment and show that it is only limited to sufficiently weak transitions with transition strength parameter $\alpha \lesssim 0.1$. More precisely, we show that this approximation stops operating at the boundary between hybrid and detonation solutions. In other words, the Jouguet velocity marks the maximum speed of the bubble wall. Furthermore we draw a comparison with unsophisticated estimates for the wall velocity in the thin- and thick-wall approximations and we found remarkable agreement with the thick-wall formula. We provide a simple derivation for this formula and make a connection with recent results of wall velocities in thermal equilibrium \cite{Ai:2021kak}. A similar derivation for the wall thickness is carried out that gives a remarkably good approximation for the Higgs wall-thickness but overestimates the singlet thickness by a factor of about $\approx 7/5$ in all cases.

Using the benchmark models mentioned above we computed the BAU employing the updated transport equations of Cline and Kainulainen \cite{Cline:2020jre} and obtained the predictions for the spectrum of stochastic background of GWs comparing them with current and future sensitivities. We underscore the importance of doing the calculation with the correct variables in front of the wall as otherwise one would not obtain the correct baryon relic. We improve upon previous studies \cite{Cline:2021iff} by using two different BSM models which makes our qualitative results more general. We also computed the percolation temperature for the GW predictions which is usually a circumvented step.

The content of this paper is organized as follows: in section~\ref{Phase_Trans} we briefly review the formalism of bubble nucleation at finite temperature including percolation and present the most important parameters for GW spectrum. An overview of the hydrodynamic treatment of the plasma is presented in section \ref{hydrodynamics}. After that, in section~\ref{semi_classical}, we review the improved transport equations and introduce the semi-classical approximation. Two separate subsections are devoted to show the most relevant improved fluid equations for the CP-even and CP-odd perturbations. We devote chapters \ref{scalar_singlet} and \ref{sec:SMEFT} to introduce the scalar singlet model and the SM effective field theory (SMEFT), respectively. Appropriate subsections in this chapter contain the results of the computations for the wall properties and for the BAU. The predictions for the GW spectrum are assembled together in chapter \ref{sec:GWs}. In chapter \ref{sec:comparison} we discuss the comparison between our results and the thin- and the thick-wall approximations. A summary of this work and our conclusions are provided in chapter \ref{sec:conclusions}. We provide the formulas used in the finite temperature effective potential for the two benchmark models in appendices \ref{sec:appendix_singlet} and \ref{sec:appendix_SMEFT}.

 %%%%%%%%%%%%%%%%%%%%%%%%%%%%%%%%%%%%%%%%%%%%%%%%%%
\section{Dynamics of the finite temperature phase transition}
\label{Phase_Trans}

  First-order phase transitions proceed via nucleation of bubbles of broken phase in the space filled with unstable phase. The probability of tunneling into the broken vacuum at temperature $T$ is~\cite{Linde:1980tt,Linde:1981zj}
\begin{equation}
\Gamma(T) = A(T)\textrm{e}^{-S},
\end{equation}
where $S$ is the Euclidean action of a critical bubble. For $O(3)$-symmetric thermal systems $S=\frac{S_3}{T}$. The prefactor $A(T)$ involves complicated functional determinants which are hard to compute. For tunneling at finite temperature it can  be approximated as 
\begin{equation}
A(T)=\left(\frac{S_3}{2\pi T}\right)^{\frac{3}{2}} T^4.
\end{equation}

The nucleation temperature $T_n$ is defined as the temperature at which the nucleation probability per horizon volume is of order $1$. It corresponds to the condition
\begin{equation}
\label{cond1}
\int_{t_c}^{t_n} dt \frac{\Gamma(t)}{H(t)^3} = 1,
\end{equation}
where $t_c$ corresponds to the time at the critical temperature when both phases are degenerate and $t_n$ the time when nucleation of bubbles begins. $H$ denotes the Hubble rate which in a radiation dominated epoch  can be expressed as
\begin{equation}
H^2=\frac{\rho_r}{3M_{P}^2}, \quad \rho_r \equiv \frac{\pi^2}{30} g_{*}(T) T^4,
\end{equation}
where $M_P = 2.4 \times 10^{18}$ GeV is the reduced Planck mass and $\rho_r$ the radiation energy density of relativistic species. To take into account the temperature dependence of the number of degrees of freedom $g_*(T)$, we use tabulated data from the estimates of Ref. \cite{Saikawa:2018rcs}.
The nucleation condition $(\ref{cond1})$ can also be approximately written as
\begin{equation}
\label{cond2}
\frac{S_3}{T_n}\approx 4\log\left(\frac{T_n}{H}\right),
\end{equation}
which for temperatures around the electroweak scale gives us the approximate condition $S_3/T_n \approx 140$. This approximation is  usually used in the literature for obtaining the nucleation temperature. For sufficiently strong transitions, however, this nucleation criteria has to be modified and in the most serious calculations one needs to include the vacuum contribution in the Hubble parameter and compute the temperature of percolation. 

It is usually assumed, that first-order phase transitions are instant and complete at the temperature $T\approx T_n$. Therefore all the parameters determining gravitational-wave signal are typically evaluated at this value. However to be more accurate, one may consider a probability, that a randomly chosen point is still in the false vacuum, given by 
\be
P(t) = e^{-I(t)},
\ee
where $I(t)$ corresponds to the fraction of the space which has already been converted to the broken phase, namely
\be
\label{space_fraction}
I(t) = \frac{4 \pi}{3} \int_{t_c}^{t} dt' \Gamma(t') a(t')^3 r(t,t')^3.
\ee
In the expression above, $r(t,t')$ denotes the comoving radius of a bubble nucleated at $t'$ propagated until a subsequent time $t$ and is given by 
\be
\label{comoving_radius}
r(t,t') = \int_{t'}^{t}  \frac{v_w(\tilde{t}) d\tilde{t}}{a(\tilde{t})},
\ee
with $a(t)$ the scale factor and $v_w(t)$ the wall velocity which, in principle, is time dependent. 

In practical calculations, it is more convenient to use temperature $T$ instead of time variable $t$ and \eqref{space_fraction} takes the form
\be
I(T) = \frac{4 \pi}{3} \int_{T}^{T_c} \frac{dT'}{H(T')} \Gamma(T') \frac{r(T,T')^3}{T'^4}.
\ee
It is usually assumed that the transition completes when $P (t) \approx 0.7$, which leads to a percolation temperature $T_p$ given by the following condition
\be
I(T_p) = 0.34.
\ee

Although the percolation time provides a more accurate estimate of when the transition completes and one should in general evaluate all physical observables at this time, its calculation presents a serious challenge given the fact that the comoving radius in \eqref{comoving_radius} depends on the velocity of the wall which is what we are aiming to achieve in this paper.  A complete solution would require an iterative method which we consider a next level of diligence and we will leave this issue as beyond the scope of this paper.

Below we will use the inaccurate condition $S_3/T_n \approx 140$ for obtaining the relevant parameter space of FOPT and in the wall velocity calculation but we will use the more accurate percolation prescription for the GW spectrum, including our results for the wall velocity in the comoving radius and taking into account the vacuum contribution to the Hubble scale. 

Another simplistic approximation which has been used to assess the viability of electroweak baryogenesis (EWBG) is the evaluation
of the sphaleron shutting off condition  after the plasma enters inside the bubble. This has become known as the sphaleron washout condition \cite{Farrar:1993hn} which translates into\footnote{In the presence of extra scalars charged under $SU(2)_L$ this condition should be modified.}
\be
\label{sphaleron_cond}
\frac{v_n}{T_n}\gtrsim 1.0,
\ee
with the numerical factor on the right hand side being a matter of some debate which could lead to slight modification in the range $1.5-0.5$~\cite{Quiros:1999jp,Funakubo:2009eg,Fuyuto:2014yia}. In this work we will use the sphaleron washout condition only for studying the shape of the parameter space
and we will show its correlation with the other parameters of the phase transition. For the BAU computation this condition is already integrated in the formula for the final asymmetry. 

To conclude this section we present the parameters which are relevant for the computation of the GW spectrum. The strength of the phase transition which,
following~\cite{Caprini:2019egz},we define as 
\be
\alpha \equiv  \frac{1}{\rho_r}\left( \Delta V_{eff}(\phi,T) - \frac{T}{4} \Delta \frac{\partial V_{eff}(\phi,T)}{\partial T} \right), \label{alpha_def}
\ee
where the $\Delta$ symbol meaning the difference between the false vacuum value and that in the true vacuum. In the above formula we write a generic $\phi$ dependence on the potential but it should be clear that it actually means dependence from all scalar fields according to the BSM model. Additionally, the factor of $1/4$ in the second term on $\alpha $ above has been ommitted in past literature. We believe that it should be included, i.e. the strength is identified with the difference of the normalized trace of the energy momentum tensor as opposed to just the difference in the normalized energy density.

The other important parameter is the inverse time duration of the phase transition which is calculated as
\be \label{eq:betaH}
\frac{\beta}{H} \equiv T_p \frac{d}{dT} \left(  \frac{S_3}{T} \right)\Big{|}_{T=T_p}.
\ee 

The parameters $\beta$ and $\alpha$ introduced above and the bubble wall-velocity play a central role in determining the GW spectrum which will be discussed in a subsequent chapter.

%%%%%%%%%%%%%%%%%%%%%%%%%%%%%%%%%%%%%%%%%%%%%%%%%%%%%%%%%

\section{Hydrodynamic treatment}\label{hydrodynamics}
When computing the dynamic properties of the bubble wall and the BAU it is of crucial importance to take into account the hydrodynamic equations which model the plasma behavior~\cite{No:2011fi}. These hydrodynamic equations, as written in the universe frame\footnote{By universe frame we mean a reference frame far away from the wall, either inside or outside.}, are given by \cite{Espinosa:2010hh}
\bea
\label{eq:Eulerx}
(\xi - v ) \frac{\partial_\xi e}{w} &=& 
2\frac{v}{\xi} + 
[1 - \gamma^2 v (\xi - v)]\partial_\xi v\ , \nn \\
(1 - v \xi ) \frac{\partial_\xi p}{w} &=& 
\gamma^2 (\xi - v ) \partial_\xi v.
\eea
where $e$ is the energy density, $p$ the pressure and $\omega$ the enthalpy of the plasma. The variable $\xi=r/t$ accounts for the self-similarity of the equations and has units of velocity.
% but can be thought of as a position in space. 
For example $v(\xi_w)$ is the fluid velocity at the location of the bubble wall and $\xi_w=v_w$ is the wall velocity. 

From these equations one can also obtain a differential equation for the temperature, simply by using the following identity 
\be
\label{eq:vvs0}
\frac{\partial P}{\partial T}  = \partial_{\xi}P \frac{\partial \xi}{\partial T}, 
\ee
from which it follows that the enthalpy can be written as 
\be
\label{eq:vvs0}
\omega \equiv T \frac{\partial P}{\partial T} = T   \partial_{\xi}P  \left(   \partial_{\xi}T  \right)^{-1} \\ ,
\ee
and then pluggig into \eqref{eq:Eulerx} one obtains
\be
\label{eq:temp}
 \frac{ \partial_{\xi}T}{T} = \gamma^2 \mu  \partial_{\xi}v,  \\ \quad \mu(\xi,v) = \frac{\xi - v}{1 - \xi v} . 
\ee

The derivatives $\partial_\xi e$ and $\partial_\xi p$ can be related 
through the speed of sound in the plasma, $c_s^2\equiv (dp/dT)/(de/dT)$,
so as to get the central equation describing the velocity profile:
\be
\label{eq:vel}
2\frac{v}{\xi} = \gamma^2 (1- v  \xi) 
\left[ \frac{\mu^2}{c_s^2} - 1 \right] \partial_\xi v.
\ee

The hydrodynamic fluid equations are supplemented by boundary conditions on both sides of the wall which follow from the conservation of the energy momentum tensor accross the interface, namely 
\be
\label{eq:vvs0}
v_+ v_- = \frac{p_+  - p_-}{e_+ - e_-}\ , \quad
\frac{v_+}{ v_-} = \frac{e_- + p_+ }{e_+ + p_-}, 
\ee
with $+$($-$) meaning in front (behind) of the bubble wall. These conditions are derived in the rest frame of the bubble wall. It is important to keep track of which reference frame one is referring to in order to obtain consistent solutions.  For the calculation of the BAU and the wall velocity one is interested in the thermodynamic properties of the plasma directly in front of the wall, i.e. $v_+$, $T_+$, $\alpha_+$ etc. Failure to use the correct variables lead to no solution for the bubble wall velocity \cite{Cline:2021iff} and a significant underestimate for the BAU. 

It is usually assumed that the system is well modeled by the bag equation of state which yields a relation between the plasma velocities ~\cite{Espinosa:2010hh}
\be
v_{+}=\frac{1}{1+\alpha}\left[\left(\frac{v_{-}}{2}+\frac{1}{6 v_{-}}\right) \pm \sqrt{\left(\frac{v_{-}}{2}+\frac{1}{6 v_{-}}\right)^{2}+\alpha^{2}+\frac{2}{3} \alpha-\frac{1}{3}}\right] \,. \label{v_plus}
\ee

All the possible solutions to the hydrodynamic fluid equations have been classified in \cite{Espinosa:2010hh} and there are three possibilities: 1) Deflagration solutions have a subsonic wall velocity and are preceded by a shock front discontinuity which allows the solution to go to zero. The velocity of the plasma behind the wall vanishes, i.e. $v_- = v_w$. This type of solution lives in the lower branch of the above formula. 2)  Detonations correspond to the opposite case with a vanishing velocity of the plasma in front of the wall and one has $v_+ = v_w$. In this case the solution is located in the upper branch of eqn. \eqref{v_plus} 3) Hybrid solutions are mixtures of the latter two and the velocity of the plasma is not vanishing on both sides of the wall. 
The thermodynamic variables in front of the wall, i.e.  $v_+$, $T_+$ and $\alpha_+$ are trivially found in the case of detonations. The cases of pure deflagrations and the deflagration component of hybrids however require some more work and in the %following subsection
remaining part of this section we summarize our procedure for them.

\begin{figure}[h]
 \centering
\includegraphics[width=0.32\textwidth]{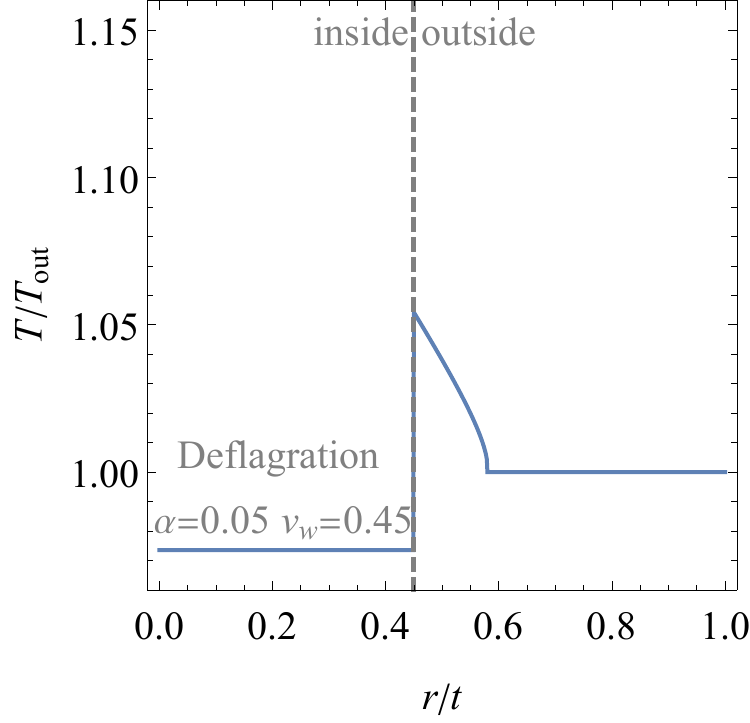}
\includegraphics[width=0.32\textwidth]{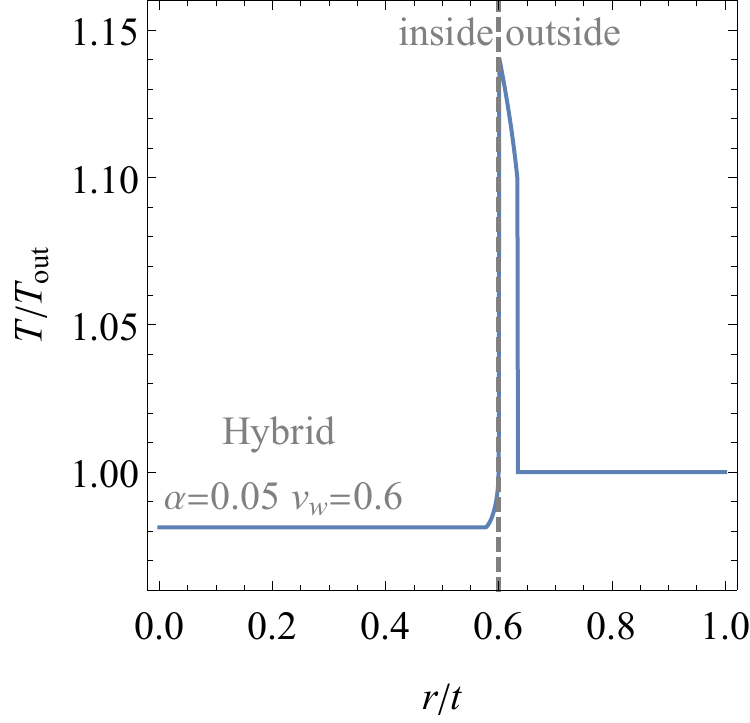}
\includegraphics[width=0.32\textwidth]{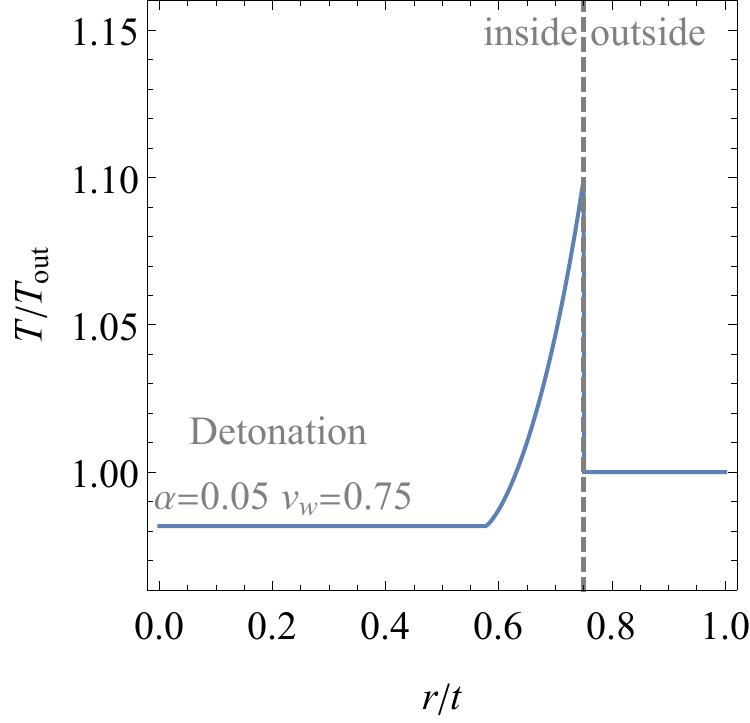}
\caption{Temperature variation across the bubble wall for a transition with fixed strength $\alpha=0.05$. The three panels correspond to $v_w=0.45$, $v_w=0.6$ and $v_w=0.75$ and three different kinds of solution corresponding to a deflagration,  a hybrid and a detonation.} \label{SteadyStatePlots}
\end{figure}

We show the temperature change across the wall for a given strength of the transition $\alpha=0.05$ for three different wall velocities realised by the three solutions discussed above in Fig~\ref{SteadyStatePlots}. The key feature of deflagrations and hybrids is that the plasma is heated and accelerated in front of the bubble wall. 
As the velocity increases for a given strength of the transition, the fluid shell around it becomes steeper and thinner. The velocity at which the shell disappears altogether and we switch to a detonation solution is given by the Jouguet velocity~\cite{Steinhardt:1981ct,Kamionkowski:1993fg,Espinosa:2010hh} 
\be \label{eq:vJ}
v_J=\frac{1}{\sqrt{3}}\frac{1+\sqrt{3 \alpha^2+2 \alpha}}{1+\alpha} \,.
\ee
This expression can be obtained from \eqref{v_plus} by taking the limit $v_- = 1/\sqrt{3}$. This quantity will be of key importance to us because if the friction does not cease the wall acceleration before this velocity the surrounding temperature drops from $T_+$ to $T_N$ . This further decreases the friction and makes finding solutions with larger velocities not possible via the semi-classical method.

%%%%%%%%%%%%%%%%%%%%%%%%%%%%%%%%%%%%%%%%%%%%%%%%%%%%%%%%%

\subsection{Deflagrations}\label{deflas_subsection}
\begin{figure}[h]
 \centering
\includegraphics[scale=0.55]{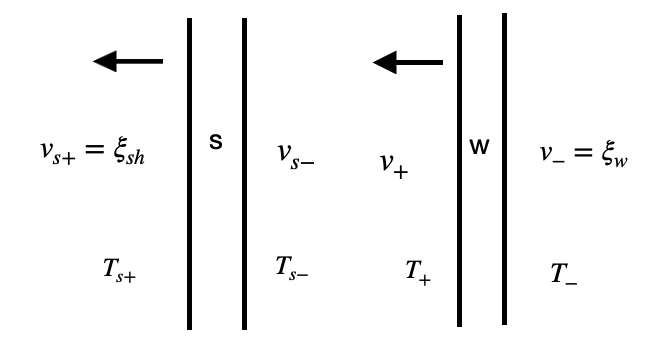}
\caption{Diagrammatic picture for a deflagration bubble (w) propagating to the left as indicated by the arrows as well as the shock front (s) propagating in front of it. The fluid velocities and temperatures measured with respect to each wall interface are also indicated.} \label{Defla}
\end{figure}

This type of solution is characterized by a subsonic bubble, i.e $v_w<c_s$ where $v_w$ is the bubble wall velocity and $c_s$ is the speed of sound in the plasma. A basic picture of this configuration is shown in figure \ref{Defla}. 

One of the consequences of a deflagration solution is the appearance of a shock front which forms in front of the bubble wall interface and causes thermodynamic quantities to suffer discontinuities. Thus one has to deal with two interface boundaries which serve as two independent inertial reference frames. The velocity of the fluid as measured in these reference frames is therefore different depending on wether the fluid is in front or behind the interface. We will denote the fluid velocities as measured with respect to the wall (shock front) frame as $v_+$ $(v_{s+})$ if the fluid is in front and as $v_{-}$ ($v_{s-}$) if the fluid is behind the front. See figure \ref{Defla} for reference.

The wall and shock front discontinuities give rise to relations between the fluid velocities. Here we  summarize them again for convenience. Across the bubble wall we have 
\be
\label{eq:wall_BC}
v_+ v_- = \frac{p_+  - p_-}{e_+ - e_-}\ , \quad
\frac{v_+}{ v_-} = \frac{e_- + p_+ }{e_+ + p_-}\ , 
\ee
while for the shock front, since the fluid on both sides of it is actually in the symmetric phase of the theory, they are more simply written by
\be
\label{eq:shock_BC}
v_{s+} v_{s-} = \frac{1}{3}\ , \quad
\frac{v_{s+}}{ v_{s-}} = \frac{T_{s+}^4 + 3 T_{s-}^4 }{T_{s-}^4 + 3T_{s+}^4}\ . 
\ee

Here it is useful to pause a moment to present a notation which was introduced in \cite{Kozaczuk:2015owa} to denote the fluid velocities as measured with respect to the center of the wall. Let us denote with $v_{\pm}$ ($v_{s\pm}$) the fluid velocities in the wall (shock front) frame while we will use tildes $\tilde{v}_{\pm}$ to refer to fluid velocities in the fluid reference frame. This has also been called the ``frame of the universe" in refs. \cite{Huber:2013kj} (and more recently in \cite{Friedlander:2020tnq}) in which the fluid far ahead  and behind of the bubble wall and shock front is at rest. Therefore the universe frame might also refer to a region of spacetime far ahead of the shock front where the fluid velocity vanishes. Here whenever we mention the universe frame we will simply mean the reference frame of the center of the bubble. 

The fluid velocities in both frames are related by Lorentz transformations, i.e.  
\be
\label{eq:vvs0}
\tilde{v}_{\pm} = \frac{v_w  - v_{\pm}}{1- v_w v_{\pm}}\ , \quad
\tilde{v}_{s\pm} = \frac{v_{sh}  - v_{s \pm}}{1- v_w v_{s\pm}} ,
\ee
where $v_w$ and $v_{sh}$ are the relative velocities between the wall and shock front and the center of the bubble, respectively.

Now we can give a more precise definition of a  deflagration. In this case the fluid velocity behind the wall is zero, i.e,
\be
\label{eq:vvs0}
\tilde{v}_{-} =0\ \quad \longrightarrow   \quad   v_w  = v_{-}\\ .
\ee

Similarly the necessity of having a shock front is that we require that the fluid velocity jumps to zero in front of it. So we get 
\be
\label{eq:vvs0}
\tilde{v}_{s +} =0\ \quad \longrightarrow     \quad  v_{sh}  = v_{s+}\\ .
\ee
Since we are interested in the fluid velocity and temperature directly in front of the wall we need to solve their corresponding hydrodynamic equations 
\be
2\frac{v}{\xi} = \gamma^2 (1- v  \xi) 
\left[ \frac{\mu^2}{c_s^2} - 1 \right] \partial_\xi v, \label{vel_eqn}
\ee
\be
 \frac{ \partial_{\xi}T}{T} = \gamma^2 \mu  \partial_{\xi}v  \label{Temp_eqn} \\ .
\ee
Given $\alpha$ and some value for the wall velocity $v_w<c_s$, the boundary condition on the velocity equation is 
\be
 \tilde{v}_+ \equiv \frac{\xi_w - v_+}{1-\xi_w v_+} = v(\xi_w),
 \ee
 with $\xi_w = v_w = v_-$ the wall velocity and $v_+$ is determined by eq. \eqref{v_plus} with the minus sign in front of the radical as is the case for deflagrations. The form of eq. \eqref{vel_eqn} does not accept analytic solutions and a standard numerical integration would stop at the singular point $\mu^2 = c_s^2$. However the solution actually stops before reaching the singularity and the final point determines the velocity of the shock front which satisfies the relation:
 \be
 \label{shock_front_location}
 \tilde{v}_{s-} \equiv \frac{\xi_s - v_{s-}}{1-\xi_s v_{s-}} = v(\xi_s)
 \ee
 with $\xi_s = v_{sh} = v_{s+}$ the position of the shock front. Notice that the equation above has two unknowns, namely $\xi_s$ and $v_{s-}$. This relation can be supplemented with the jump in velocities at the shock front eqn. \eqref{eq:shock_BC} as follows
  \be
v_{s-} =  \frac{\xi_s - v(\xi_s)}{1-\xi_s v(\xi_s)} = \frac{1}{3 \xi_s}, \quad \rightarrow \quad  \mu(\xi_s, v(\xi_s)) \xi_s= \frac{1}{3} = c_s^2
 \quad  (\text{shock front position}),
 \ee
 where on the left we Lorentz transformed \eqref{shock_front_location} in favor of $v_{s-}$.
 Having solved for the velocity profile, the fluid velocities behind and in front of the wall and the shock front are completely specified. To determine the corresponding temperatures one needs to solve the equation for the temperature profile, eqn. \eqref{Temp_eqn}. One plugs in the solution $v(\xi)$ into the temperature equation and integrate. The boundary conditions relate the values of the temperature outside the wall $T_+$ with the temperature inside the shockfront $T_{s-}$. In reality what is fixed is their ratio, i.e.
 \be
  \label{T_integral}
\frac{T_+}{T_{s-}} =\exp{\left[ \int_{v_w}^{v_{sh}} d\xi \gamma^2 \mu \partial_{\xi}v \right]} =  \exp{\left[  \int_{\xi_{w}}^{\xi_{\text{shock}}}d\xi \frac{2c_s^2 v (\xi - v )}{\xi \left( (\xi-v)^2 - c_s^2 (1-v \xi)^2  \right)}  \right]}.
\ee
In addition one has to satisfy the boundary conditions at the shock front eqn. \eqref{eq:shock_BC}. From these one can eliminate $v_{s+}$ to find 
\be
\label{shock_bc}
\frac{T_{s-}^4 }{T_{N}^4}= \frac{3(1-v_{s-}^2)}{9v_{s-}^2-1} ,
\ee
where we notice that the temperature in the region in front of the shock front corresponds to the nucleation temperature of the transition, i.e. $T_{s+} = T_N$. Using \eqref{T_integral} and \eqref{shock_bc} we can write
\be
\frac{T_{+} }{T_{N}} = \left(  \frac{3(1-v_{s-}^2)}{9v_{s-}^2-1}  \right)^{1/4}  \exp{\left[  \int_{\xi_{w}}^{\xi_{\text{shock}}}d\xi \frac{2c_s^2 v (\xi - v )}{\xi \left( (\xi-v)^2 - c_s^2 (1-v \xi)^2  \right)}  \right]}.
\ee
Notice that in the region of integration in the above formula the latent heat is given by its value in front of the bubble wall, $\alpha_+$ which is different from $\alpha_N$ that is calculated from the phase transition properties . To find the correct $\alpha_+$ we iterate the above procedure for $T_+$ until the condition $\alpha_+ T_+^4 = \alpha_N T_N^4$ is satisfied. This is a good approximation for the bag equation of state with equal number of degrees of freedom on each side of the wall. Once $\alpha_+$ is found, $v_+$ is fixed by~\cite{Espinosa:2010hh}
\be
v_{+}=\frac{1}{1+\alpha_{+}}\left[\left(\frac{v_{-}}{2}+\frac{1}{6 v_{-}}\right) \pm \sqrt{\left(\frac{v_{-}}{2}+\frac{1}{6 v_{-}}\right)^{2}+\alpha_{+}^{2}+\frac{2}{3} \alpha_{+}-\frac{1}{3}}\right] \, .
\ee

\subsection{Detonations}
Detonations constitute the opposite case to deflagrations, namely the bubble wall is supersonic $v_w>c_s$ and the fluid velocity in front of the wall vanishes, i.e. $\tilde{v}_{+} =0$ which means $v_w = v_+$. That is, the wall velocity equals the fluid velocity in front of the wall. For the nucleation temperature and strength of the transition it also follows that $T_N = T_+$ and $\alpha_N = \alpha_+$.  For our purposes this case represents the most  trivial one
in regard to the BAU and the wall velocity calculation. We briefly review this case for completeness.

The fluid velocity behind the wall can be obtained by inverting \eqref{v_plus} in favor of $v_-$,
\be
v_- =  \frac{ 1-3 \alpha + 3 v_+^2 (1+\alpha) + \sqrt{ -12 v_+^2 + (1-3 \alpha  + 3 v_+^2(1+\alpha))^2 }  }{6v_+},
\ee
where we took the plus sign in the radical by definition. Contrary to the case of deflagrations which do not require further thinking in applying formula \eqref{v_plus} for finding the fluid velocity in front of the wall, for detonations the above formula hides a non trivial constraint on $\alpha$ and $v_w$; we must require that the velocity is positive and that the term inside the square root do not become negative. It can be proven that the necessary and sufficient condition that satisfy these constraints is given by 
\be
\alpha \leq \frac{(1-\sqrt{3} v_+)^2}{3(1-v_+^2)}, \quad (\text{consistency condition}).
\ee
This equation can be interpreted as an upper bound for the transition strength given the wall velocity. However we find this explanation to be  counter-intuitive given the pipeline structure of our calculations where we first find the strength of the transition and after that we calculate the wall velocity. Thus we prefer to express the last equation in a form that is most suitable for our motives. Using $v_+ = v_w$ we can write it as
\be
0 \leq (v_w -v_J^+)(v_w - v_J^-), \quad 
v_J^{\pm}=\frac{1}{\sqrt{3}}\frac{1\pm\sqrt{3 \alpha^2+2 \alpha}}{1+\alpha} \label{consistency condition_2},
\ee
and since $v_J^+> v_J^-$ we thus see that the Jouguet velocity encountered at the beginning of this section, eqn. \eqref{eq:vJ} indicates the lower bound for the wall velocity for which detonations can be found.  Within the range $c_s<v_w<v_J$ neither deflagrations nor detonations can be obtained. This is the range of velocities for hybrids which we discuss next.

\subsection{Hybrids}
Now we discuss the computation of the thermodynamic variables in front of the wall $v_+$, $T_+$, $\alpha_+$ for the case of hybrid solutions. As their name suggests, this type of solution has features of both deflagrations and detonations. The wall velocity in this case is not identified with neither $v_+$ nor $v_-$ and it falls within the range  $c_s< v_w \leq v_\text{J}$, the reason they are also called supersonic deflagrations. 

The hybrid case is therefore a superposition of a deflagration and a detonation. For the detonation part we need to fix $v_+ = v_J^-$ in order to satisfy the consistency condition in eqn. \eqref{consistency condition_2}, doing so automatically fixes $v_-=c_s$. The initial condition for this part is found by Lorentz transforming to the fluid rest frame, i.e. $v(\xi_w)=\mu(\xi_w , v_-)$, see eqn. \eqref{eq:temp}. The deflagration component is solved by fixing as initial condition the alternative Lorentz transformation, i.e. $v(\xi_w)= \mu(\xi_w , v_+)$. The total solution is found by patching these two. 

To obtain the thermodynamic variables in front of the wall in the hybrid case we apply the same procedure explained in section \ref{deflas_subsection} to iteratively solve for $\alpha_+$, $T_+$ and $v_+$.

\begin{figure}[h]
 \centering
\includegraphics[scale=0.35]{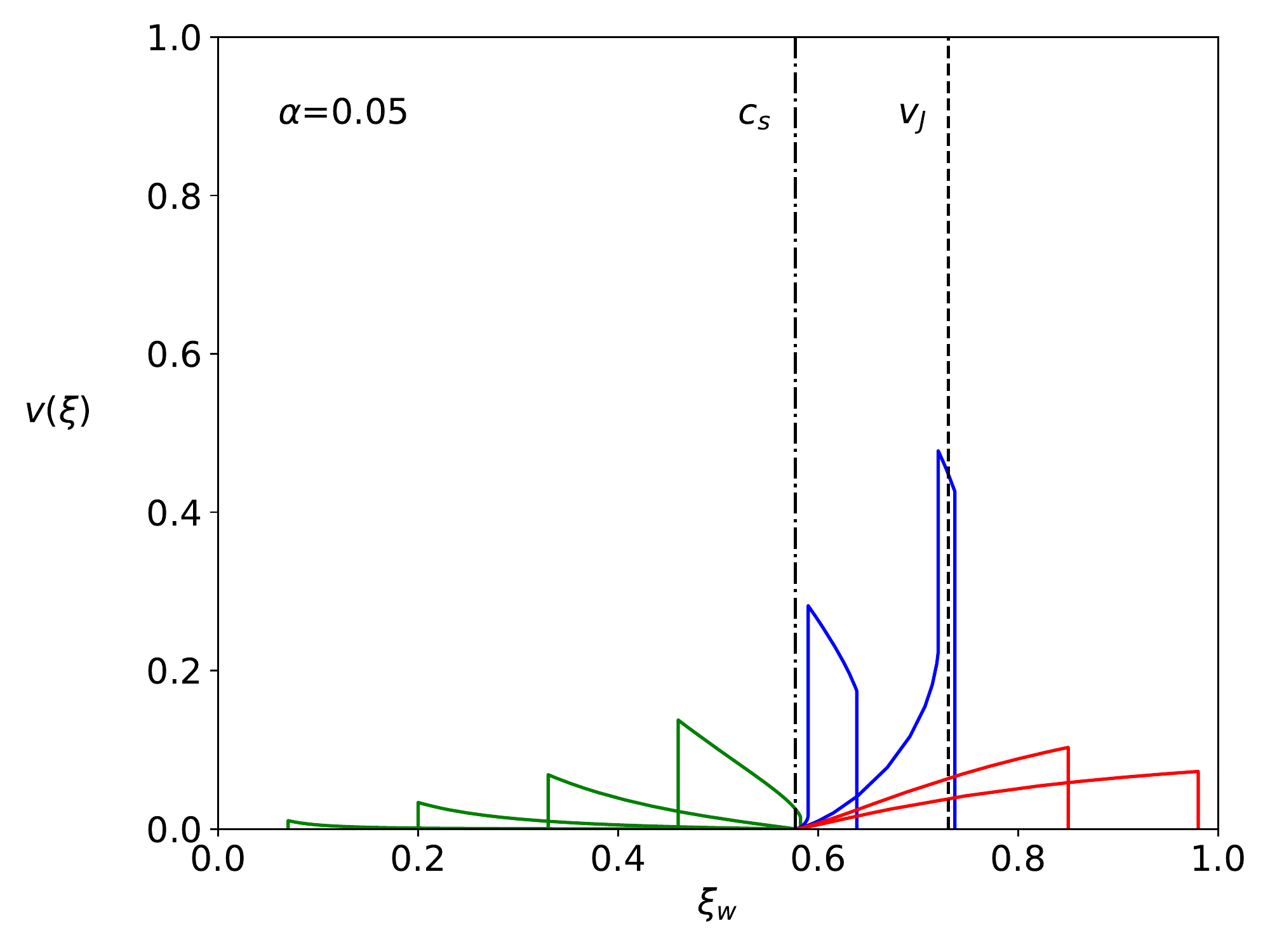}
\includegraphics[scale=0.35]{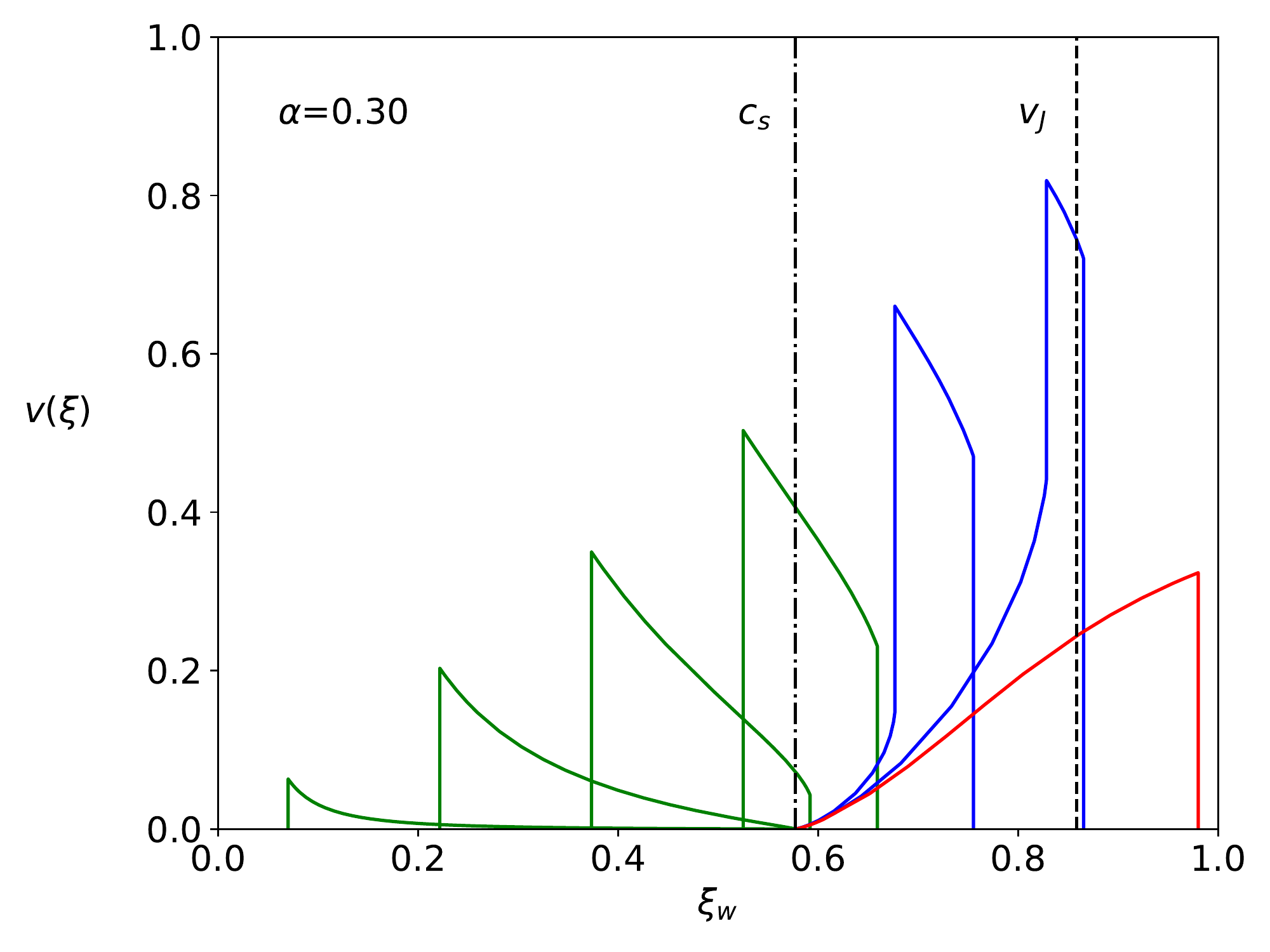}
\includegraphics[scale=0.35]{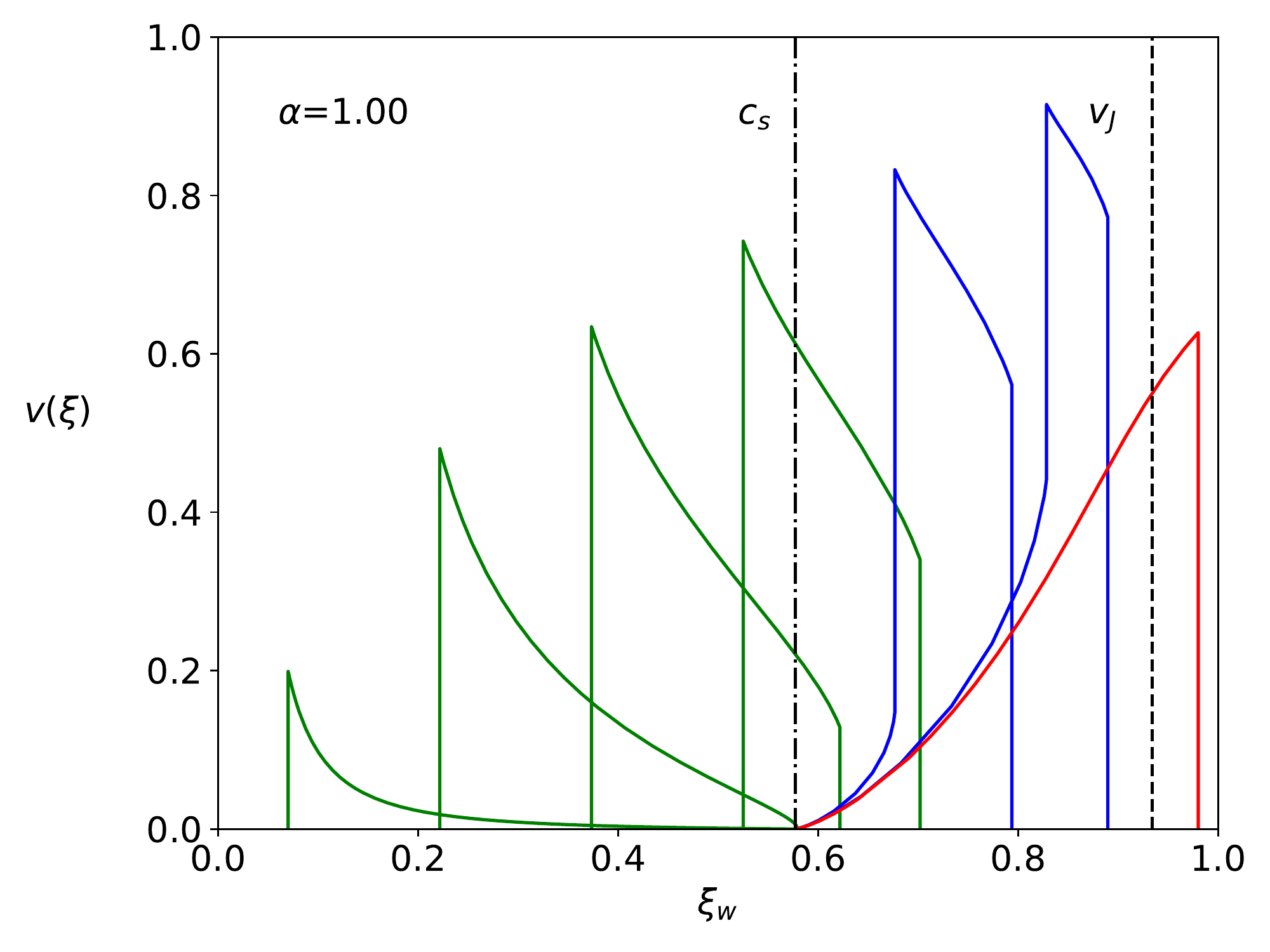}
\caption{The hydrodynamics solutions for the velocity profile of the plasma around the bubble wall for $\alpha = 0.05$ in the upper left, $\alpha = 0.3$ in the upper right and $\alpha = 1$ in the bottom. Green curves are deflagrations, blue colored curves hybrids and red curves correspond to detonations.} \label{Hydro_dynamic_solutions_velocity_profile}
\end{figure}

To conclude this section we show examples of all three types of solutions for different values of $\alpha$ in figure \ref{Hydro_dynamic_solutions_velocity_profile}.  The wall interface is located at the maximum of each curve.  For reference we show in each figure the speed of sound and the Jouguet velocity as vertical dot-dashed and dashed lines respectively 

%%%%%%%%%%%%%%%%%%%%%%%%%%%%%%%%%%%%%%%%%%%%%%%%%%%%%%%%%%%%%%%%%%%%%%%%%%%%%%%%%%%%%%%%%%%%%%%%%%%%%%%%%%%%%%%%%%
\section{Electroweak baryogenesis and bubble wall expansion} \label{semi_classical}

\subsection{Review on updated transport equations}
One of the most popular ways to account for the BAU is realised through EWBG, for a complete review see Ref.~\cite{Morrissey:2012db}. In EWBG a first order phase transition is needed to provide the necessary out of equilibrium condition. 
The phase boundary exerts a force on the particles in the plasma perturbing their distribution functions from their equilibrium forms. 
Due to CP violation in the underlying theory, the form of this force depends on which particles it is acting upon, e.g. particles and anti-particles (as well as different helicity states) experience the same force but with opposite signs. Thus the effect of CP violation is to cause an asymmetry in the population densities of particles in front of the wall. This asymmetry biases the non-perturbative sphaleron rates to produce more baryons than anti-baryons. Then as the phase transition completes and the plasma is swept inside the bubble the sphaleron processes turn-off and the total baryon number  freezes to its present value.

For quite some time it was commonly accepted that EWBG could only work if the bubble wall velocity is small $v \ll c_s$, with $c_s$ the velocity of sound in the plasma, since otherwise the diffusion of the particle asymmetries would not be efficient. The claim~\cite{Moore:1995si}, that the wall velocity would have been subsonic in the SM, further motivated the derivation of the transport equations for particles asymmetries using the small velocity approximation $v \ll 1$. This approximation leads to transport equations which breakdown for wall velocities close to the speed of sound \cite{Fromme:2006wx}. Recently, Cline and Kainulainen \cite{Cline:2020jre} have re-derived improved transport equations for arbitrary wall velocities arguing that the process of particle diffusion is not dependent on the propagation of sound waves in a plasma and thus even for highly relativistic walls one would expect a significant fraction of particles traveling faster than the wall and contributing to the diffusion tail.  Using a scalar singlet model they showed that the baryon to entropy ratio is actually a continuous function of the wall velocity asymptotically reaching zero as $v \rightarrow 1$. Furthermore the transport equations are well behaved for velocities close to the speed of sound. 

The Boltzmann transport equations for the out of equilibrium distribution functions can be separated into CP-even and CP-odd equations which can be solved independently of each other. While the CP-odd equations are suitable for computing the BAU, their CP-even counterparts can be used to predict the bubble wall velocity and its shape. These equations have also been recently improved for arbitrary wall velocities by Cline and Laurent in ~\cite{Laurent:2020gpg}. One crucial point of Refs.~\cite{Cline:2020jre, Laurent:2020gpg} has been to treat the velocity perturbations differently than the chemical potential and temperature fluctuations. In this case one does not assume a specific form for the velocity perturbation and instead uses a~$\textit{factorization}$ assumption which allows one to factor out the velocity perturbation $u$ from non-trivial integrals which appear when taking moments of the Liouville term. Also, in the case of the CP-odd equations, the authors used a $\textit{truncation} $ scheme which related higher moments of $u$ to the first moment in a linear fashion.  As recognized in~\cite{Cline:2020jre, Laurent:2020gpg}, the factorization and truncation prescriptions, although leading to transport equations which do not breakdown for velocities close to the speed of sound,  are still arbitrary.  These results, however, have been recently corroborated in~\cite{Dorsch:2021ubz} that used a fluid ansatz with higher orders in the momentum expansion and found that baryogenesis could be realized even if the wall velocity is supersonic. 

%%%%%%%%%%%%%%%%%%%%%%%%%%%%%%%%%%%%%%%%%%%%%%%%%%%%%%%%%%%%%%%%%%%%%%%%%%%%%%%%%%%%%%%%%%%%%%%%%%%%%%%%%%%%%%%%%%%%%%%%%%%%%%%%%%%%%%%%%%%%%%%%%%%%%%%%%%%%%%%%
\subsection{The semi-classical fluid approximation}
\label{Sec:FluidApproximation}

In the semiclassical treatment as introduced in the seminal papers by Moore and Prokopec  \cite{Moore:1995ua, Moore:1995si} the Higgs EOM receives an extra contribution due to deviations from equilibrium in the particle distribution functions. This extra term is interpreted as a friction component and the distribution functions are calculated by means of Boltzmann equations. 

The fluid approximation relies on the following assumptions: 1) that the deviations from equilibrium are small enough and the system of equations can be linearized, 2) the time scale for the phase transition to complete is much smaller than the inverse Hubble rate and one is entitled to neglect the expansion of the universe which justifies the use of ordinary derivatives instead of the full covariant form of general relativity, 3) the de-Broglie thermal wavelength of particles in the system is smaller  than the bubble wall thickness, this guarantees that the WKB approximation is valid for calculating the dispersion relations and group velocities.  In addition we take it for granted that a steady state regime is achieved such that the wall has a well defined rest frame and furthermore that the bubble becomes sufficiently large for it to be treated as a planar surface. 

 In the following we will introduce the modified fluid equations as derived in  \cite{Cline:2020jre, Laurent:2020gpg} for completeness. Starting from the Boltzmann equation which dictates the time evolution of the particle distribution of species $a$ as
\be
\label{Boltzmann}
\frac{df_a}{dt} = \partial_t f_a  + \dot{\vec{x}} \cdot \partial_{\vec{x}} f_a +  \dot{\vec{p}} \cdot \partial_{\vec{p}} f_a   = C[ f_a ],
\ee
where the dots denote time derivative. In the steady state the bubble has expanded sufficiently so that it can be viewed as a planar interface and the problem can be reduced to one dimension, the direction of the wall propagation . We work in the rest frame of the bubble wall where the first term on the left hand-side of the Boltzmann equation vanishes as the solution becomes static. The $\textit{fluid ansatz } $ for the distribution function is written as 
\be
\label{ansatz}
f \approx f_v - f_v' \delta \bar{X}   + \delta f_u + \mathcal{O}(\delta f^2),
\ee
with
\be
 f_v= \frac{1}{e^{\beta \gamma (E - v p_z) } \pm 1}, \quad f_v' \equiv \frac{d f_v}{ d \beta \gamma E},
\ee
where this form makes manifest the Lorentz transformation to the wall frame and 
\be
\delta \bar{X} = \mu  + \beta \gamma \delta \tau (E - v p_z), 
\ee
encodes the perturbations from equilibrium. The variable $\mu$ is the chemical potential and $\delta \tau$ is the temperature perturbation. The extra term $\delta f_u$ gives rise to the velocity perturbation and its form remains completely undetermined. The factorization prescription amounts to  
\be
\int d^3p \ Q \ \delta f_u \rightarrow u  \int d^3p \ Q \ \frac{E}{p_z} f_v
\ee
for any prefactor $Q$ that may appear when we take moments of the Boltzmann equation. As we mentioned above in the introduction, this prescription is arbitrary but it can be justified a posteriori by obtaining transport equations which are well behaved for all velocities. 

The velocity and the force on the particle follow classical Hamiltonian equations of motion. For models where CP violation can be written as complex phases in fermionic mass terms, i.e., $ m(z)  = |m(z)| e^{i \gamma_5 \theta(z)}$, the dispersion relation computed using the WKB approximation \cite{Cline:2000nw} gives
\be
\label{Dispersion}
\dot{z}  \equiv \frac{\partial \omega}{\partial p_z } = \frac{p_z}{E} + s \frac{m^2 \theta'}{2 E^2 E_z}    , \quad \quad \dot{p_z}  \equiv -\frac{\partial \omega}{\partial z} =- \frac{(m^2)'}{2 E} + s \frac{(m^2 \theta')'}{2E E_z} ,   
\ee 
where primes denotes derivatives with respect to the transverse direction to the wall, $\omega$ is the energy of the WKB wave packet and $E_z^2 \equiv p_z^2 + m^2$. The variable $s=0$ ($1$) for particles (anti-particles). In the derivation of the above equations an expansion in gradients $\partial_z$ was assumed. Thus we see that CP violation appears at higher order in gradients and one can separate the Boltzmann equation as well as perturbations into CP even and CP odd components which can be independently solved.  Solution to the CP even equations are connected to the bubble wall properties while the CP odd ones are useful for the computation of the BAU. In the following two subsections we present the most relevant formulas for each case.

\subsubsection{CP-even equations: bubble wall properties}\label{Wall_eqns}

 The derivation of the bubble wall properties entails solving the transport equations for the CP even perturbations. These can be obtained by inserting the force and group velocity of eq. \eqref{Dispersion} into the Boltzmann equation \eqref{Boltzmann}  
 \be
\left[  \frac{p_z}{E} \partial_z   -   \frac{(m^2)'}{2 E}   \partial_{p_z} \right] \left(  f_v   -f_v' \delta \bar{X} + \delta f_u \right)  = C[f],
\ee
where we used the fluid ansatz of eq. \eqref{ansatz}. Since this is a partial integro-differential equation with momentum and space-time dependence some form of massaging is necessary to obtain a tractable system of equations. First notice that in the fluid ansatz of eq. \eqref{ansatz} three parameters were introduced, namely $\mu$, $\delta \tau $ and $u$. Therefore it is customary to take moments of the Boltzmann equation for each variable that is introduced. The choice of weight factors for these moments is also somewhat arbitrary. In the improved fluid equations of \cite{Laurent:2020gpg} these were chosen as 
\be 
\int d^3p \frac{1}{T^3}, \quad \int d^3p \frac{E}{T^4}, \quad \int d^3p \frac{1}{T^3}\frac{p_z}{E}.
\ee
After some algebra and by grouping the perturbations in a vector object as $q =(\mu ,\delta \tau, u)^T$ the simplified transport equations take the form 
\be
\label{Transport}
A_v \vec{q} \ '  + \Gamma \vec{q} = S,
\ee
with 
\be
 A_v = \begin{pmatrix}
    C_v^{1,1} & \gamma v C_0^{-1,0} & D_v^{0,0}  \\
    C_v^{0,1} & \gamma(C_v^{-1,1}-v C_v^{0,2}) & D_v^{-1,0} \\
    C_v^{2,2} & \gamma(C_v^{1,2}-v C_v^{2,3}) & D_v^{1,1}
\end{pmatrix},
\ee
and 
\be
S =
\gamma v\frac{(m^2)'}{2T^2} \begin{pmatrix}
C_v^{1,0} \\
    C_v^{0,0} \\
    C_v^{2,1} 
    \end{pmatrix},
\ee
being the source term. The coefficients in the $A$ matrix and in the source term are defined as non-trivial integrals of the particle distribution functions. They are given in equation $(8)$ of \cite{Laurent:2020gpg}.  The $\Gamma$ matrix  arises from treating the collision term and the numerical expressions were recomputed in the $\textit{leading-log}$ approximation in the same reference. For more details we refer the reader to this reference.  

In principle, the linearized transport equations presented above have to be solved for every particle in the plasma that is expected to contribute significantly to friction.  For the SM extensions considered in this paper, the particles which contribute the most to the friction term are the heaviest SM particles, namely the top quark and the $W$ and $Z$ bosons. The remaining particles do not couple significantly to the Higgs and their interactions are assumed to be very efficient so that they equilibrate quickly and form a thermal background with a $z$ dependent temperature and velocity. The effect of this background is taken into account by taking $\delta \tau \rightarrow \delta \tau  + \delta \tau_{\text{bkgnd}}$, $u \rightarrow u  +u_{\text{bkgnd}}$, where a $z$ dependence on each term should be understood. This effect gives rise to additional transport equations for the background perturbations. 

The Higgs EOM in the presence of out of equilibrium particle populations is given by \cite{Moore:1995ua, Moore:1995si, Konstandin:2014zta, Hindmarsh:2020hop}
\be
\label{Higgs_EOM}
 E_h \equiv \Box \phi   + \frac{dV_{\text{eff}}(\phi,T)}{d\phi} + \sum_{i} \frac{d m_i^2}{d \phi} \int \frac{d^3 p }{(2\pi)^3}\frac{ \delta f_i(p,x) 
}{2 E} = 0,
\ee
where the sum in the last term is over all particles that receive mass from the Higgs condensate. This term is interpreted as a frictive force which must be carefully estimated for an accurate prediction of the bubble wall properties. Equations \eqref{Transport} and \eqref{Higgs_EOM} form a consistent set of constraints for the out equilibrium particle distributions and for the Higgs condensate.
 In the presence of additional scalars, as in the case of extensions of the SM, one has to supplement this system with the corresponding equations of motion for each extra scalar. For the scalar singlet model considered in this paper one has to satisfy, in addition to \eqref{Higgs_EOM}, 

\be
\label{Singlet_EOM}
 E_s \equiv - s''   + \frac{\partial V_{\text{eff}}(h,s,T)}{\partial s}  = 0,
\ee
which doesn't have a frictive term because the singlet doesn't contribute to the Higgs mechanism of mass generation. 

Calculation of the bubble wall velocity in the scalar singlet extension of the SM, see Ref.  \cite{Friedlander:2020tnq}, has shown that the scalar profiles can be well approximated by $\tanh$ functions. This form is well justified since the instanton solutions for bubble nucleation follow this shape so that scalar fields interpolate continuously across the bubble interface.   Thus we use the expressions  
\be
h(z) = \frac{h_0}{2 }\left[ \tanh{\left( \frac{z}{L_h} \right)} + 1\right],
\ee
\be
s(z) = \frac{s_0}{2 }\left[1-  \tanh{\left( \frac{z}{L_s} - \delta_s \right)} \right],
\ee
with $L_h$, $L_s$ denoting the bubble wall thicknesses and  $\delta_s$ an extra off-set factor. The field values $h_0$ and $s_0$ do not necessarily correspond to the vevs obtained by the bubble nucleation calculation since the form of the bubble after it has reached a steady state velocity is in principle different than its form during nucleation. We do expect however these latter values not to be too different in magnitude from the former. 

By using the $\tanh$ ansatz in the field profiles one cannot expect that the EOMs to be satisfied everywhere in space. Instead one can impose that the EOM has  two vanishing moments, i.e. 
\be
\label{moment_1}
M_1 \equiv \int dz E_h h' dz = 0,
\ee
\be
\label{moment_2}
M_2 \equiv \int dz E_h h' [ 2 h(z) - h_0 ] dz = 0. 
\ee
The form of $M_1$ above has physical intuition as it corresponds to the total pressure acting on the wall. It must vanish for a steady state wall \cite{Moore:1995ua,Moore:1995si}. The second condition\footnote{This choice of $M_2$ differs from the one first introduced in Refs. \cite{Moore:1995ua,Moore:1995si} but we believe that the final solution for $v_w$ should be only mildly dependent on which form of $M_2$ one adopts. } is interpreted as pressure gradients which mostly depend on the wall thickness. 

To obtain the wall-velocity we follow the methodology introduced in \cite{Cline:2021iff}. Although this was introduced for the scalar singlet extension, it can however be applied to other (not very exotic) SM extensions as well. Here we summarize the workflow :

\begin{itemize}
\item{} For a given set of model parameters;  $m_s$, $\lambda_{hs}$ and $\lambda_s$ we calculate the thermodynamic quantities of the nucleation process using the ComoTransitions code  \cite{Wainwright:2011kj}.
\item{} Having the nucleation temperature $T_N$ and the latent heat of the transition $\alpha_N$ we perform a grid scan in the $v_w$, $L_h$ plane. At each point in this grid we perform the next steps: 

\subitem{1.}  Solve the hydrodynamic equations for the energy budget of the FOPT, see Sec~\ref{hydrodynamics}. %Ref. \cite{Espinosa:2010hh}. 
This allows us to obtain the thermodynamic parameters evaluated in front of the wall, that is $T_+$, $\alpha_+$, $v_+$.

\subitem{2.}  Modify the field amplitudes so that they satisfy the correct minimization conditions with the effective potential evaluated at the temperature in front of the wall, with 
\be
\label{minmization_1}
\frac{dV_{\text{eff}}(h,0,T_+)}{dh} \Big{|}_{h=h_0}   =   \frac{dV_{\text{eff}}(0,s,T_+)}{ds}\Big{|}_{s=s_0} = 0.
\ee 

\subitem{3.} Solve the transport equations \eqref{Transport} for the perturbations from equilibrium. This step determines the friction force in the Higgs EOM. 

\subitem{4. } Update the Higgs amplitude $h_0$ by requiring the Higgs EOM to be satisfied deep inside the bubble, that is
\be
\label{minmization_2}
    \frac{\partial V_T(h,0, T_+)}{\partial h} \Big{|}_{h=h0}  + \sum_{i} \frac{d m_i^2}{d h} \int \frac{d^3 p }{(2\pi)^3}\frac{ \delta f_i(p,z) 
}{2 E} \Big{|}_{z \rightarrow \infty}= 0.
\ee

\subitem{5.}  Satisfy the singlet EOM by minimizing its action with respect to $L_s$ and $\delta_s$,  
\be
\label{minmization_3}
S_{\text{scalar}} = \frac{s_0^2}{6 L_s}  +  \int dz \left[  V_{\text{eff}}(h,s,T_+)   -    V_{\text{eff}}(h,s^*,T_+) \right],
\ee
where $s^*$ is the scalar singlet profile with its parameters fixed as $L_s^* = L_h$ and $\delta_s^* = 0$.

\subitem{6.} Recompute the perturbations \eqref{Transport}  and calculate $M_1$ and $M_2$.
\end{itemize} 

\begin{figure}[h]
\label{Moment_grid}
\includegraphics[scale=.49]{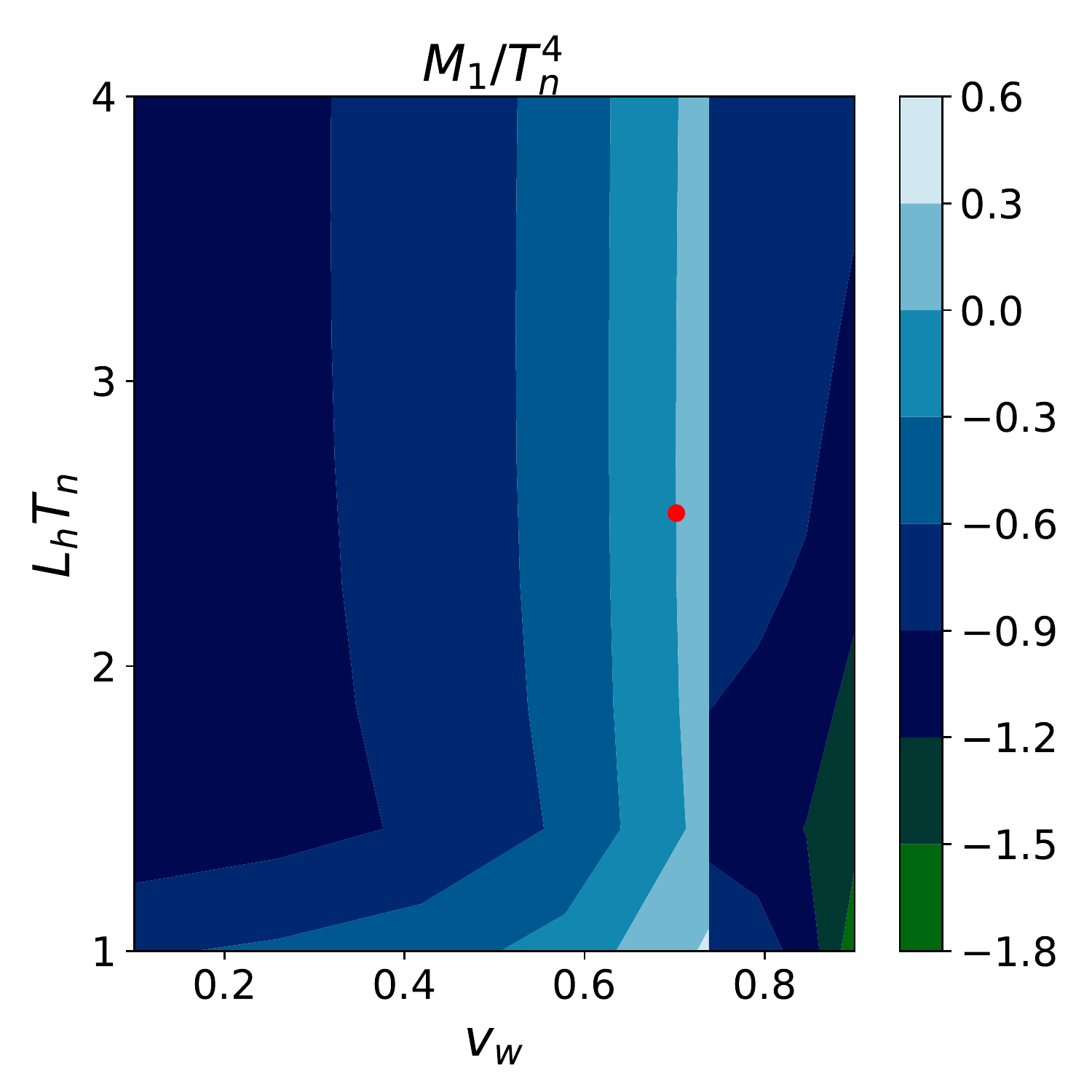}
\includegraphics[scale=.49]{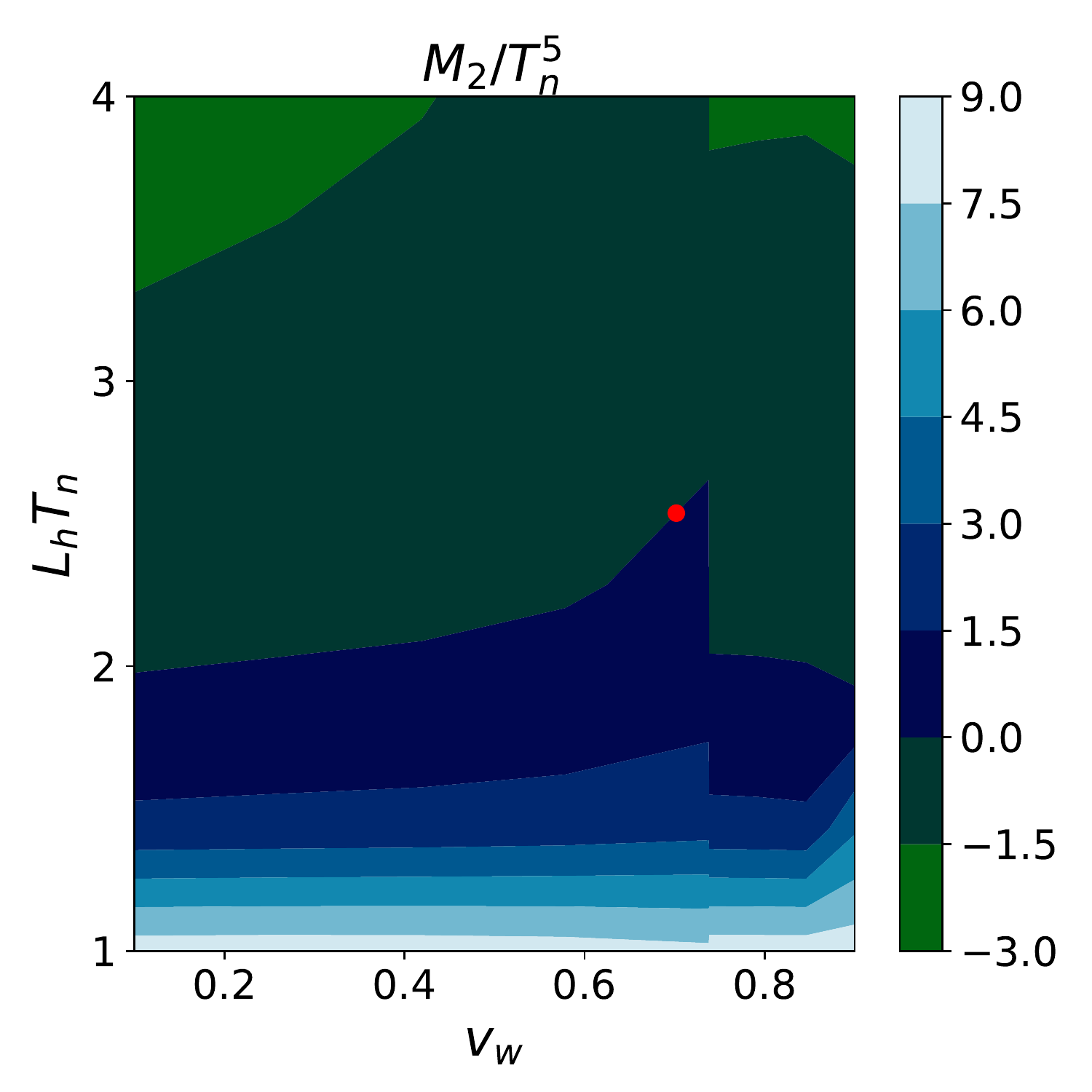}
\centering
\caption{Moment grid for a model with $m_s = 80$ GeV, $\lambda_{\text{hs}} = 0.84$ and $\lambda_{\text{s}}=1 $. The red point gives the minimum of the quadrature $f(v_w, L_h) = M_1^2 + M_2^2$.}
\end{figure}

The outcome of this algorithm for a single parameter space point is presented in figure \ref{Moment_grid} for a model with $m_s = 80$ GeV, $\lambda_{\text{hs}} = 0.84$ and  $\lambda_{\text{s}}=1 $. It shows filled contour plots with color maps for $M_1$ and $M_2$ normalized by appropriate factors of temperature. The red point on both plots shows the minimum of the scalar function $M_1^2 + M_2^2$.

The method of mapping the moments presented above is reliable in determining if there is a solution to eqns. \eqref{moment_1} and \eqref{moment_2}. Unfortunately to obtain an acceptable level of precision it sometimes requires too many iterations which are time consuming. In order to achieve better precision one can use this mapping only to produce an initial guess for $v_w$ and $L_h$ and then feed this guess into a root finding algorithm.

%%%%%%%%%%%%%%%%%%%%%%%%%%%%%%%%%%%%%%%%%%%%%%%
\subsubsection{CP-odd equations: BAU }
As it has been mentioned in previous chapters, the CP-odd contribution to the Boltzmann equations is what is needed to compute the BAU. In this subsection we present the simplified set of equations that one has to solve to achieve this goal. Here we only describe how the equations are obtained. For a full derivation we refer the reader to Cline and Kainulainen \cite{Cline:2020jre}.

The starting point is \eqref{Boltzmann} and now one includes in the group velocity and force the CP-violating contribution, i.e. the terms proportional to $\theta'$ in eqns.\eqref{Dispersion}. At the same time the energy in the distribution function receives a CP-violating contribution of the form $E \rightarrow E  - s \frac{m^2 \theta'}{2E E_z}$, with $E\equiv \sqrt{\vec{p}^2 + m^2}$, the usual relativistic dispersion relation and $E_z \equiv \sqrt{p_z^2  + m^2}$. Since CP-violation is a higher order effect in an expansion in derivatives we separate the chemical potential and the velocity perturbation in the following form
\bea
\label{CP_odd_transport}
\mu &\equiv& \mu_e + s \mu_o ,
\nn\\
\delta\! f &\equiv& \delta\! f_e + s \delta\! f_o ,
\eea
with the subscript $e$ ($o$) standing for even (odd). To obtain the CP-odd transport equations, moments from the Boltzmann equations were taken as follows 
\be
\frac{1}{N_1}\int d^3 p, \quad \ \frac{1}{N_1} \int d^3p \frac{p_z}{E}, 
\ee
with $N_1 \equiv -\frac{2}{3}\pi^3 T^2 \gamma_w$. Additionally, for the BAU one does not need to include a temperature fluctuation and the linearized Boltzmann equations for a particle species take the form 
\be
A w' + (m^2)'Bw = \! S + \delta C,
\label{CP_odd}
\ee
where $w = (\mu,u)^T$ and 
\be
  A = \left( \!\! \begin{array}{cc}-D_1 &  1\\
	                               -D_2 &  R \end{array}\!\right)
\,,\quad 
  B = \left( \! \begin{array}{cc}-v_w\gamma_w Q_1&   0 \\
	                               -v_w\gamma_w Q_2 &  \bar R \end{array}\!\right)
\label{eq:AB},
\ee
with coefficients that depend non-trivially on velocity and are given by integrals of the distribution function. The source terms $S_l = (S_1, S_2)^T$ are given by 
\be
S^o_{h\ell} = v_w \gamma_w s \big[ (m^2\theta^\prime)^\prime Q^{8o}_\ell - (m^2)^\prime m^2\theta^\prime 
Q^{9o}_\ell \big] ,
\label{sterms}
\ee
with non-trivial functions $Q^{8o}_\ell $ and $Q^{9o}_\ell $. We notice that the derivation of the transport equations in \cite{Cline:2020jre} assumed that the wall is propagating from left to right. In this paper however we are considering the opposite case and we replace $v_w \rightarrow - v_w$. 

The collision terms are generically given by 
\be
\delta C_1  = \sum_{ij} \Gamma_{i} s_{ij}\mu_j, \quad \delta C_2 = \Gamma_{\text{total}}u - v_w K_0 \delta C_1.
\ee
In the SM case we have 
\begin{align}
\delta \mathcal{C }_1^{t_L} &= \Gamma_y (\mu_{t_L} -\mu_{t_R}  +\mu_{h} )  + \Gamma_m (\mu_{t_L} - \mu_{t_R})   + \Gamma_W (\mu_{t_L} - \mu_{b_L}) + \tilde{\Gamma}_{ss}[\mu_i], \nonumber\\
\delta \mathcal{C }_1^{b_L} &= \Gamma_y (\mu_{b_L} - \mu_{t_R}  +\mu_{h})   + \Gamma_W (\mu_{b_L} -  \mu_{t_L})  + \tilde{\Gamma}_{ss}[\mu_i],  \nonumber \\
\delta \mathcal{C }_1^{t_R} &=  - \Gamma_y (\mu_{t_L} + \mu_{b_L}   - 2\mu_{t_R} + 2 \mu_h)   + \Gamma_m (\mu_{t_R} -  \mu_{t_L})  - \tilde{\Gamma}_{ss}[\mu_i], \nonumber \\
\delta \mathcal{C }_1^{h} &= \Gamma_y (\mu_{t_L} + \mu_{b_L}   - 2\mu_{t_R} + 2 \mu_h)  + \Gamma_h \mu_h,
\end{align}
with
\be
\tilde{\Gamma}_{ss}= \Gamma_{ss} \left(  (1+ 9 D_0^t) \mu_{t_L}  + (1+ 9 D_0^b) \mu_{b_L}  -  (1- 9 D_0^t) \mu_{t_R} \right).  
\ee
In the above expressions 
$\Gamma_{\text{sph}} = 10^{-6}\ T
$, $\Gamma_{\text{ss}} = 4.9 \times 10^{-4}\ T
$, $\Gamma_{\text{y}} = 4.2 \times 10^{-3}\ T
$, $\Gamma_{\text{m}} = \frac{m_t^2}{63\ T}
$, $\Gamma_{W} = \Gamma_{h,\text{total}}
$, $\Gamma_h = \frac{m_\text{W}^2}{50\ T}$ are the sphaleron, strong, top Yukawa, helicity-flips, $W$ boson and Higgs number violation rates \cite{Fromme:2006cm}. The total interaction rates are given in terms of diffusion constants, $\Gamma_{i,\text{total}} = D_2/(D_0 \tilde{D}_i) $ with $\tilde{D}_h =20/T$ and $\tilde{D}_q = 6/T$.

When studying the improved transport equations of Cline and Kainulainen \cite{Cline:2020jre} we noticed that there is a mismatch in mass dimensions between their equations (42) and (53); one needs to divide $\delta \bar{C}$ by $1/T$. To fix this mismatch we identified that their definition of the normalization factor $K_0$ given in their eq. (43) should carry an extra factor of $1/T$ as the same authors pointed out that $K_0 \cong 1.1$ for a massless fermion. On the other hand, for $v_w=0.5$, it gives us $K_0 \cong 110$ without the extra factor of $1/T$ ($T=100$ GeV in their fiducial model).  Additionally we believe there is a typo in their eqn. (A5) for the helicity eigenstates, it should read 
\be
V = \frac{\tilde{p}_z^2}{\tilde{p}_z^2 + x^2 } \frac{1}{\sqrt{1-\frac{x^2}{\tilde{E}^2}}}.
\ee
We take into account these modifications throughout our calculations. 

By solving the system \eqref{CP_odd} one obtains the total left-handed baryonic chemical potential
\be
\mu_{B_L} = \frac{1}{2}(1 + 4 D_0^t)\mu_{t_L} + \frac{1}{2}(1 + 4 D_0^b)\mu_{b_L} + 2D_0^t \mu_{t_R},
\ee
and the BAU can be calculated as \cite{Cline:2020jre}
\be
	\eta_B = {405\,\Gamma_{\text{sph}}\over 4\pi^2 v_w\gamma_w g_* T}\int dz\, 
	\mu_{\!B_{\rm L}}f_{\rm sph}\,e^{-45\Gamma_{\rm sph}|z|/4v_w},
\label{etab}
\ee
with $f_{\rm sph}(z) = {\rm min}(1,2.4\frac{T}{\Gamma_{\rm sph}}e^{-40h(z)/T})$ 
introduced so that the integral interpolates smoothly the sphaleron contribution between false and true vacua.

%%%%%%%%%%%%%%%%%%%%%%%%%%%%%%%%%%%%%%%%%%%%%%%
\section{The scalar singlet extension}
\label{scalar_singlet}

It is known that in the SM the phase transition is a crossover and not suitable for explaining the BAU.  Let us first consider the simplest extension of the SM which is the addition of an extra gauge singlet with a $Z_2$ symmetric potential, for other studies on this model see Refs.~\cite{Espinosa:1993bs, Espinosa:2007qk, Profumo:2007wc,Espinosa:2011ax, Barger:2011vm, Curtin:2014jma,Kurup:2017dzf}.

\subsection{Model notation and assumptions}

 The scalar potential at tree-level reads
\begin{equation}
V_0(\Phi, s) = -\mu_h^2 \Phi^{\dagger} \Phi +\lambda ( \Phi^{\dagger} \Phi)^2 + \left( m_s^2  -\frac{ \lambda_{hs}v^2}{2} \right) \frac{s^2}{2} +\frac{ \lambda_{s}}{4} s^4 +\frac{ \lambda_{hs}}{2} s^2 \Phi^{\dagger} \Phi,
 \label{V0}
\end{equation}
where  $\Phi = (G^+ , \frac{h + i G^0}{\sqrt{2}})^\text{T}$ is the Higgs doublet with $v=246 \ $GeV the SM Higgs vev.  The mass squared term for the scalar singlet has been written in such a way that it is easy to see that  
$m_s$ is the physical mass for the scalar singlet at the vacuum $(h,s) = (v,0)$  which corresponds to the electroweak symmetry breaking (EWSB) vacuum.

We assume that at zero temperature the electroweak symmetry is broken by the vev of the Higgs doublet while the complex singlet does not develop a vev, see Ref. \cite{Carena:2019une}, for a scenario that breaks the $Z_2$ spontaneously.  As we will show below, the scalar singlet develops a vev at finite temperature and the $Z_2$ symmetry is broken at some point during the thermal history. However we will assume, in the context of baryogenesis, that higher dimensional operators break the discrete symmetry explicitly and the occurrence of a domain wall problem is avoided. This explicit breaking also means that the singlet is not stable and doesn't contribute to the dark matter relic density at all. See Refs.~\cite{Cline:2012hg,Beniwal:2017eik} for analyses of this case.

 Imposing the minimization conditions on the potential, namely that the first partial derivatives with respect to the fields vanish in the $(v,0)$ vacuum, we obtain the familiar SM relation $ \lambda =\frac{ \mu_h^2}{v^2}$ and the Higgs mass is given by $m_h^2 = 2 \mu_h^2$. Thus there are three free parameters coming from the tree-level scalar potential: $m_s^2$, $\lambda_s$ and $\lambda_{hs}$. 
 
 We notice that the constraints imposed above are necessary conditions for the vacuum $(v,0)$ to be a minimum of the tree-level potential. For some parameter values a second minimum in the $s$-field direction can develop, this situation corresponds to the case of $2 m_s^2< \lambda_{hs}v^2$ and this minimum might coexist with the EWSB minima. In this case one needs to check that the EWSB is indeed the global minimum. 

The parameters of the tree-level potential are also constrained by the positivity requirement, i.e. that the potential doesn't become unbounded from below at large field values. Therefore one must have $\lambda>0$ and $\lambda_s>0$ and notice that if $\lambda_{hs}>0$ the potential is automatically positive at large field values thus we only care about the case $\lambda_{hs}<0$. A simple analytic analysis of the potential shows that the positivity requirement entails
\begin{equation}
 \lambda_{hs} > - \sqrt{2\frac{m_h^2}{v^2} \lambda_s}.
\end{equation}

Negative values of $\lambda_{hs}$ do not lead to FOPTs and hence we will not consider this region of parameter space further in this paper. Another simplification that we will adopt in this work is to consider scalar singlet masses above the Higgs mass threshold, in other words we will only consider the case $m_s>2 m_h$ so that no exotic Higgs decays are predicted. This has been famously referred as the $\textit{nightmare}$ scenario in \cite{Curtin:2014jma} because of its difficult prospects of detection at colliders. Moreover, since we are considering the case where the scalar singlet has vanishing vev at zero temperature, there is no eigenstate mixing between singlet and Higgs field and the two-loop Barr-Zee contributions to electron and neutron electric dipole moment (EDM) are absent. 

The above simplifications allow us to put negligible focus on the collider constraints on this model, which have been studied elsewhere. Here we are primarily interested in calculating the bubble wall properties and their repercussions on the BAU and on the energy density of GWs.  

To study the thermal history of the model one needs the one-loop Coleman-Weinberg contribution as well as the finite temperature potential for which we include the relevant formulas in the appendices. For the computation of the instanton solution of the Euclidean action we have made use of the publicly available CosmoTransitions code \cite{Wainwright:2011kj}.
We modified the CosmoTransitions module for a generic potential to make it compatible with the on-shell renormalization prescription.  
Additionally we have written our own code with a simplified calculation that instead of calculating the trajectory in field space that minimizes the Euclidean action, uses overshooting along the path that minimizes the potential energy (as in Refs. \cite{Chung:2011it,Vaskonen:2016yiu} ). The EOM was solved for temperatures from the range in which two minima coexist. Sampling different values and checking the condition \eqref{cond2} at every step, we used a bisection algorithm to find the nucleation temperature. In a similar way, $\alpha$ and $\frac{\beta}{H}$ were evaluated. We cross-checked the output from the CosmoTransitions package with our own code and we found reasonably good agreement between the two methods.

\begin{figure}[h] 
\includegraphics[scale=.4]{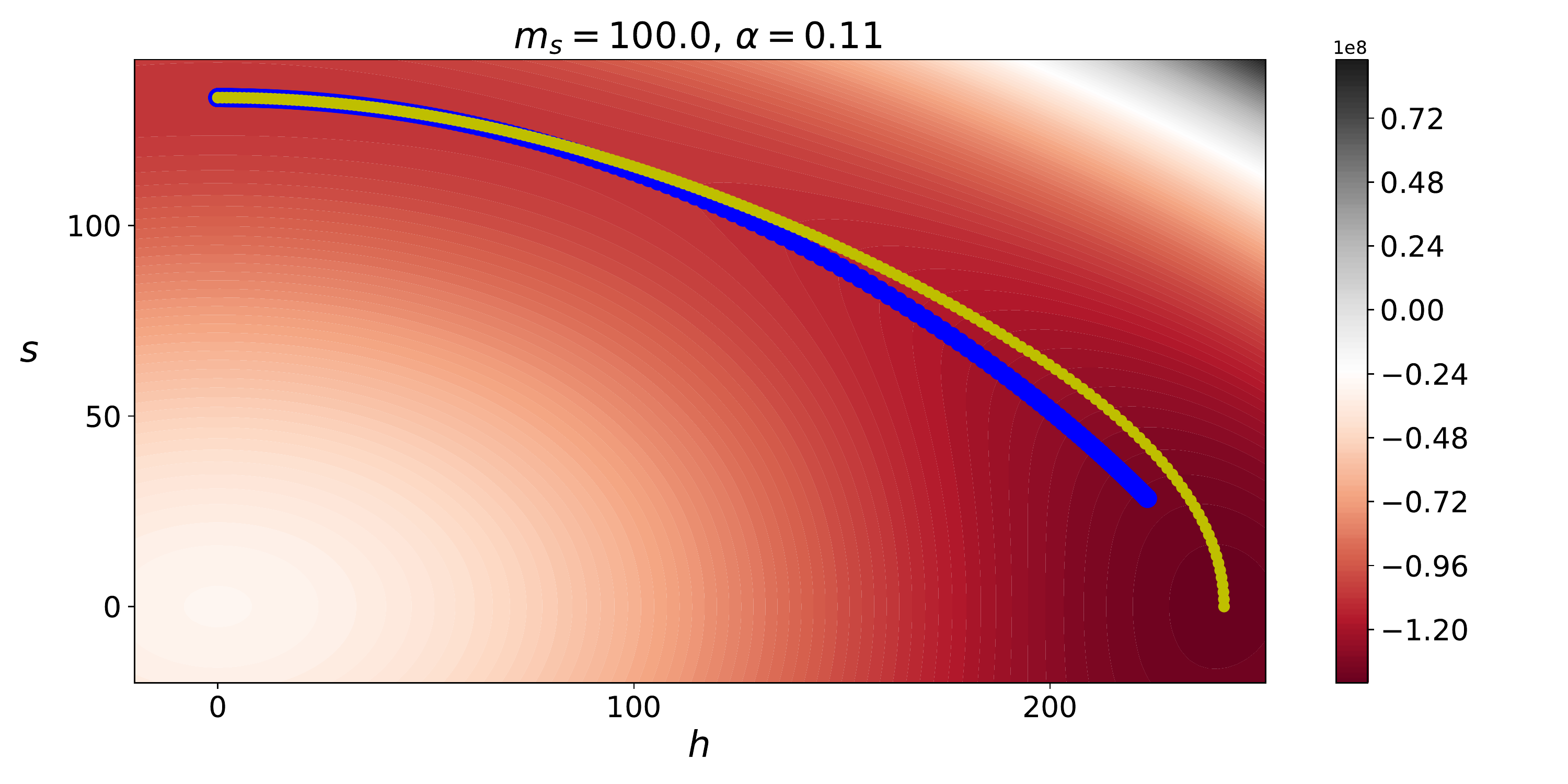}
\centering
\caption{Contour plot of the effective potential for the scalar singlet model with parameters: $m_s = 100$ GeV, $\lambda_{hs} = 0.93 $, $\lambda_s = 1$. The blue dots represent the trajectory in field space of the instanton solution as obtained with CosmoTransitions while the yellow points show the path of minimum energy. The critical and nucleation temperatures are given by $T_c = 94$ GeV and $T_n = 59$ GeV, respectively. These parameter space values yield a phase transition strength $\alpha = 0.11$.}
\label{Contour_pot}
\end{figure}

An example of a parameter set that gives rise to a FOPT is presented in figure \ref{Contour_pot}. This parameter space point exemplifies what is commonly called a two-step FOPT;  at very high temperatures the only stable minimum of the effective potential lies at the origin of field space. Then as the universe evolves and the temperature drops a second minimum in the $s$ field direction starts to form while the origin starts to become a local maximum. The transition in this case is second order. After more cooling the electroweak vacuum appears, eventually becoming degenerate with the $s$-vacuum. Afterwards the electroweak vacuum becomes the global minimum and a FOPT occurs at the nucleation temperature when the probability of bubble formation becomes comparable to the Hubble volume at that time. 

On figure \ref{Contour_pot} we indicate the trajectory in field space as found by CosmoTransitions with the blue points while the yellow points show the path of minimum energy. One can easily observe a difference between the two paths in this case. While studying the parameter space of the model we noticed that for a given strength of the transition, bigger masses tend to deviate more from the minimum energy path.  Furthermore, the range of FOPT strengths narrows down as the mass becomes smaller and the thin wall approximation becomes worse because the escape point is significantly far away from the global minimum around $(v,0)$. This is why the blue points do no reach the other end in the figure. The parameter space point chosen in this case illustrate these two effects.

In this paper we will restrict our attention to two-step FOPTs as these type of transitions can provide CP violating mass terms for quarks at high temperature. In our case we consider a complex mass term for the top quark via the following dimension-5 operator 
\be
\mathcal{L}_{\text{Yukawa}} \supseteq y_t \bar{Q}\Phi t_R \left( 1  + \frac{i s}{\Lambda_{\text{CP}}} \right) + \text{h.c.},\label{singlet_topmass}
\ee
where $\Lambda_{\text{CP}}$ is some high energy cutoff scale and is treated as a free parameter. In principle one can also add similar operators for all SM fermions, however one expects their effect to be proportional to their Yukawa couplings which are suppressed relative to the top quark Yukawa coupling. We will ignore them completely in this paper. In the above expression we have set the Wilson coefficient of the dimension-5 operator equal to one imaginary unit. This is done for simplicity as it corresponds to maximal CP-violation.

\begin{figure}[h]
\includegraphics[scale=.36]{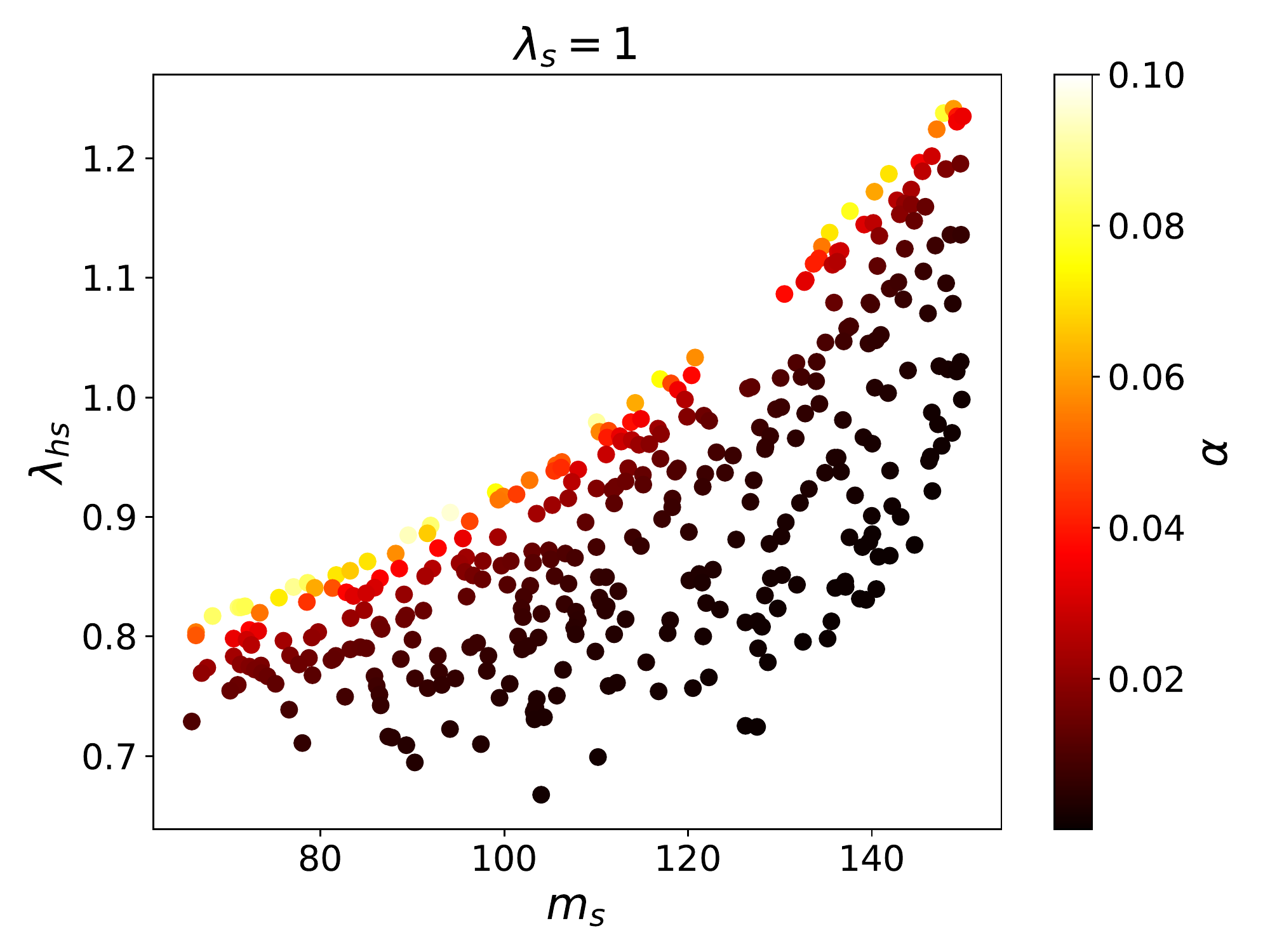}
\includegraphics[scale=.36]{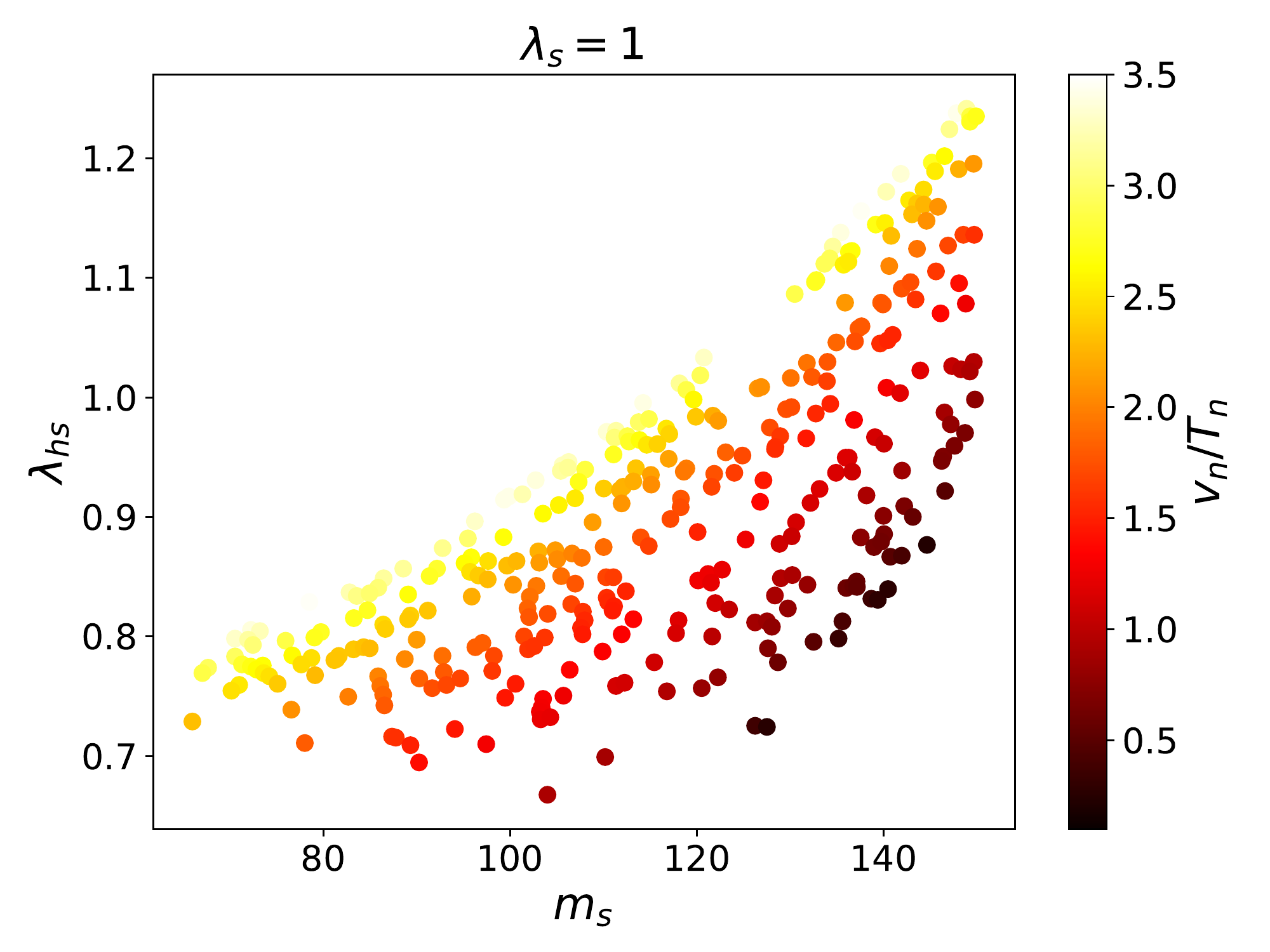}
\centering
\caption{Parameter scan of the scalar singlet model featuring two-step FOPTs. $\textit{Left}$: The color map indicates the value of the latent heat released during the transition. $\textit{Right}:$ the color shows the value of the sphaleron washout condition. }
\label{singlet_scan}
\end{figure}

We present, in figure \ref{singlet_scan}, the result of a parameter scan of the scalar singlet model featuring two-step FOPTs  with the singlet quartic coupling fixed to $\lambda_s=1$ for simplicity. Changing the value of $\lambda_s$ will not change the qualitative results and will only shift the points up or down. On the left, the color bar represents the latent heat released during the phase transition. This is also a measure of the strenght of the phase transition. The colorbar on the right shows the sphaleron washout condition. 

The relevant ``$\textit{arm}$'' of parameter space has a very well delineated structure: for a given mass, stronger transitions prefer larger values of the Higgs portal parameter. The same applies for the sphaleron condition. The white region (with no colored points) above the $\textit{arm}$ is excluded by the requirement of $(v,0)$ to be the global minimum. The white region below the $\textit{arm}$ has too low values of $\lambda_{hs}$ and there is no FOPTs there.

%%%%%%%%%%%%%%%%%%%%%%%%%%%%%%%%%%%%%%%%%%%%%%%%%%%%%%
\subsection{Bubble wall properties}
As we have discussed in chapter \ref{Wall_eqns}, the bubble wall properties in the scalar singlet extension consist in the determination of the parameters $v_w$, $L_h$, $L_s$, $h_0$, $s_0$, $\delta_s$ which satisfy the set of equations \eqref{moment_1}, \eqref{moment_2},
\eqref{minmization_1},\eqref{minmization_2} and \eqref{minmization_3}.

\begin{figure}[h]
\includegraphics[scale=.35]{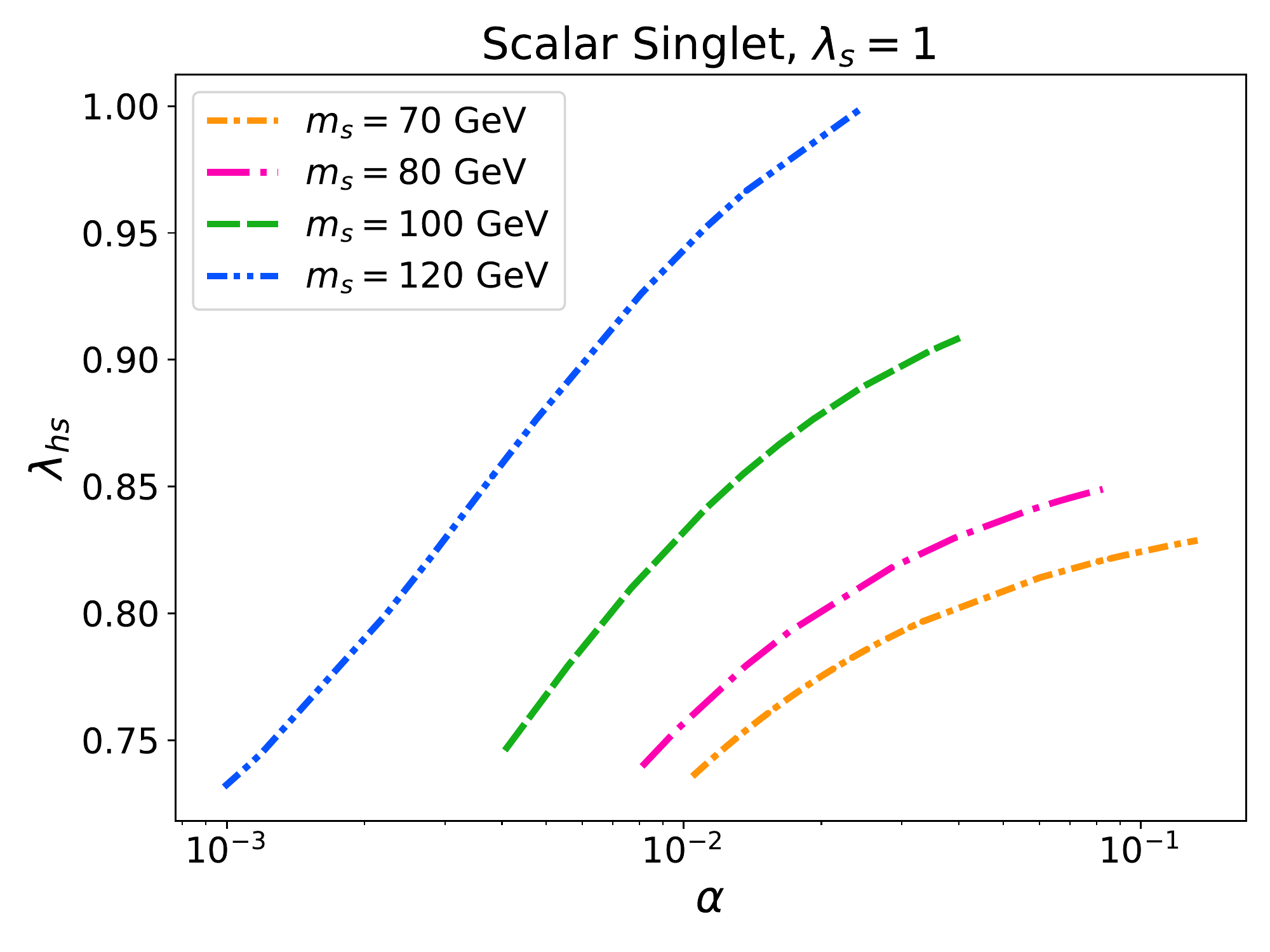} 
\includegraphics[scale=.35]{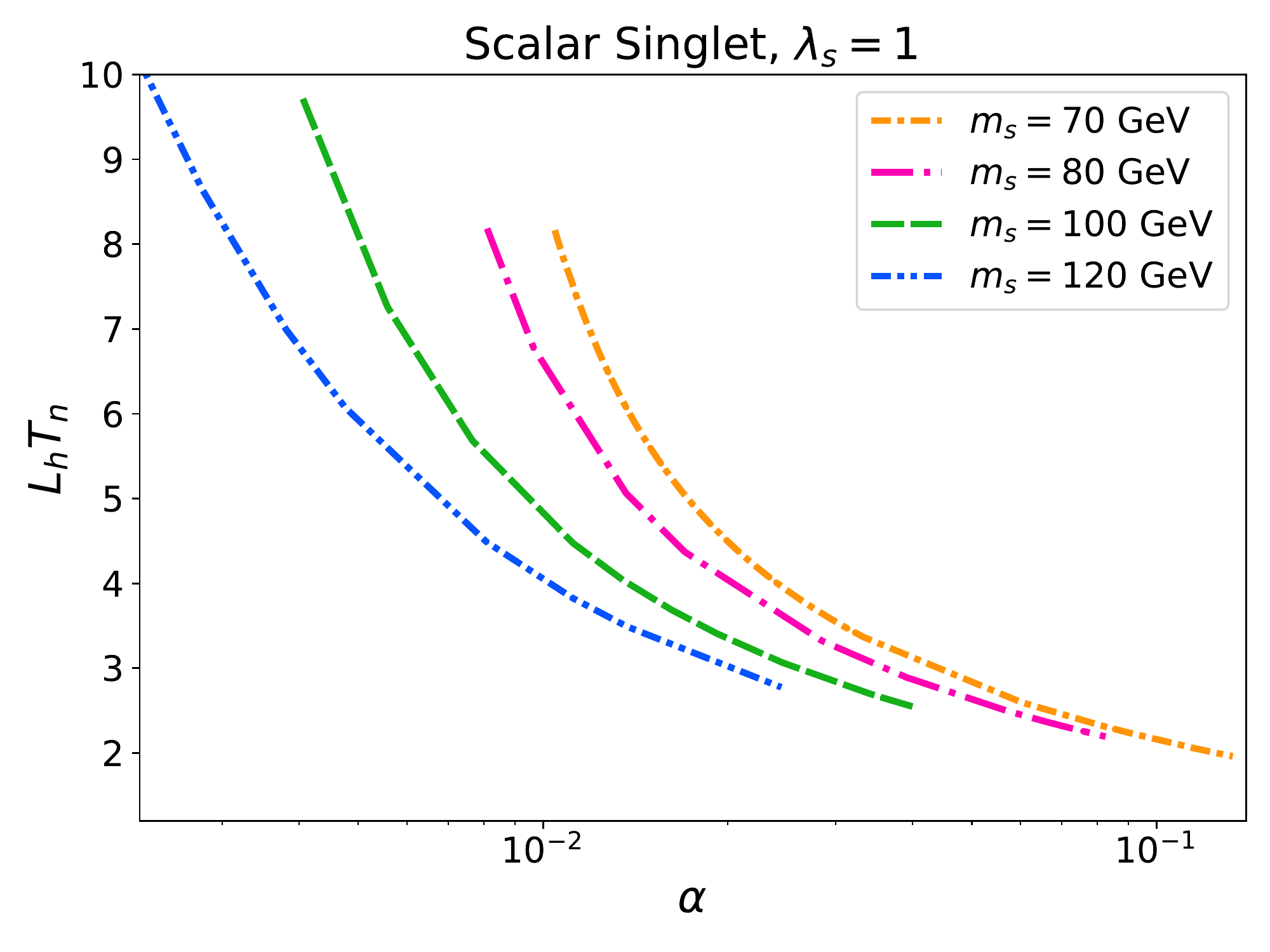}
\includegraphics[scale=.35]{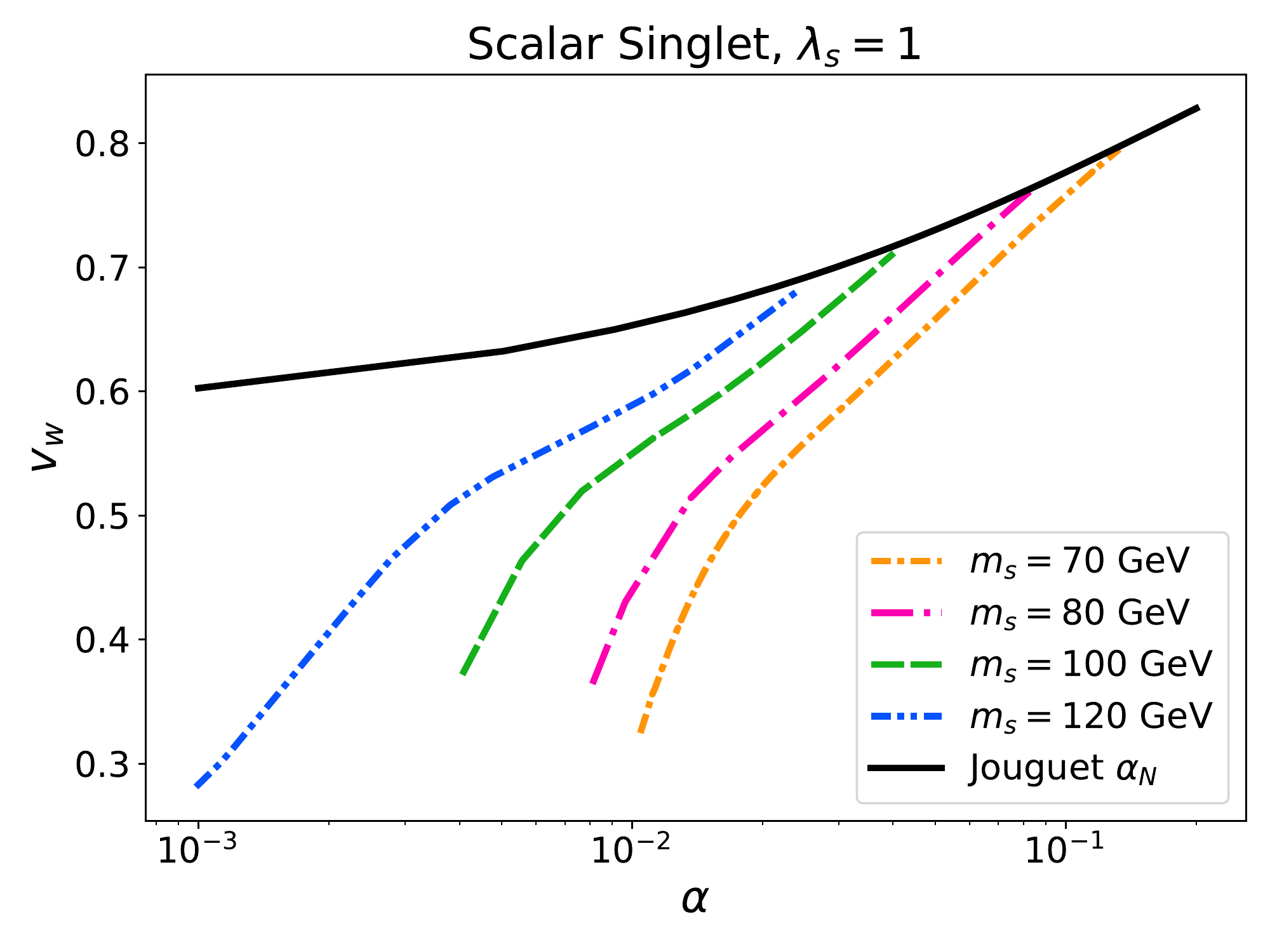}
\includegraphics[scale=.35]{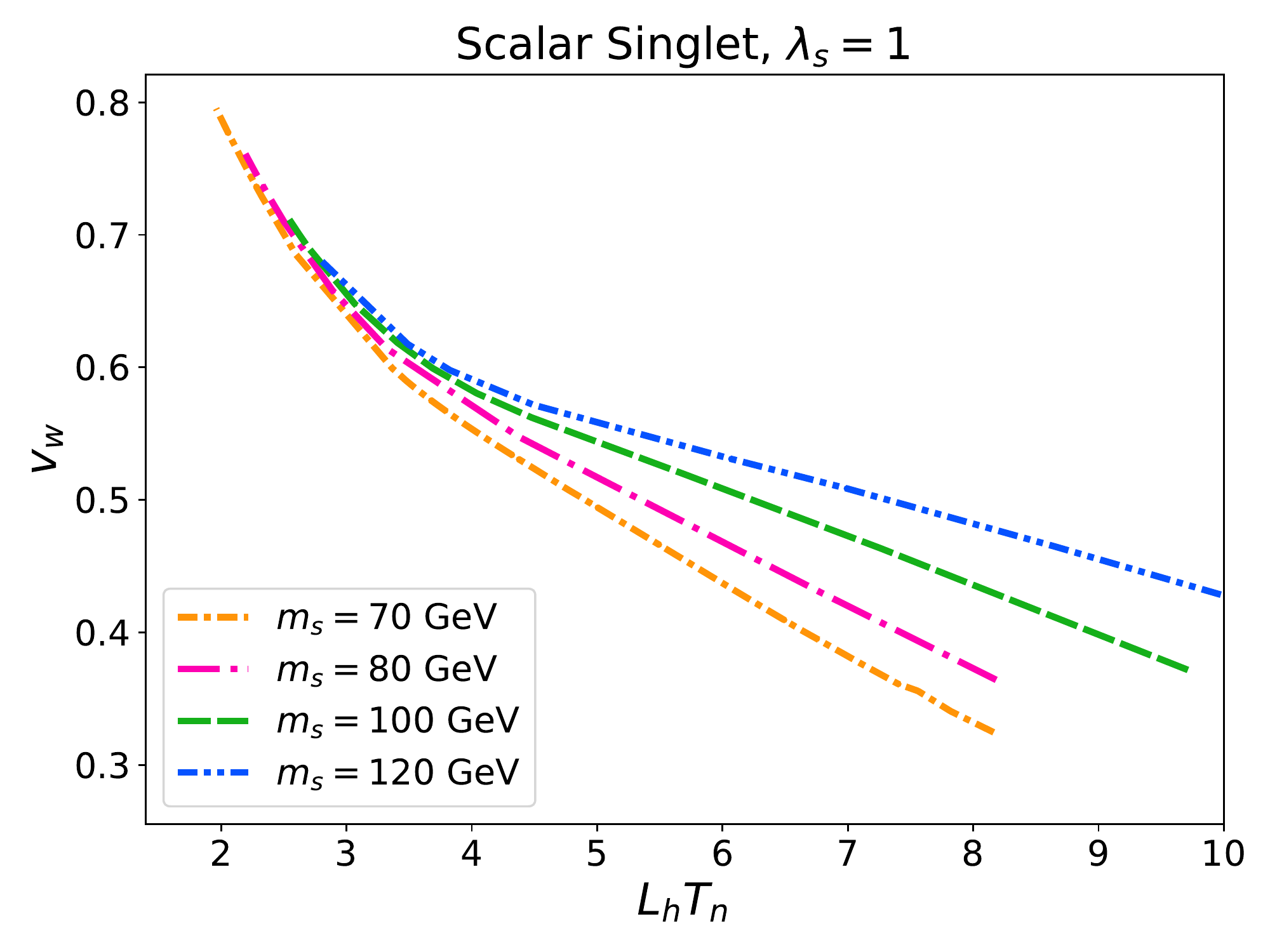}
\centering
\caption{Results of the calculation for the bubble wall properties in the scalar singlet model for different $m_s$ benchmarks. We have set $\lambda_{s} = 1$ everywhere and the qualitative description for each figure is provided in the text. }
\label{singlet_bubble}
\end{figure}

In order to produce easily readable results we choose four benchmark values for the scalar singlet mass $m_s= 70,\ 80, \ 100, \ 120$ GeV and we scan the Higgs portal parameter $\lambda_{hs}$ with values compatible with two-step FOPT. The singlet quartic coupling is set $\lambda_{s} = 1$ throughout the paper as we expect that varying its value will not have any qualitative repercussions on the results.  This simplification has the added benefit that for a given mass the only micro-physical parameter from the Lagrangian is the Higgs portal coupling $\lambda_{hs}$. The value of this parameter then completely determines the properties of the FOPT and of the bubble wall. The results of the calculation are presented in figure \ref{singlet_bubble}. We show on the upper left plot the relation between the strength of the FOPT and the portal coupling. This is clearly the same pattern from fig. \ref{singlet_scan} which indicates that stronger transitions are given by the largest possible value of $\lambda_{hs}$. On the upper-right the dimensionless wall thickness $L_h T_n$ is plotted against $\alpha$, evidencing the inverse proportionality between the two. In the lower plots we show the dependence of the velocity on the strength (left) and on the wall thickness (right). These plots corroborate our intuition that: 1) stronger transitions lead to faster moving walls and 2) faster walls are thinner. 
It is important to mention that the lines of the plots in fig. \ref{singlet_bubble} terminate at some maximum value of the strength of the transition $\alpha_{\text{max}}$. 
As we see in the lower left panel this corresponds to strong transitions for which the walls reach the Jouguet velocity (see Eq.~\eqref{eq:vJ}). As we discussed in Sec~\ref{hydrodynamics}, if the acceleration of the wall is not stopped by the friction below that velocity the solution changes into a detonation. As a result the heated plasma shell around the bubble disappears and the effective temperature determining the friction drops to the background temperature of the unbroken phase. This lowers the friction considerably and we never find solutions with $M_1 = M_2 = 0$ with higher velocities. This result agrees with~\cite{Cline:2021iff} which also was not able to find the wall properties for detonation solutions for the same reason.

To conclude this section we explore the relationship between the rest of the bubble wall properties, namely the field amplitudes $h_0$, $s_0$, the scalar singlet thickness parameter $L_s$ and the offset constant $\delta_s$. These are shown in figure \ref{singlet_bubble_1} for $m_s=100$ GeV. However, it is important to mention we have verified that the same qualitative behaviour follows for different masses. The two wall thicknesses are positively correlated as shown on the upper left plot with thicker walls requiring a bigger offset factor. This positive correlation was also reported in fig. 7.c of ref. \cite{Cline:2021iff}, however, due to the multiple parameter scan in that reference, it is hard to see the relation with the offset parameter. The field amplitudes $h_0$ and $s_0$ are also positively correlated and larger amplitudes are associated with faster walls. The plots on the bottom show the wall thicknesses $L_h$ and $L_s$ as function of $h_0$ and $s_0$, respectively. In both cases the larger the field amplitude the thinner the wall.

\begin{figure}[h]
\label{singlet_bubble_1}
 \includegraphics[scale=.35]{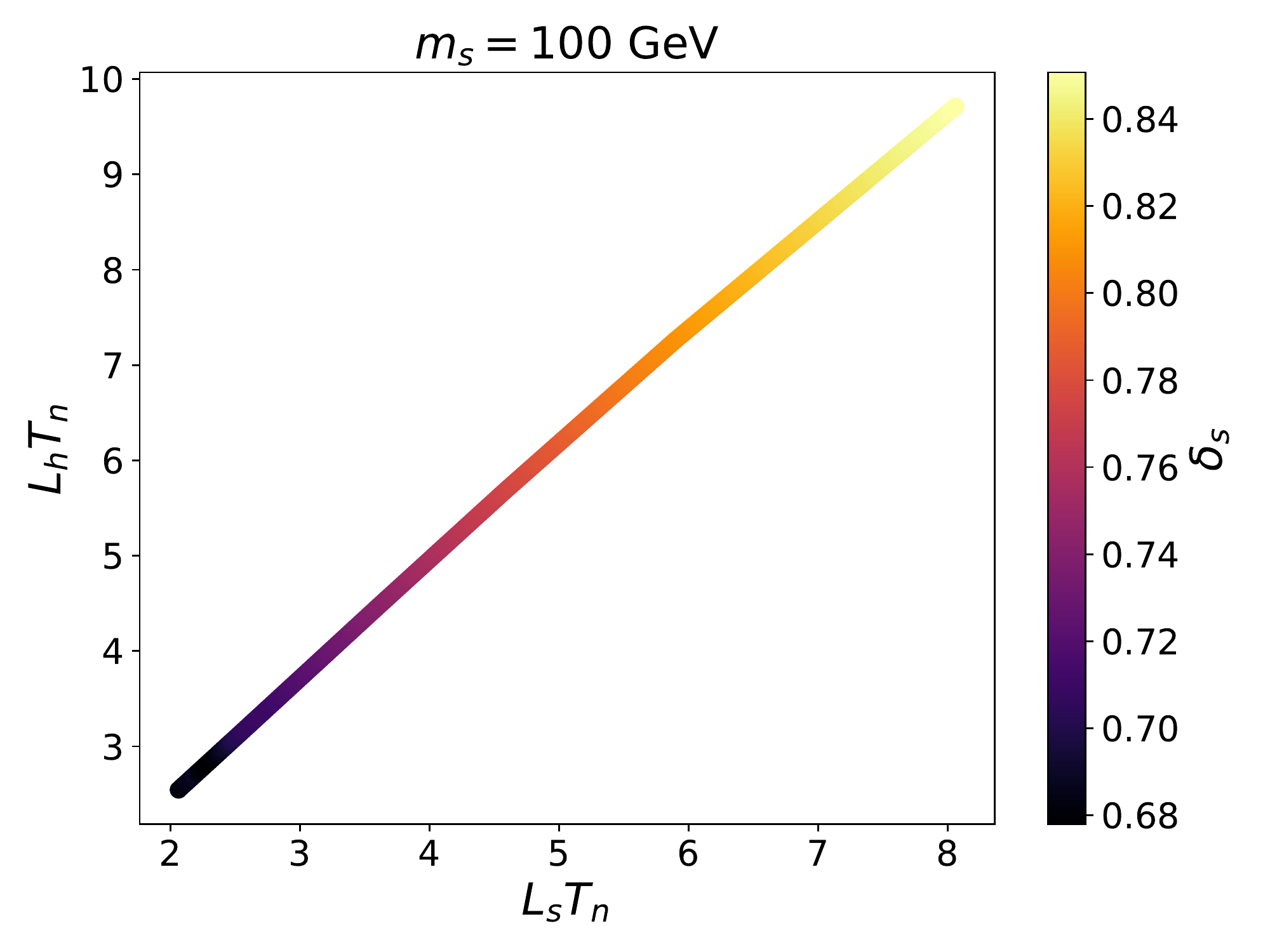} 
\includegraphics[scale=.35]{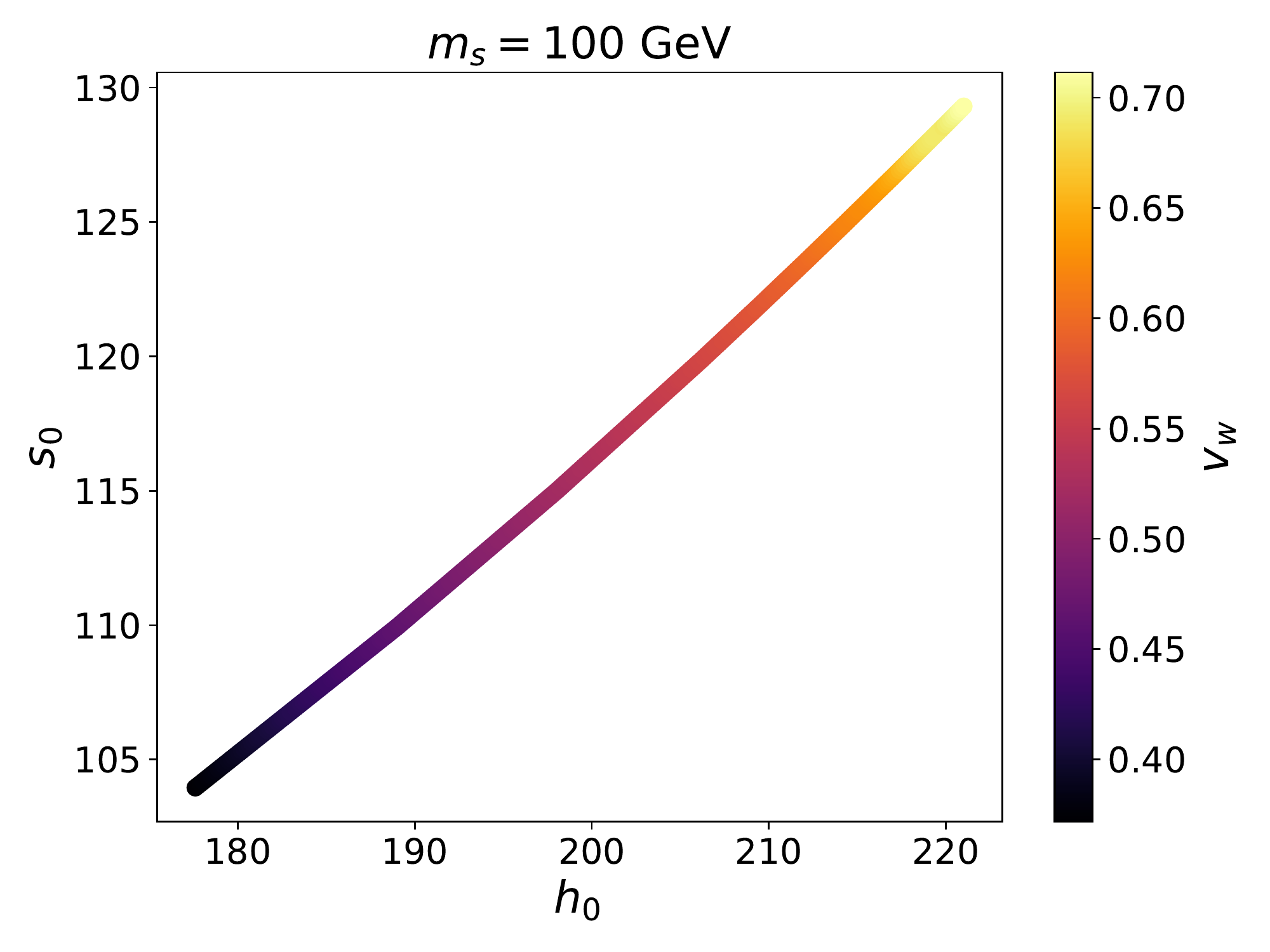} 
\includegraphics[scale=.35]{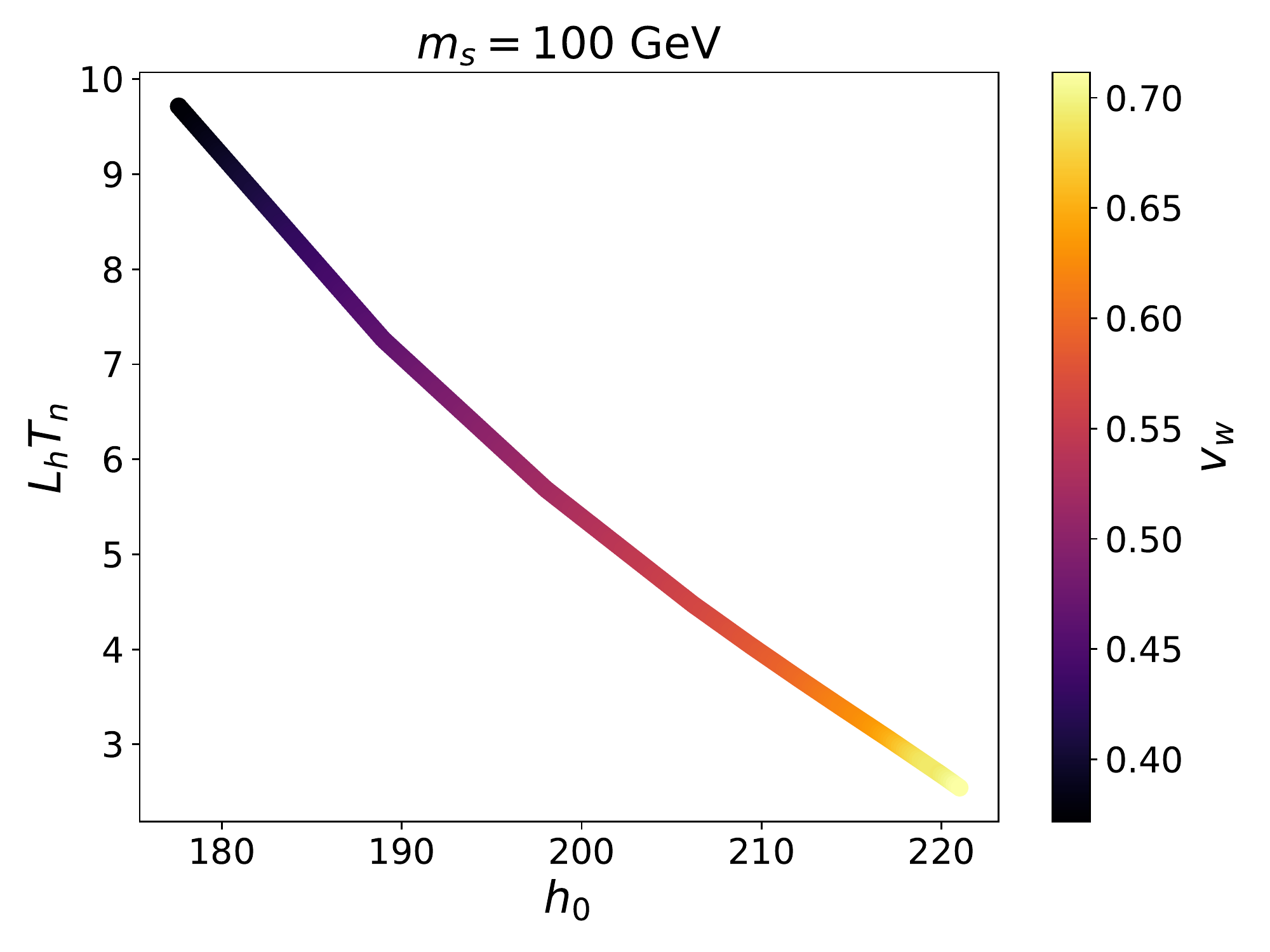}
\includegraphics[scale=.35]{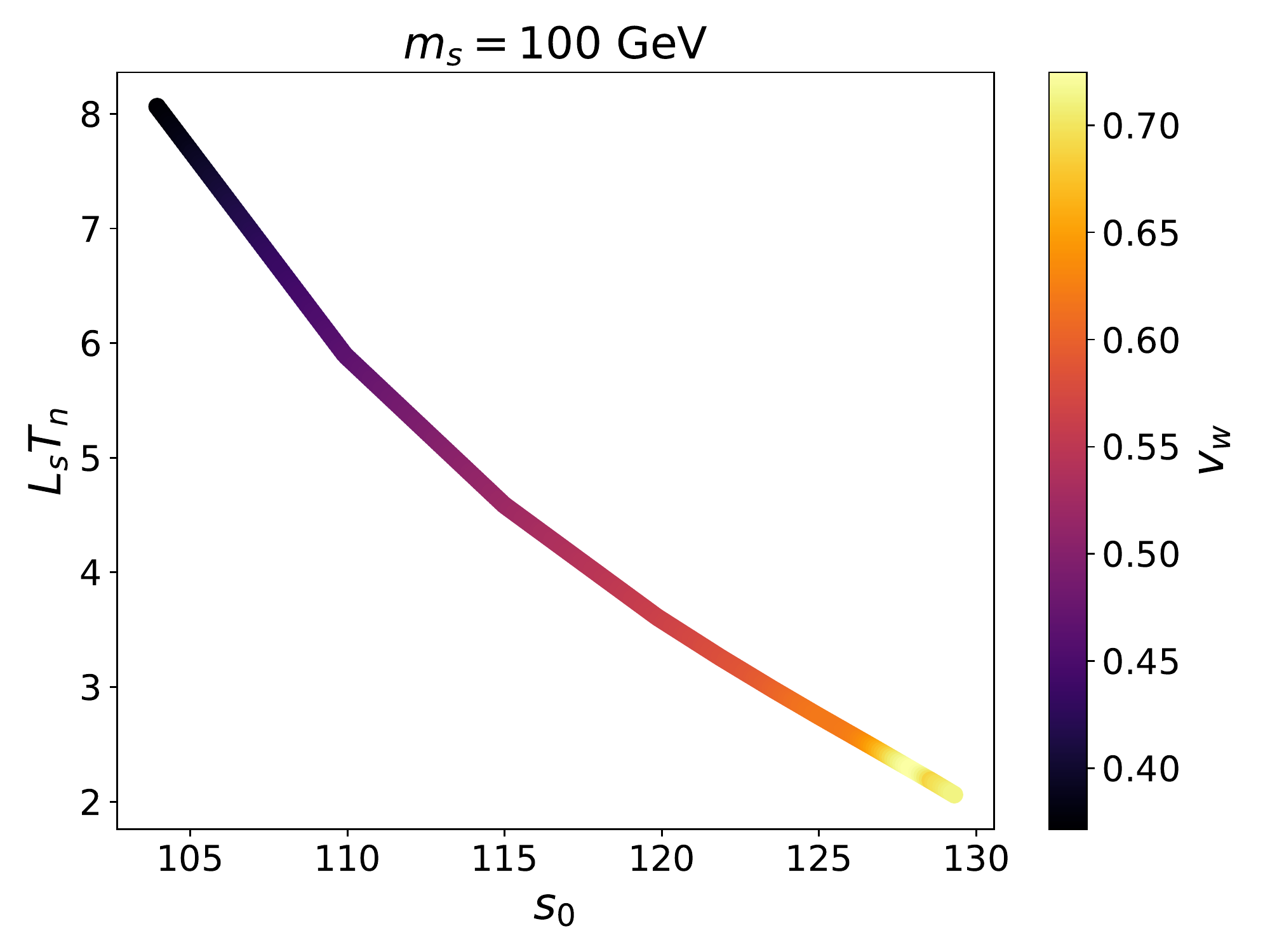}
\centering
\caption{Correlation between parameters of the bubble wall. Color maps show the value of the offset parameter for the upper-left plot and the wall velocity for the rest of them. $\lambda_s = 1$ as in the rest of the paper.}
\label{singlet_bubble_1}
\end{figure}

\subsection{Baryogenesis}

As a consequence of the higher dim-5 operator in \eqref{singlet_topmass} the top quark acquires a complex mass during the two-step FOPT. This mass term violates CP and is the source of the baryon assymmetry. 
 The space-time dependent mass term in the Dirac equation of the top can be written as $m_t(z) e^{i \theta_t(z)}$, with 
\be
m_t(z) \equiv    \frac{y_t h(z)}{\sqrt{2}}  \sqrt{  1 + \frac{s(z)^2}{ \Lambda_{\text{CP}}^2}  },
\label{Singlet_top_mass}
\ee
and the CP-violating phase
\be
\theta_t (z) = \arctan{\left[   \frac{s(z)}{ \Lambda_{\text{CP}}}  \right]   }. 
\ee
We can see that the cutoff scale can affect the value of the top quark mass during the FOPT however we expect that this effect is not significant for the thermodynamic properties of the transition as well as the wall speed computation. Thus $\Lambda_{CP}$ is taken as a free parameter which can be fixed to accommodate the final BAU.

\begin{figure}[h]
\includegraphics[scale=.36]{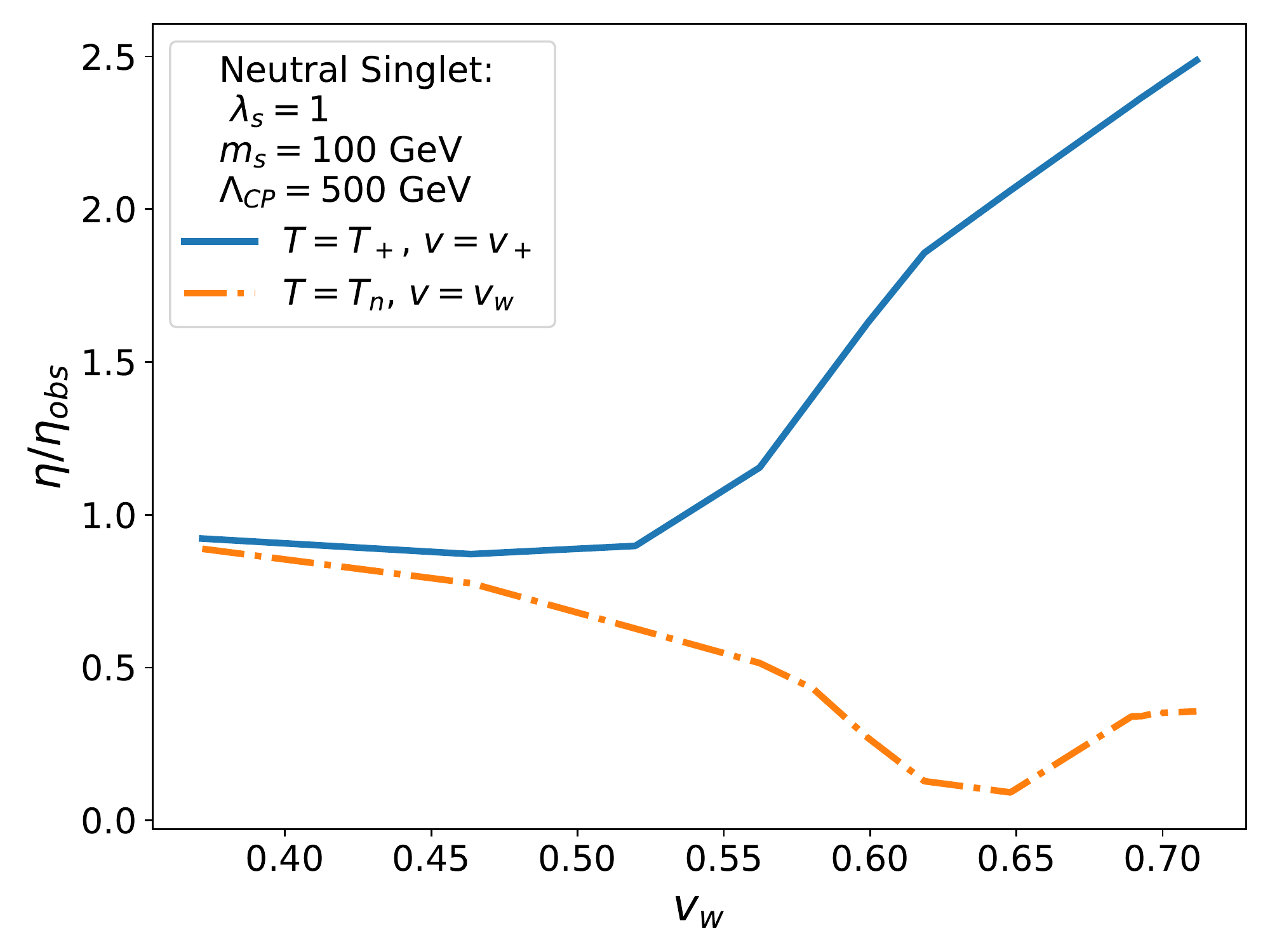}
\includegraphics[scale=.36]{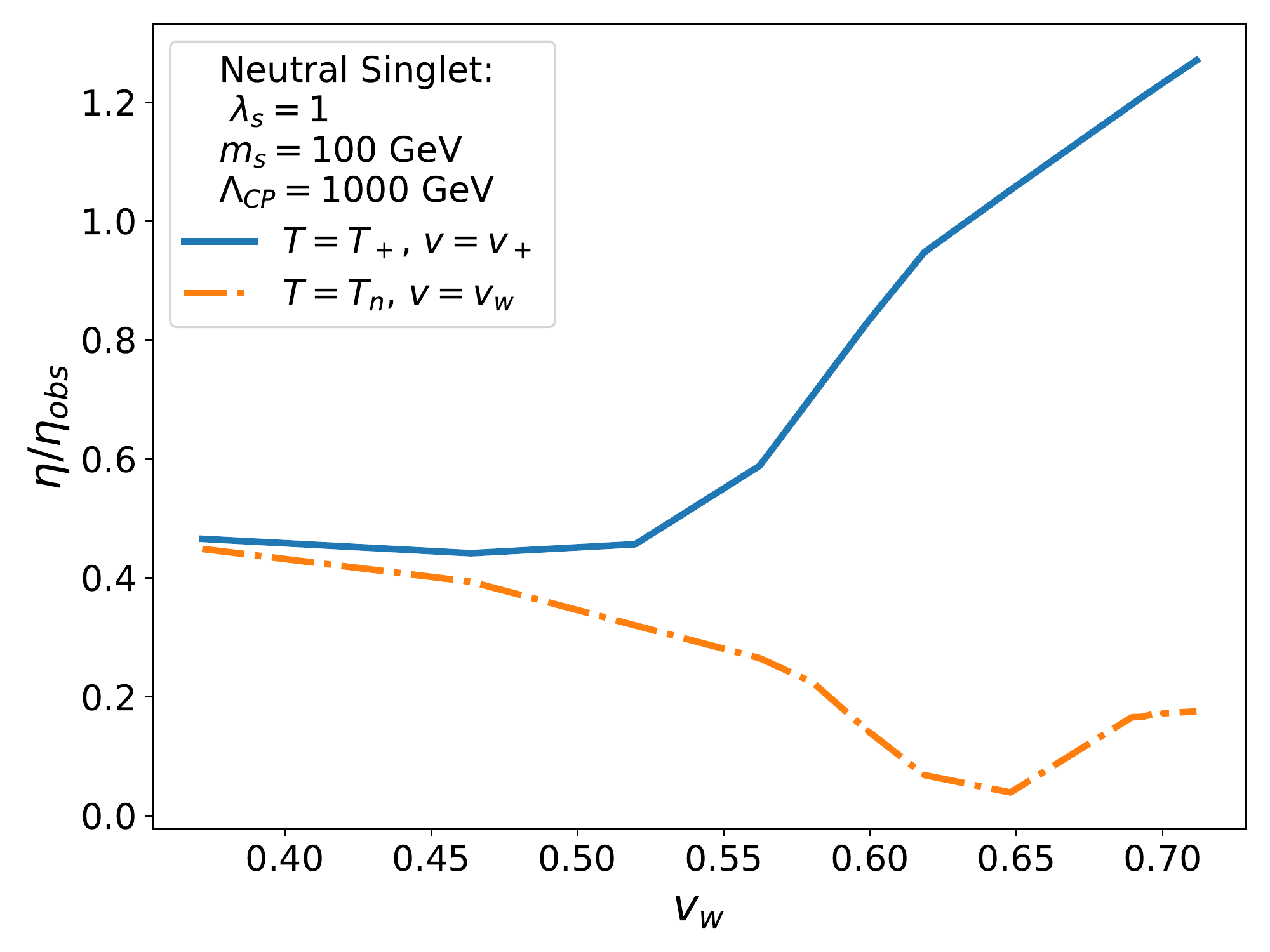} 
\centering
\caption{Prediction for the BAU normalized to its observed value in the scalar singlet model using the correct variables in front of the wall $T_+$, $v_+$ (blue solid) and using the naive variables $T_n$, $v_w$ (orange dash-dot). The parameters have been fixed as indicated. The horizontal axis gives the wall velocity of each parameter set.   }
\label{BAU_hydro_care}
\end{figure}

In ref.~\cite{Cline:2021iff} it was found that use of variables $T_+$ and $v_+$ including heating of the plasma around the bubble had a significant impact on computation of the wall properties.
In this paper we emphasise this issue and show that this can also lead to considerable different estimates for the BAU. We illustrate this in figure \ref{BAU_hydro_care} where we show the BAU, normalized to its observed value $\eta_{\text{obs}} \approx 8 \times 10^{-11}$ \cite{Cooke:2013cba, Planck:2015fie}, computed using $v_w$, $T_N$ in orange and with $v_+$ $T_+$ in blue. With the former variables one would conclude that the BAU yield decreases at higher wall velocities and none of the shown transitions can give the right amount of asymmetry. This is however not the case, as shown by the blue curves, the model with a velocity of about $v_w \approx 0.55$ ($\approx 0.64$) on the left (right) yields the right value of BAU. Therefore the inclusion of the hydrodynamic effects is key for correct baryogenesis predictions. 

The variation of the BAU with the cut-off scale can also be deduced from fig.~\ref{BAU_hydro_care}. For the cut-off values chosen one can directly observe the qualitative scaling $\eta_B \sim 1/\Lambda_{\text{CP}}$. Increasing the value of $\Lambda_{\text{CP}}$ requires faster walls to explain the matter asymmetry. Furthermore the final BAU is well behaved for all the velocities computed which provides a consistency check of the possibility of  baryogenesis for supersonic walls. 

\begin{figure}[h]
 \includegraphics[scale=.36]{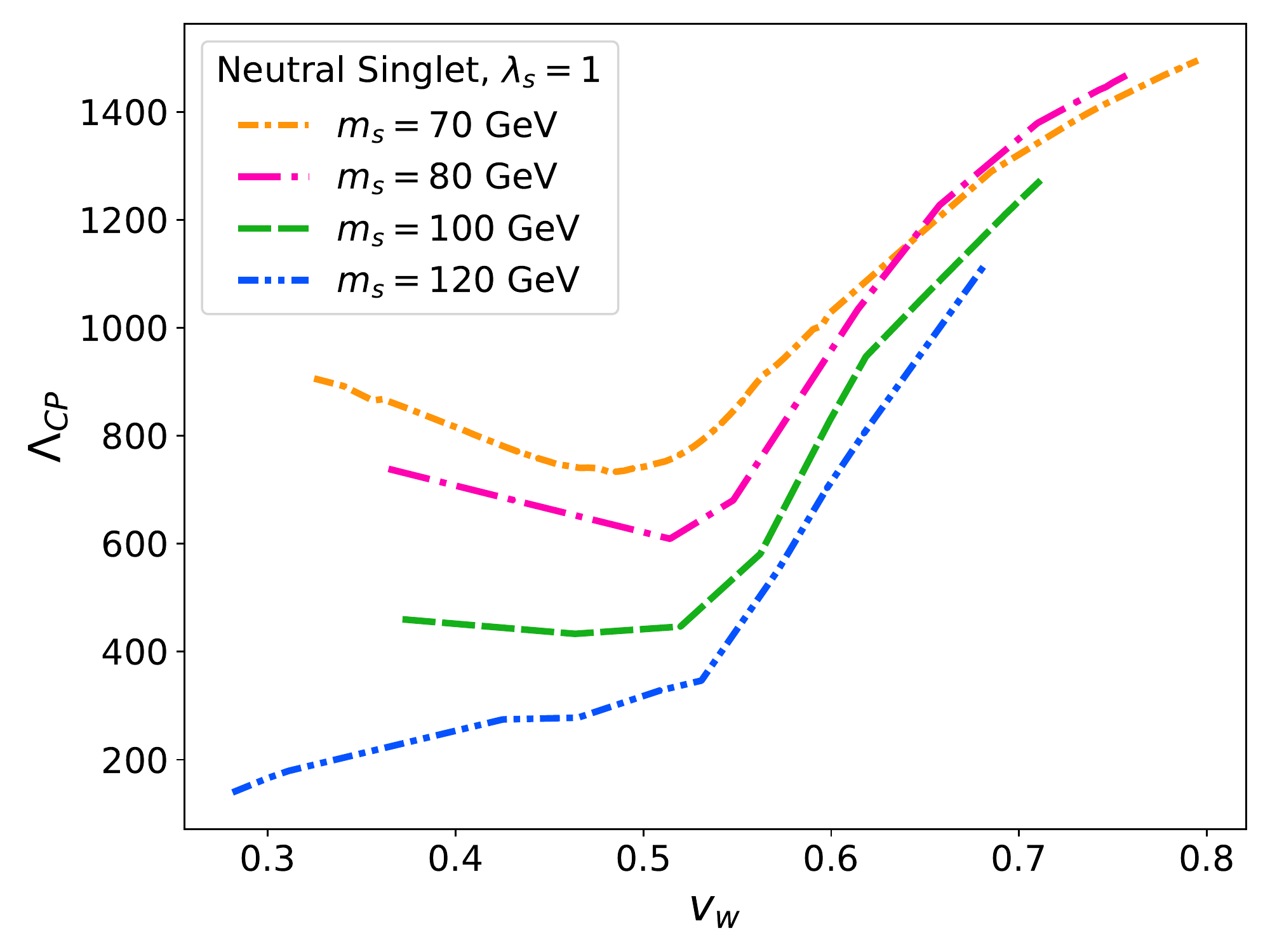} 
\includegraphics[scale=.36]{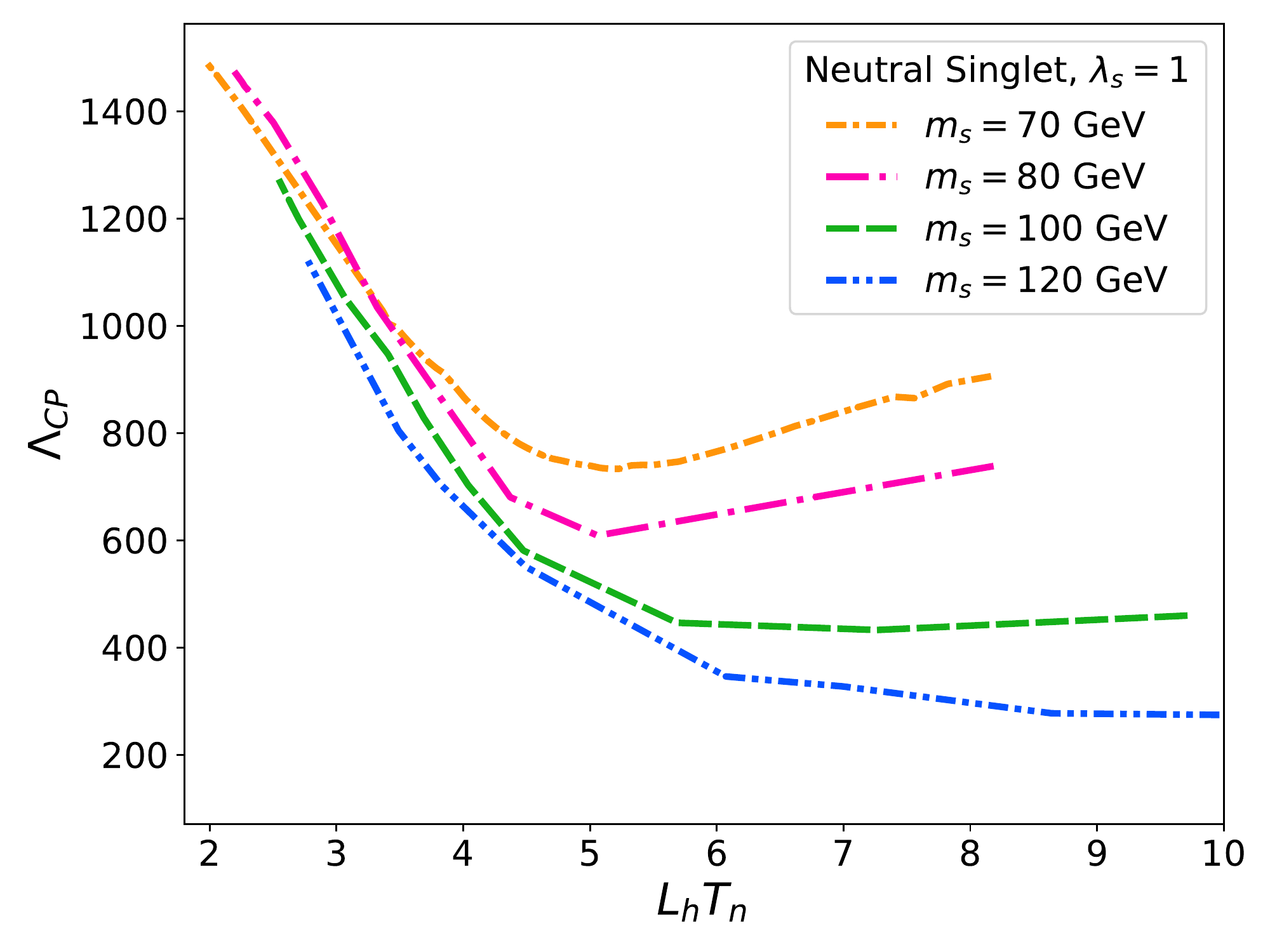} 
\centering
\caption{Cut-off scale $\Lambda_{CP}$ required to obtain the observed baryon asymmetry as a function of the wall velocity (left panel) and  wall thickness (right panel) in three benchmark values of the mass $m_s = 70$, $80$, $100$, $120$ GeV in the scalar singlet model.
}
\label{fig:BAULambdaPlot_singlet}
\end{figure}

Fig.~\ref{fig:BAULambdaPlot_singlet} shows the cut-off scale $\Lambda_{CP}$ necessary to obtain observed baryon asymmetry for our chosen range of masses $m_s=70-120$ GeV. According to ~\cite{Cline:2021iff}, the value  $\Lambda_{\text{CP}} \gsim 540$ GeV is consistent with experimental bounds on scalar singlet production at the LHC and agreeing with that reference we conclude it is not difficult to find plausible scenarios predicting the correct baryon asymmetry. Higher wall velocities (and stronger transitions) predict a larger asymmetry and the CP scale predicting the observed value can be as large as $\Lambda_{\text{CP}} \approx 1.5$ TeV.

\section{SMEFT}
\label{sec:SMEFT}
The SM effective field theory is a low energy representation of a possibly more UV completed theory. To parametrize our ignorance to the physics at high energy scales one adds a tower of higher dimensional operators that are consistent with the gauge symmetries of the SM and suppressed by a high energy scale $\Lambda$. In our case we consider the scalar potential augmented by a dimension six operator, namely
\be
\label{SMEFT_potential_tree}
V_0 = -m^2 \Phi^{\dagger} \Phi+ \lambda( \Phi^{\dagger} \Phi)^2  +   \kappa (\Phi^{\dagger} \Phi)^6,
\ee
in this notation $\kappa \equiv 1/\Lambda^2$. 
The vacuum stability conditions at tree-level are 
\begin{equation}
V'(h=v)=0, \quad V''(h=v)=m_h^2 \, ,
\end{equation} 
which allow us to express the parameters in the potential as
\begin{equation}
m^2=\frac{m_h^2}{2}-\frac{3 v^4}{4 \Lambda^2}, \quad \lambda=\frac{m_h^2}{2v^2}-\frac{3 v^2}{2 \Lambda^2}, 
\end{equation}
notice that for sufficiently low values of the cutoff scale, the quartic coupling can have negative values. This allows the formation of a tree-level potential barrier giving rise to FOPT. The relevant formulas for the one-loop corrections and the temperature dependent masses are relegated to an appendix. It is also important to mention that including higher dimensional operators would not change the possible results for transition parameters instead simply mapping our one variable $\Lambda$ onto a certain combination of more operators~\cite{Chala:2018ari}.

A comment about the use of the effective field theory should be provided;  it has been shown~\cite{Damgaard:2015con} that the $\textit{projection}$ between the scalar singlet model (both with and without $Z_2$ symmetry) and SMEFT up to the dimension six operator in the potential is not always one-to-one with regards to the character of the phase transition. In other words, some regions of parameter space which show FOPT within the scalar singlet model do not always manifest the same type of transition when mapped onto the low energy effective field theory (EFT). The main problem lies on the fact that strongly FOPTs in the scalar singlet model are located in a region of parameter space with lower masses and large mixing quartic $\lambda_{hs}$. This is in direct tension with the premises of effective field theory which requires a large separation of scales (large masses in this case) so that heavy physics are sufficiently decoupled and the EFT remains valid. 

The possibility of EWBG within the SMEFT has been scrutinized in ref.~\cite{Balazs:2016yvi} where the authors directly tested one of the fundamental properties of effective field theories; the use of the EOM to remove redundant operators. This means operators that are related by the EOM should lead to the same physical prediction up to higher order effects in the perturbative EFT expansion. In that reference two different types of CP violating operators, connected by the EOM, were used to predict the EDM contributions and the BAU. While it was found that both operators give the same prediciton for the EDM, the prediction for the BAU was significantly different unless higher order effects are included. Their results thus contradict the hypothesis of EFTs about the redundancy of operators connected by the EOM in the context of EWBG. A very similar of study appeared in~\cite{deVries:2017ncy} by some of the same authors which further confirmed the breakdown of the EFT for the purposes of EWBG calculations. 
In this paper we do not interpret eq.~\eqref{SMEFT_potential_tree} as coming from any particular UV completion and we treat it simply as a $\textit{toy}$ model for which we can compute the properties of the phase transition and apply the semi-classical treatment for the bubble wall velocity and the BAU.

%%%%%%%%%%%%%%%%%%%%%%%%%%%%%%%%%%%%%%%%%%%%%%%%%%%%%%
\subsection{Bubble wall properties}
Calculating the wall velocity in SMEFT is relatively easier  than with the scalar singlet scenario. In this case the grid scan method presented in section \ref{Wall_eqns} has less steps since one only needs to satisfy the Higgs EOM. Furthermore in this model the cutoff scales $\Lambda$ and $\Lambda_{CP}$ are the only parameters that have to be fixed and they are disentangled with respect to the bubble wall properties and the BAU, respectively \footnote{The top quark field dependent mass in \eqref{SMEFT_top_mass} has a direct dependence on $\Lambda_{\text{CP}}$ and this might affect the calculation of the FOPT quantities. For the values of $\Lambda_{\text{CP}}$ considered here we expect this effect to be negligible.}. 
 
\begin{figure}[h]
 \includegraphics[scale=.35]{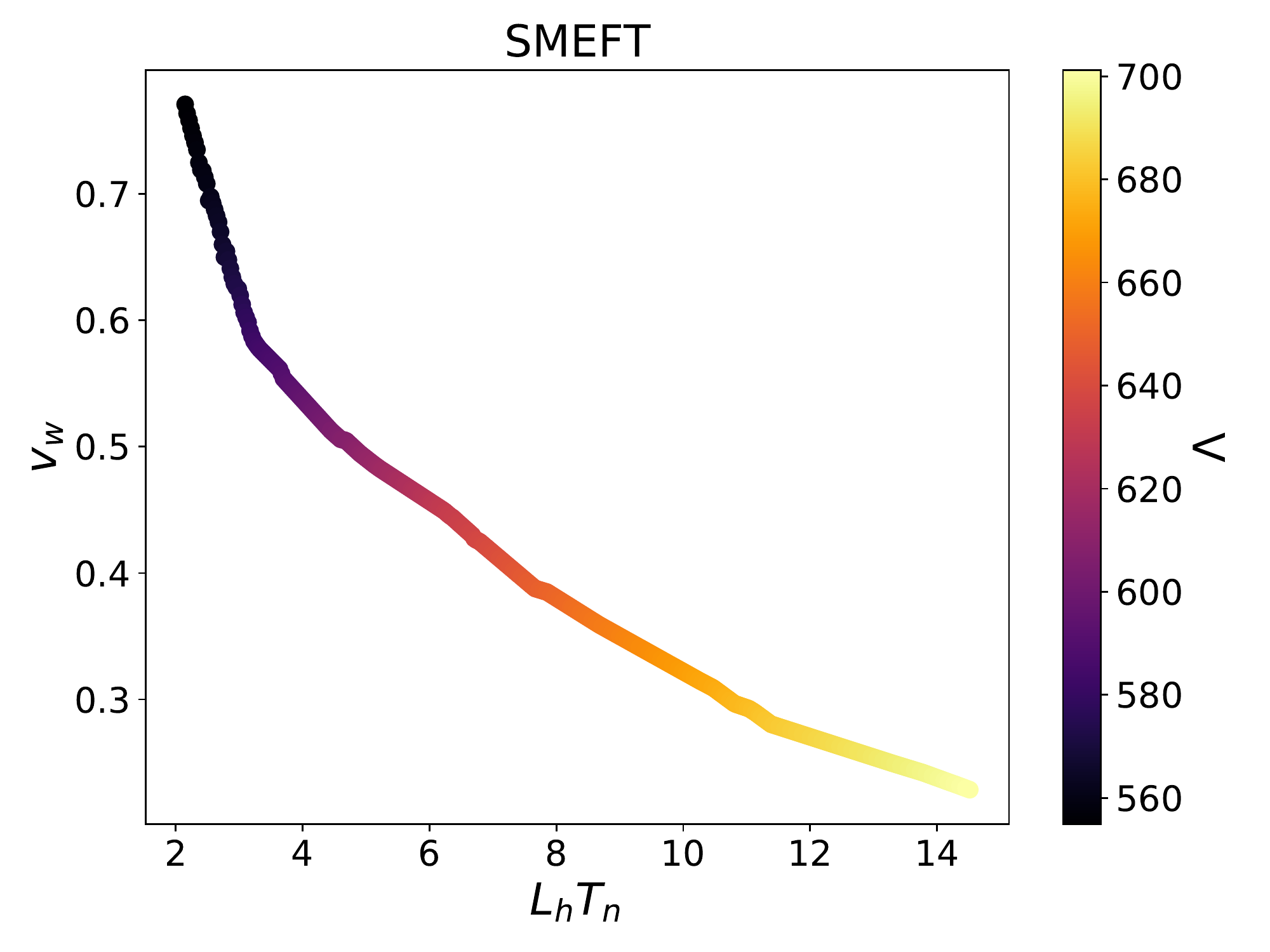} 
\includegraphics[scale=.35]{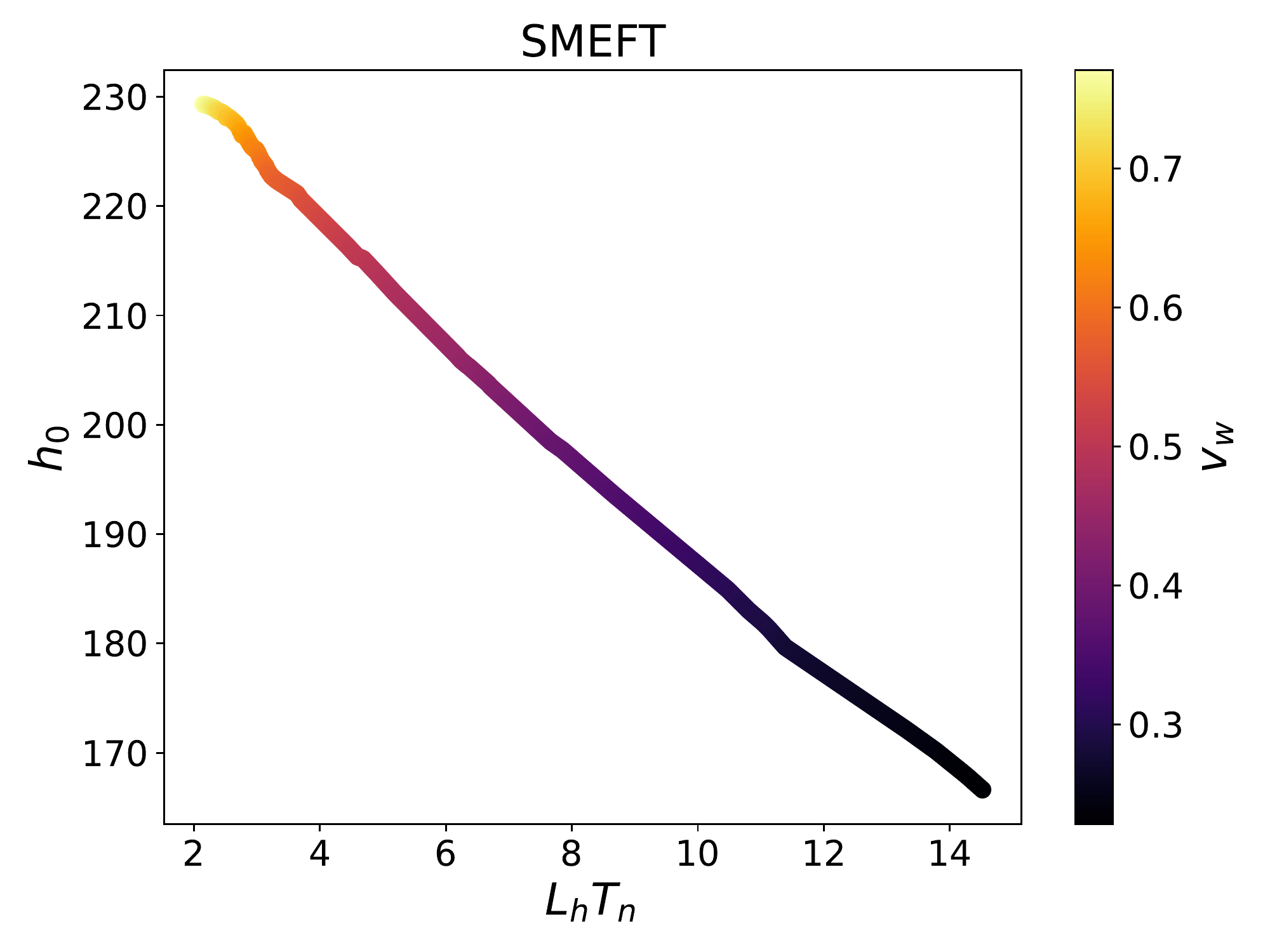}
\includegraphics[scale=.35]{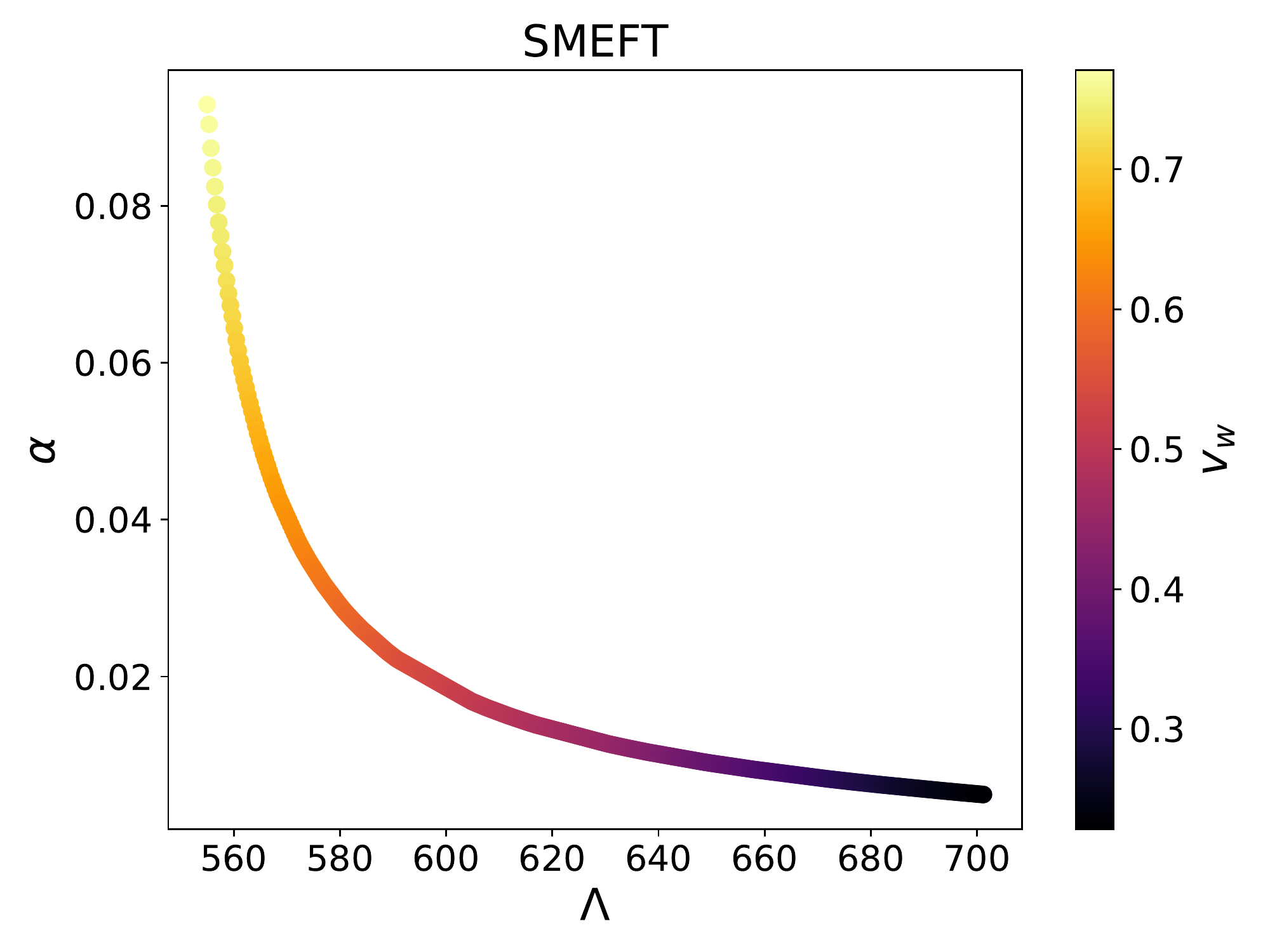} 
\includegraphics[scale=.35]{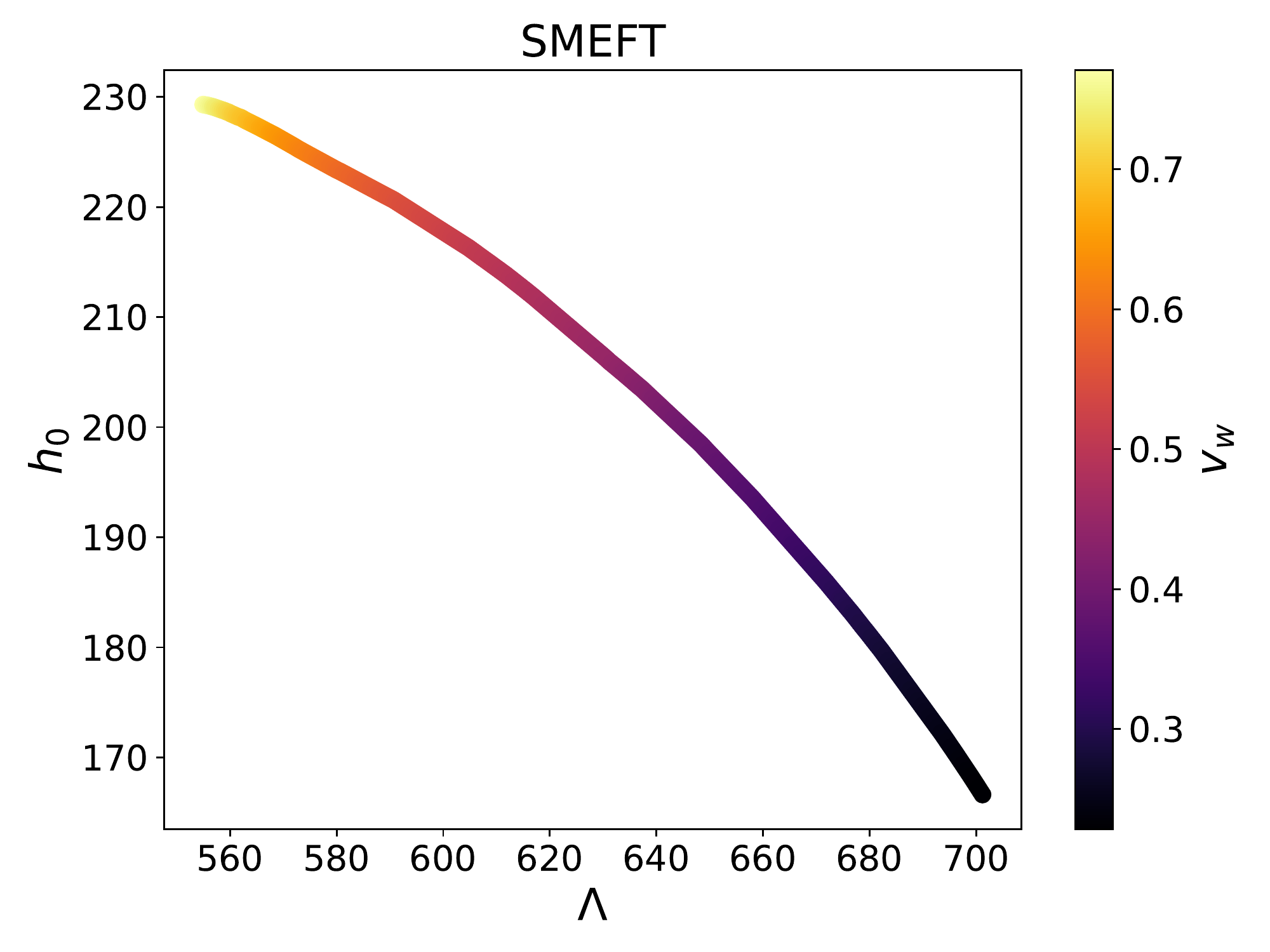}
\centering
\caption{Bubble wall properties in the SMEFT. Color bar shows the value of $\Lambda$ in the upper left figure and of $v_w$ in the rest of them. }
\label{SMEFT_bubble_1}
\end{figure}

 The outcome of the computation of the bubble wall shape and speed is displayed in figure \ref{SMEFT_bubble_1} which shows different cross sectional views of the velocity, thickness and  amplitude as well as their dependence on $\Lambda$. The main qualitative results found for the scalar singlet model also hold here; thicker walls are slower and larger field amplitudes correspond to faster walls, this is represented by the upper plots.
 
 The cutoff $\Lambda$ in the SMEFT completely determines the character of the FOPT with smaller values of $\Lambda$ giving the strongest possible transitions. The points scanned for $\Lambda$ terminate for low values because the $h=0$ becomes the deepest minimum of the potential while for very high values one starts recovering the SM for which a FOPT is not possible.

%%%%%%%%%%%%%%%%%%%%%%%%%%%%%%%%%%%%%%%%%%%%%%%%%%%%%%%%%%%%%%%%%%%%%%%
%%%%%%%%%%%%%%%%%%%%%%%%%%%%%%%%%%%%%%%%%%%%%%%%%%%%%%%%%%%%%%%%%%%%%%%
\subsection{Baryogenesis in SMEFT}

In order to compute the BAU in this model we introduce CP violation in the following form 
\be
\label{CP_SMEFT}
\mathcal{L}_{\text{Yukawa}}  \supseteq \ y_t \bar{Q} \Phi t_R +  \frac{y'}{\Lambda_\text{CP}^2} \bar{Q} \Phi t_R  (\Phi^{\dagger} \Phi)   + \text{h.c.}
\ee
where we assume the coefficient of the higher dimensional operator is purely complex $y' = i$, which corresponds to maximal CP violation. It is also assumed that the only CP violating operator is the one presented above. We also ignore the existence of other higher dimensional operators that might have phenomenological constraints. For a study of operators with constraints from precision electroweak observables see~\cite{Grojean:2004xa}.

As a consequence of \eqref{CP_SMEFT}, a space-time dependent complex mass term for the top quark appears, that is, $m_t(z) e^{i \theta_t(z)}$, with 
\be
m_t(z) \equiv    \frac{y_t h(z)}{\sqrt{2}}  \sqrt{  1 + \frac{h(z)^4}{4 \Lambda_{\text{CP}}^4}  },
\label{SMEFT_top_mass}
\ee
and the CP-violating phase
\be
\theta_t (z) = \arctan{\left[   \frac{h(z)^2}{2 \Lambda_{\text{CP}}^2}  \right]   }. 
\ee

 The WKB approximation for computing the BAU in SMEFT has been investigated in~\cite{Bodeker:2004ws}. In this reference the authors considered the two thresholds to be correlated, i.e. $\Lambda = \Lambda_{\text{CP}}$, and showed the the observed BAU could be obtained. When this reference appeared the Higgs boson had not yet been discovered and the physical Higgs mass was taken as a free parameter. At the same time the authors examined the properties of the bubble wall using the analytic estimates of Ref.~\cite{Moore:2000wx}. The study of the bubble wall properties and EWBG within the SMEFT in this paper thus provides an updated and improved analysis compared to that reference. Moreover we consider the case of decoupled thresholds, i.e. $\Lambda \neq \Lambda_{\text{CP}}$ as it has been shown in Ref.~\cite{Huber:2006ri} that the parameter space is ruled out for $\Lambda = \Lambda_{\text{CP}}$ unless one includes extra CP-violating higher dimension operators to provide a cancellation for EDM contributions. 

\begin{figure}[h]
\includegraphics[scale=0.72]{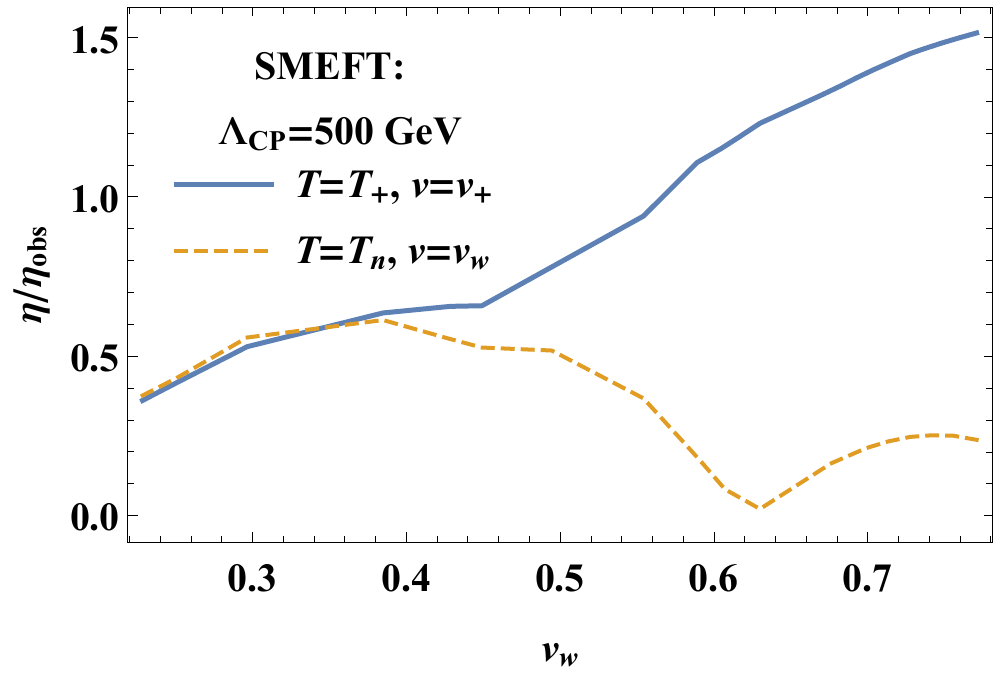}
\includegraphics[scale=0.72]{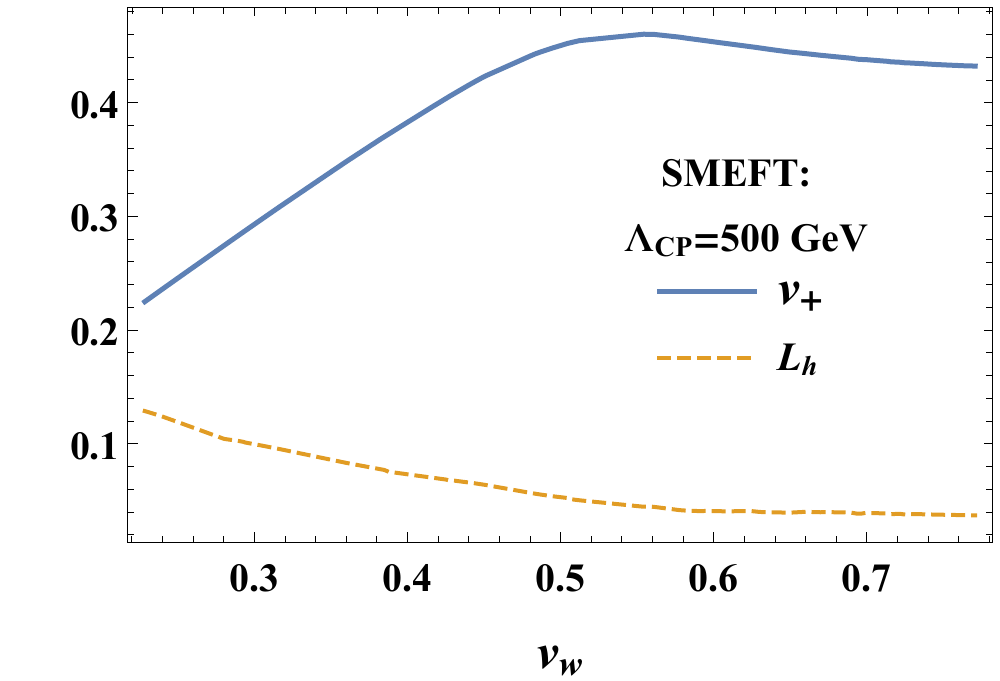}
\centering
\caption{Left: Normalized BAU against the wall velocity computed for the SMEFT.
Right: The variation of the wall properties as function of velocity. The CP-violating cut-off scale was fixed to $\Lambda_{\text{CP}} = 500$ GeV in both cases. }
\label{fig:BAU_SMEFT}
\end{figure}

Fig.~\ref{fig:BAU_SMEFT} shows the baryon yield normalised to the observed value as a function of the wall velocity for fixed $\Lambda_{CP}=500$ GeV. We again highlight the impact of using the parameters $T_+$ and $v_+$ in front of the wall in comparison with the  naive ones $T_n$ and  $v_w$. As we can see, for the latter case, the model can never produce the observed BAU. In the right hand side panel of fig.~\ref{fig:BAU_SMEFT} we show values of relevant parameters again as a function of the wall velocity. The only thing that varies in this plot is the cut-off scale of the dim-6 operator, all other quantities have been computed from first principles.

\begin{figure}[h]
\includegraphics[scale=0.36]{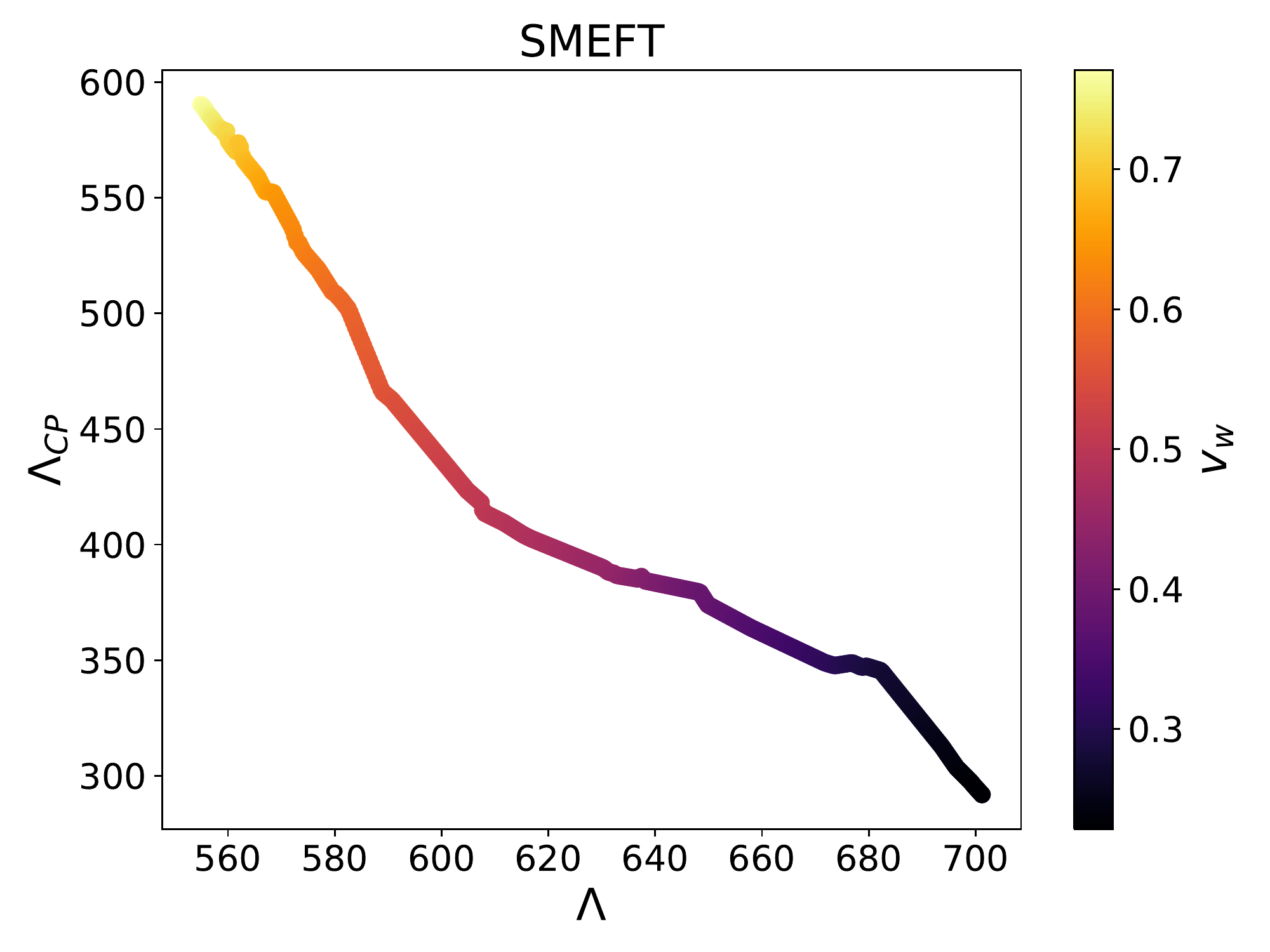}
\includegraphics[scale=0.36]{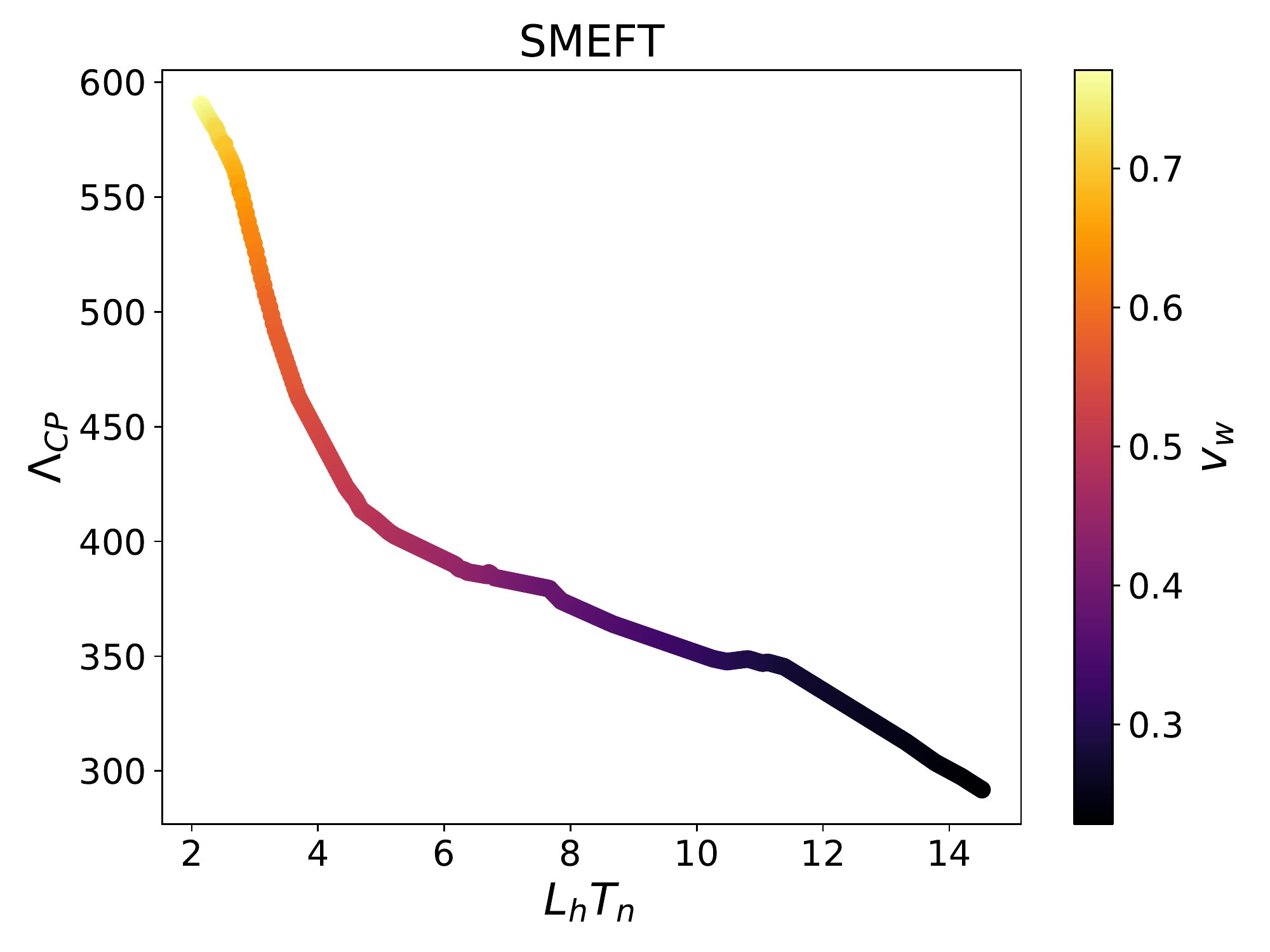}
\centering
\caption{Cut-off scale $\Lambda_{CP}$ necessary to produce the observed BAU as a function of the cut-off scale of the scalar potential (left) and  of the bubble wall thickness (right) in SMEFT}
\label{fig:BAU_SMEFT_Lambda}
\end{figure}

Fig~\ref{fig:BAU_SMEFT_Lambda} shows the cut-off scale $\Lambda_{CP}$ necessary to produce the observed baryon asymmetry as a function of the cut-off scale in the scalar potential (left) and  of the bubble thickness (right) in SMEFT. The colorbar shows the value of the wall velocity.  One can see that the allowed scale for BAU has negative correlation with the cut-off scale in the scalar potential, i.e., smaller $\Lambda$ values are associated with stronger phase transitions and faster walls pushing the scale $\Lambda_{CP}$. The same result we found in the singlet model. At the same time, thicker walls yield the lowest $\Lambda_{CP}$ necessary for the correct asymmetry. 

Even though $\Lambda_{CP}$  reaches only a few hundred GeV for the stronger transitions,  the bound from EDM constraints on CP violating cutoff has been estimated to be in the multi TeV range, i.e.,~\cite{deVries:2017ncy} $\Lambda_{\text{CP}} > 2.5 \text{ TeV}$ and we conclude that SMEFT cannot be responsible for the baryon asymmetry unless some additional mechanism is added to hide the CP violating interactions from EDM experiments.

\section{Gravitational wave signals} \label{sec:GWs}
In this section we will discuss stochastic gravitational wave backgrounds produced by the first order phase transitions our models predict. Given the transitions in question are relatively weak and the walls never reach very relativistic velocities we will exclusively focus on GWs sourced by plasma motion~\cite{Caprini:2019egz,Caprini:2015zlo}. Thus we will neglect the bubble collision contribution which would necessarily require a much stronger transition~\cite{Ellis:2019oqb,Lewicki:2019gmv,Lewicki:2020jiv,Lewicki:2020azd} not feasible in our models featuring polynomial potentials~\cite{Ellis:2018mja}.
Further, despite recent progress concerning GWs produced by turbulence~\cite{RoperPol:2019wvy,Kahniashvili:2020jgm,Pol:2021uol} the overall size of this contribution sourced by a phase transition remains uncertain and following~\cite{Caprini:2019egz} we will neglect it.
Finally for the range of wall velocities we compute it seems crucial to use updated hybrid calculations of GW generation through sound waves in the plasma~\cite{Hindmarsh:2016lnk,Hindmarsh:2019phv,Jinno:2020eqg,Gowling:2021gcy} predicting a non-trivial dependence of the spectral shape on the wall velocity. However, we have checked these modifications have a negligible impact on the observational prospects of upcoming experiments simply because for weak signals these are dominated by the peak abundance which is not significantly modified.  

As a result we will use the results of lattice simulations for the GW signal from sound waves~\cite{Hindmarsh:2013xza,Hindmarsh:2015qta,Hindmarsh:2017gnf} as summarised in~\cite{Caprini:2019egz,Caprini:2015zlo}. The abundance of the signal can be expressed as
\bea
&\Omega_{\rm sw}(f)h^2 = 4.13\times 10^{-7} \, \left(R_* H_*\right)  \left(1- \frac{1}{\sqrt{1+2\tau_{\rm sw}H_*}} \right)  \left(\frac{\kappa_{\rm sw} \,\alpha }{1+\alpha }\right)^2 \left(\frac{100}{g_*}\right)^\frac13 S_{\rm sw}(f) \,, \\
& S_{\rm sw}(f)=\left(\frac{f}{f_{\rm sw}}\right)^3 \left[\frac{4}{7}+\frac{3}{7} \left(\frac{f}{f_{\rm sw}}\right)^2\right]^{-\frac72} \,,
\eea
with the peak frequency given by
\begin{equation}
 f_{\rm sw} \,=2.6\times 10^{-5} {\rm Hz} \left(R_* H_*\right)^{-1} \left(\frac{T_p}{100 {\rm GeV}}\right)\left(\frac{g_*}{100}\right)^\frac16 \,.  
\end{equation}
where $g_*$ is the number of degrees of freedom at temperature $T_p$ for which we use the results of~\cite{Saikawa:2018rcs}.
The duration of the sound wave period normalised to Hubble can be approximated as~\cite{Hindmarsh:2017gnf,Ellis:2018mja,Ellis:2019oqb,Ellis:2020awk,Guo:2020grp}
\begin{equation}
\tau_{\rm sw}H_* =\frac{H_* R_*}{U_f}\,, \quad U_f\approx \sqrt{\frac34 \frac{\alpha}{1+\alpha} \kappa_{\rm sw}}\,.
\end{equation}
The average bubble radius normalised to Hubble rate can be approximated as
\begin{equation}
H_*R_* \approx (8\pi)^\frac13 \, {\rm Max}(v_w,c_s)\left(\frac{\beta}{H}\right)^{-1} \, ,
\end{equation}
using duration of the transition from eq.~\eqref{eq:betaH}.
Finally we calculate the sound wave efficiency factor $\kappa_{sw}$ using the fluid profiles velocity and temperature profiles as explained in Sec~\ref{hydrodynamics}.
This quantity can be approximated as the energy converted into bulk fluid motion, given by~\cite{Espinosa:2010hh}
\be
\kappa_{sw} = \frac{3}{\alpha\, \rho_{\mathrm{R}} \,v_w^3} \int w\, \xi^2 \frac{v^2}{1-v^2} d\xi= \frac{4}{\alpha \,v_w^3} \int \left( \frac{T(\xi)}{T_p} \right)^4\, \xi^2 \frac{v^2}{1-v^2} d\xi \, .
\label{eq:kappa_eff}
\ee

\begin{figure}[t]
\includegraphics[height=7cm]{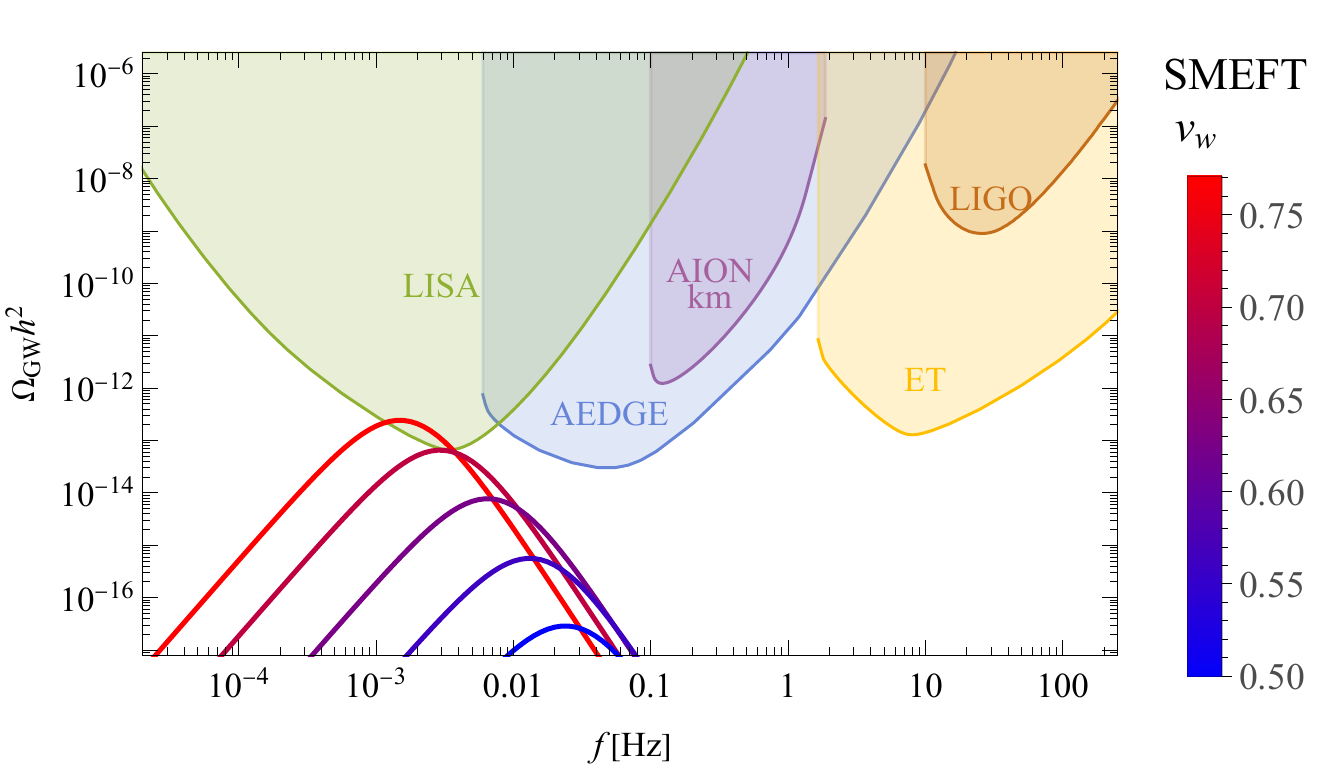} 
\centering
\caption{Gravitational wave signals for a range of parameters starting with the strongest transitions in the SMEFT model together with power-law integrated sensitivities of current and upcoming experiments}
\label{fig:SMEFTGWplot}
\end{figure}

\begin{figure}[t]
\hspace*{1.6cm}\includegraphics[height=7cm]{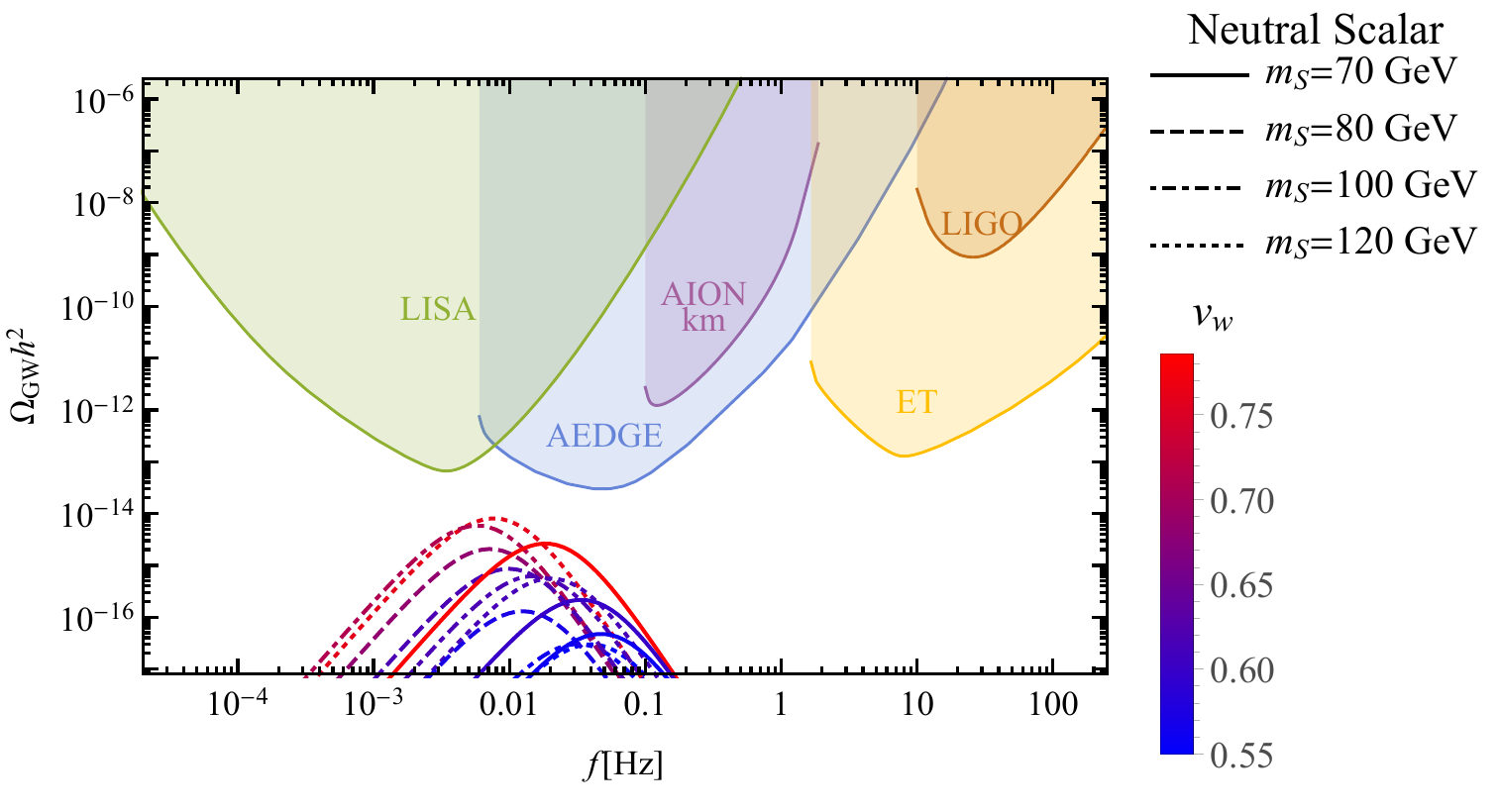} 
\centering
\caption{Gravitational wave signals for a range of parameters starting with the strongest transitions in the SM plus neutral scalar model together with power-law integrated sensitivities of current and upcoming experiments}
\label{fig:NeutralScalarGWplot}
\end{figure}

\begin{figure}[t]
\includegraphics[width=7cm]{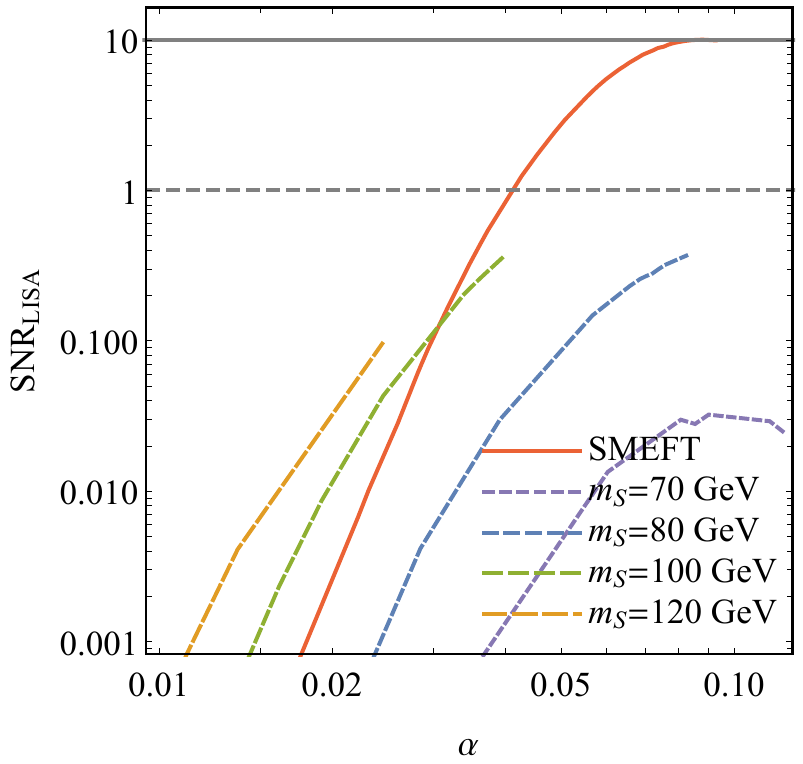} 
\includegraphics[width=7cm]{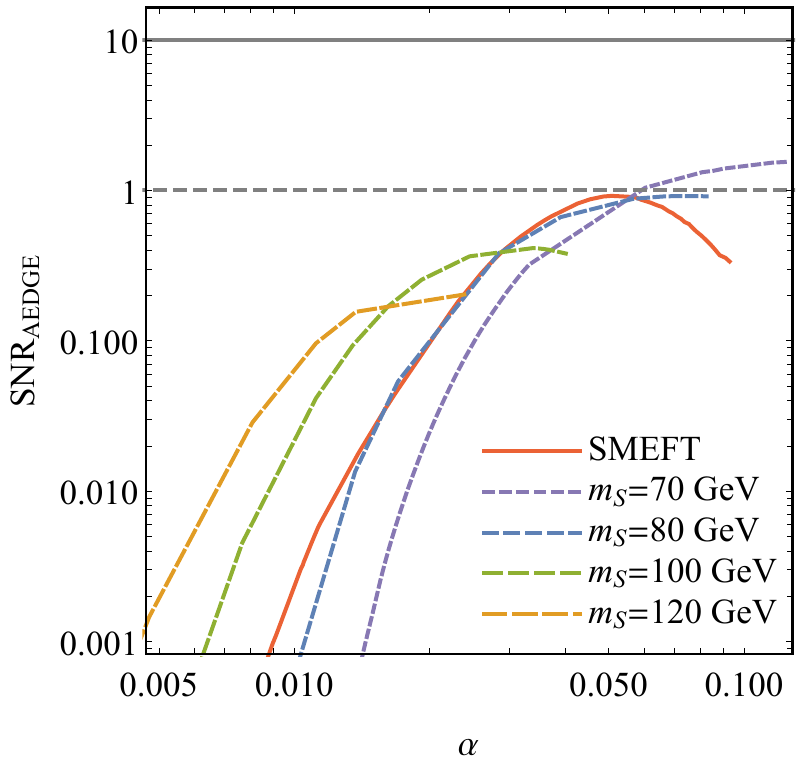} 
\centering
\caption{Signal to Noise Ratio for the GW backgrounds predicted by the singlet scalar and SMEFT models.}
\label{fig:SNRs}
\end{figure}

Figures~\ref{fig:SMEFTGWplot} and~\ref{fig:NeutralScalarGWplot} show GW spectra produced in the SMEFT and neutral singlet models together with power-law integrated sensitivity of
LIGO~\cite{LIGOScientific:2014pky,Thrane:2013oya,LIGOScientific:2016fpe,LIGOScientific:2019vic}
together with upcoming laser interferometer experiments LISA~\cite{Bartolo:2016ami,Caprini:2019pxz} and
ET~\cite{Punturo:2010zz,Hild:2010id} as well as future devices based on atom interferometry \cite{ 
Badurina:2021rgt}  AEDGE~\cite{AEDGE:2019nxb} and AION-1km~\cite{Badurina:2019hst}. Figure~\ref{fig:SNRs} shows the corresponding signal to noise ratio for LISA and AEDGE. We find that only a very small fraction of the parameter space of SMEFT predicts a transition strong enough to be observed while in the scalar singlet model none of the transitions for which we can verify baryogenesis can be observed~\cite{Cline:2021iff}.

\begin{figure}[t]
 \hspace*{2.6cm} \includegraphics[width=12cm]{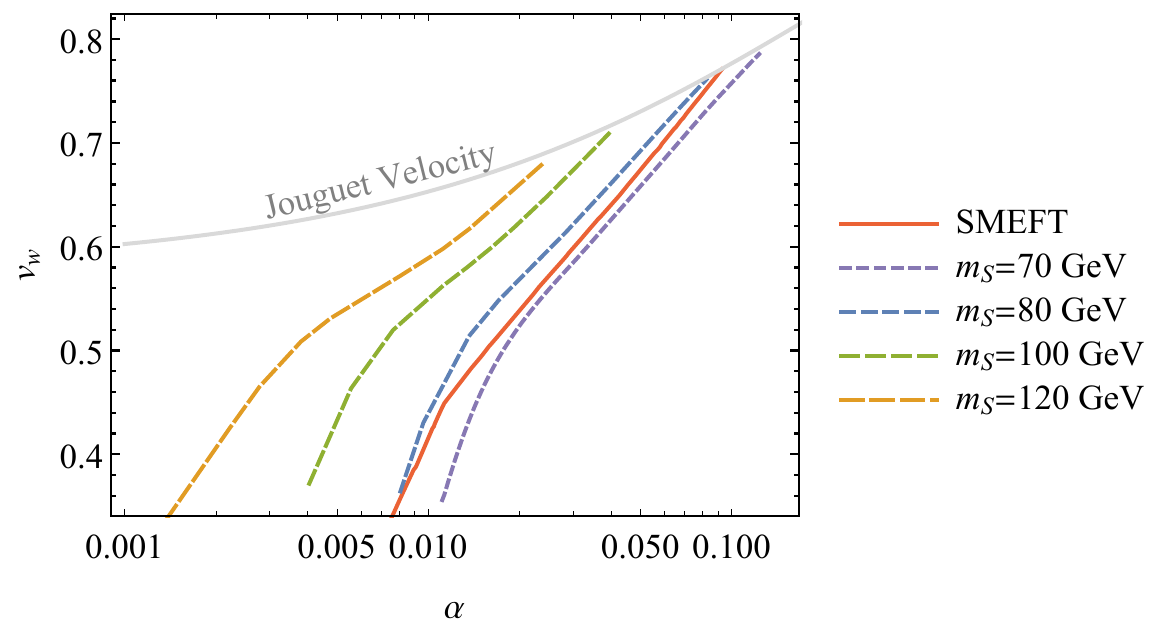} 
\centering
\caption{Bubble wall velocity solutions as a function of the phase transition strength $\alpha$ for all the models tested. }
\label{fig:vel_Jouguet}
\end{figure}

While both models are of course capable of supporting much stronger transitions with clearly visible signals~\cite{Caprini:2015zlo,Caprini:2019egz} we focus only on the cases in which friction of the plasma is large enough for the walls to cease accelerating below the Jouguet velocity. Only in those cases we are able to compute the wall velocity and width and calculate the final baryon yield, see Fig. \ref{fig:vel_Jouguet}. If the acceleration of the wall is not stopped below the Jouguet velocity we obtain a detonation solution in which plasma in front of the wall is no longer heated up. As a result above this threshold the friction on the wall drops significantly and we don't find any solutions until the fluid approximation breaks down at $v_w\approx1$ rendering our calculation inadequate and viability of baryogenesis uncertain. 

%%%%%%%%%%%%%%%%%%%%%%%%%%%%%%%%%%%%%%%%%%%%%%%%%%%%%%%%%%%%%%%%%%%%%%%%%%%%%%%%%%%%%%%%%%%%%%%%%%%%%%%%%%%%
\section{Simple estimate for the wall velocity and thickness}%Comparison of thin-wall and thermal equilibrium results} 
\label{sec:comparison}

\begin{figure}[h]
\includegraphics[scale=0.37]{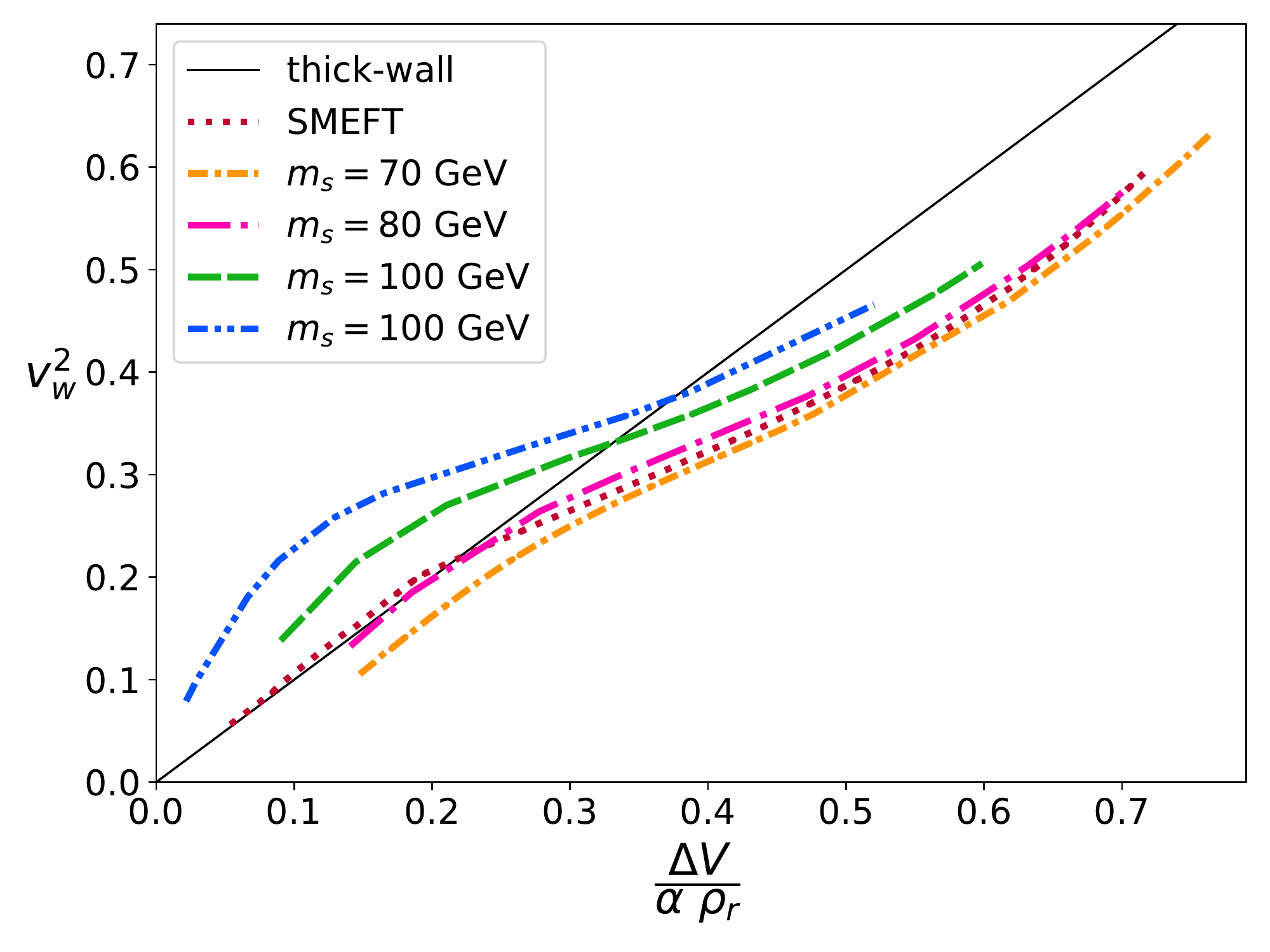}
 \centering\includegraphics[scale=0.36]{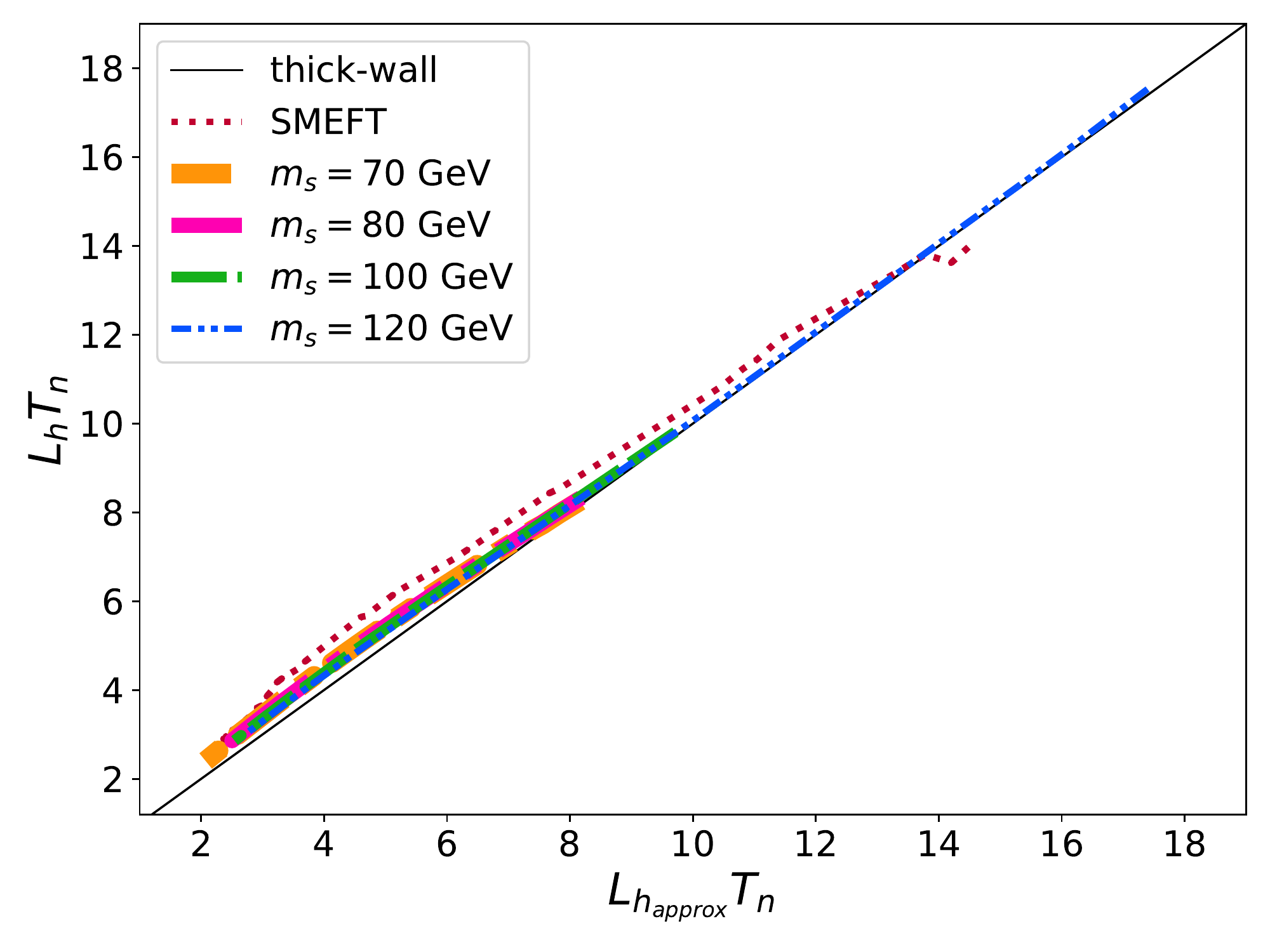}
\caption{Left: Comparison between the wall velocity calculation using the semi-classical method and the thermal equilibrium formula, eqn. \eqref{eqn:approx_velocity}. Right: Comparison between our results for the wall thickness in the Higgs direction with the estimated formula eqn. \eqref{thickness}. Both: The solid black line has slope one and is plotted for visual guidance.} \label{Comparison}
\end{figure}

In this section we will discuss methods that can approximate the wall velocity in a simple manner. Starting with the thermal equilibrium formula~\cite{Dine:1990fj} corrected for the cases in which the transition is too strong and the fluid approximation breaks down
\be \label{eqn:approx_velocity}
v_w=
\begin{cases}
\sqrt{\frac{\Delta V}{\alpha \rho_r}} \quad \quad {\rm for} \quad \sqrt{\frac{\Delta V}{\alpha \rho_r}}<v_J(\alpha)
\\
1 \quad \quad \quad \quad {\rm for} \quad  \sqrt{\frac{\Delta V}{\alpha \rho_r}} \geq v_J(\alpha)
\end{cases}
\ee
where $v_J(\alpha)$ is the Jouguet velocity from Eq.~\eqref{eq:vJ}. 
The lower case corresponds to transitions too strong for the fluid approximation to find a solution as discussed in Sec~\ref{Sec:FluidApproximation}. 
We show in the left panel of Fig.~\ref{Comparison} the comparison of this simple approximation with the results we find in all the SM extensions we studied.
Although the level of agreement between the naive formula and our results might seem extraordinary at first, it is not difficult to understand. Reminding ourselves that the velocity is determined by finding the roots of the moments introduced in eqns. \eqref{moment_1}, \eqref{moment_2} and that $M_1$ mostly fixes $v_w$ as can be appreciated in Fig. \eqref{Moment_grid}, it is natural to ask what is the relation between  $M_1$ and the approximation in Eq.~\ref{eqn:approx_velocity}.  The answer is that in thermal equilibrium the condition $M_1 =0 $ is simply equivalent to that formula. First we notice that if the temperature remains constant we can write \eqref{moment_1} as
\be
M_1 = \int dz \frac{\partial T_{\phi}^{zz}}{\partial z} = \Delta T^{zz}_{\phi},
\ee
where $T_{\phi}^{zz}$ is the $zz$ component of the energy momentum tensor of the scalar field, i.e. the momentum flux in the $z$ direction. Now if one replaces the scalar field contribution, in the above formula, by that of a perfect fluid in thermal equilibrium one obtains instead
\be
M_1 = \Delta \left(   \omega  \gamma^2 v^2 + p \right) = 0 , \quad \rightarrow \quad -\Delta p = \gamma^2 v^2 \Delta \omega,
\ee
where we used that in the steady state the wall has reached a constant velocity. Then using the relation $p = -V(\phi,T)$ together with the thermodynamic identity $\omega = T \frac{\partial p }{\partial{T}}$ and solving for the velocity we arrive at
\be
v^2 = \frac{\Delta V(\phi,T)}{\Delta \left( V(\phi,T) - T \frac{\partial V(\phi,T)}{{\partial T}} \right) } \equiv \frac{ \Delta V}{\alpha \rho_r} \label{thick_wall_formula},
\ee
where the definition of the strength, eqn. \eqref{alpha_def} was used and we omitted the factor $1/4$ in the denominator.
This gives us the first case in Eq.~\ref{eqn:approx_velocity}.
The fact that the above formula can provide a proxy for the wall velocity in some cases is tied to the assumption of small departure from equilibrium.
If for a given strength, the above formula yields a velocity above the Jouguet value then the transition is too strong and there cannot be thermodynamic equilibrium. 
Let us also note that the above expression can be derived from recent results on the wall velocity in local equilibrium \cite{Ai:2021kak}. In particular, equating the two pressures in  formula $(17)$ of that reference and assuming constant velocity with thermal equilibrium one arrives at the same result.  

Having obtained a usable formula for the wall velocity in thermal equilibrium, it is instructive to obtain an expression for the wall thickness using the second moment. Assuming again constant temperature we can write eqn. \eqref{moment_2} as follows
\be
M_2 =  \int dz \ \partial_z T_{\phi}^{zz}  \phi(z)   = - V(\phi_0,T)\phi_0 - \int dz \frac{1}{2}(\partial_z\phi)^3 + \int dz V(\phi,T) \partial_z \phi,
\ee
where in the last equality we simply integrated by parts. The second integral above simplifies to $\int \frac{1}{2}(\partial_z\phi)^3 = \phi_0^3/30L^2 $ when using the $\tanh$ ansatz, then one can solve for the thickness
\be
L^2 = \frac{\phi_0^3}{30\left[ \int V(\phi,T) d\phi   - V(\phi_0,T) \phi_0 \right] }.
\ee
In the above derivation we have assumed the dependence on a single field $\phi$, which in our case is identified with the Higgs but this expression can be generalized for multiple fields. We do not pursue the generalization here but instead notice that the expression in the denominator can be traded for the the height of the potential barrier\footnote{To convince oneself, one can use as an example the simplest looking potential $V= \lambda/4 \phi^2 (\phi - \phi_0)^2$, See~\cite{Huber:2013kj}.} thus $L^2 \propto \phi_0^2/V_h$ where $V_h$ is the height. The proportionality constant is model dependent and we found that for the scalar singlet model and for SMEFT the best approximation is given by 
\be 
L^2 = \frac{\phi_0^2}{4 V_h} .\label{thickness}
\ee

To quantify the discrepancy of this approximation with our results for the scalar singlet we calculate the height of the potential barrier using the maximum value of the potential along the path of minimum energy at the critical temperature. In the numerator we used the expectation value of the field at the true vacuum. 
As we show in Fig.~\ref{Comparison}, Eq.~\eqref{thickness} provides an extremely good approximation for the width of the wall of the Higgs field.

\begin{figure}[h]
 \centering \includegraphics[scale=0.37]{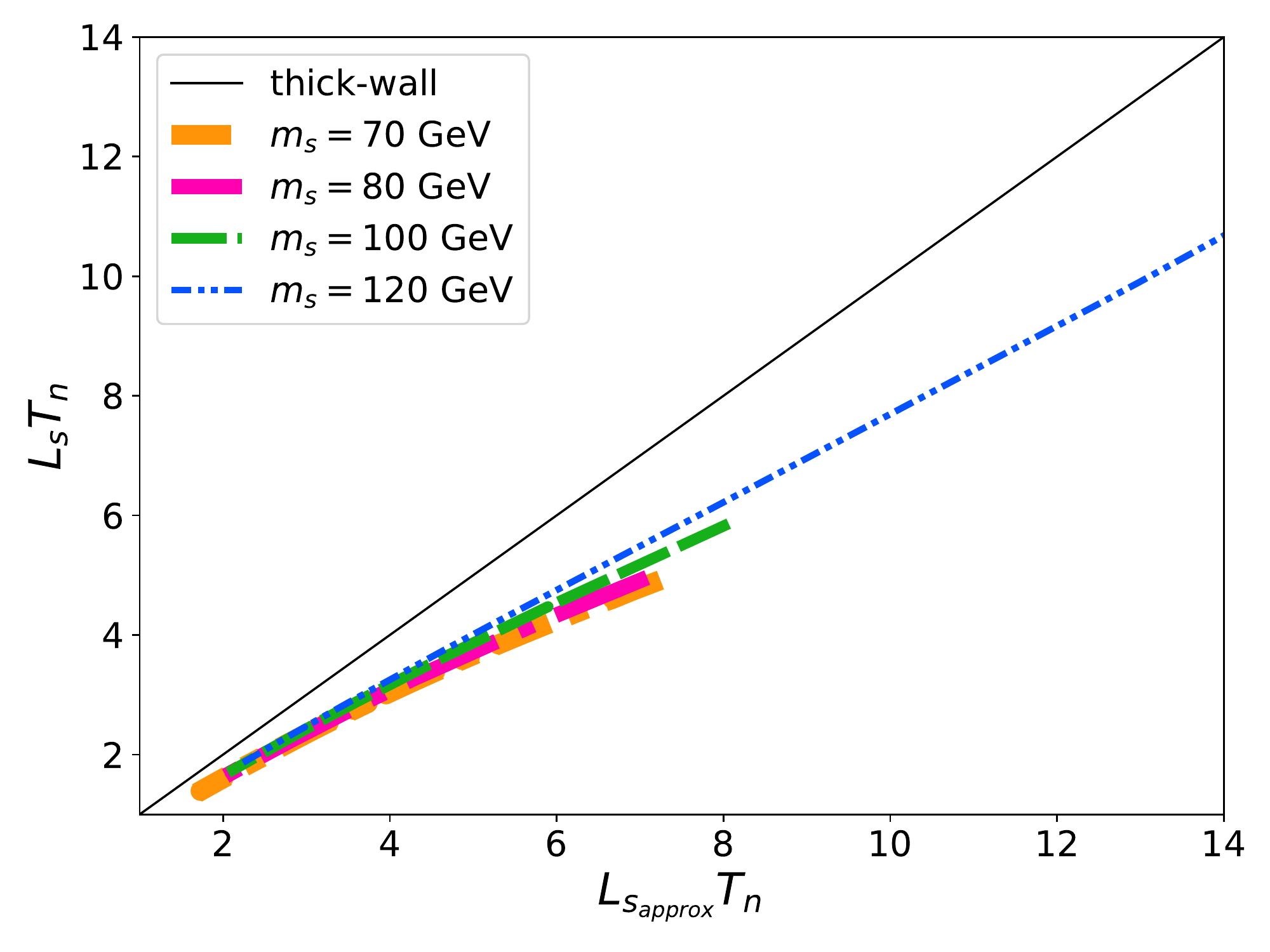}
\caption{Comparison between our results for the wall thickness in the singlet direction with the estimated formula eqn. \eqref{thickness}. The solid black line has slope one and is plotted for visual guidance.} \label{Comparison_Ls}
\end{figure}

In Fig.~\ref{Comparison_Ls} we show the comparison of results from Eq.~\eqref{thickness} (with the numerator evaluated at the false vacuum, $s_0 \neq 0 $ ) with our numerical results for the scalar singlet wall. Clearly the appropriate numerical constant in Eq.~\eqref{thickness} for this case depends on the mass of the scalar and so reproducing the details of the scalar singlet wall is more difficult than in the Higgs direction making full reproduction of the BAU from these simplifications difficult.

%%%%%%%%%%%%%%%%%%%%%%%%%%%%%%%%%%%%%%%%%%%%%%%%%%%%%%%%%%%%%%%%%%%%%%%%%%%%%%%%%%%%%%%%%%%%%%%%%%%%%%%%%%%%
\section{Summary and conclusions} \label{sec:conclusions}

In this paper we have investigated the FOPTs arising within the scalar singlet extension of the SM with parity symmetric potential and in the SMEFT with a dimension six operator. The bubble profile and its velocity were computed from first principles using the recently improved semi-classical fluid equations of Cline and Laurent~\cite{Laurent:2020gpg}. The parameter space of the scalar singlet model has been thoroughly studied in Ref.~\cite{Cline:2021iff}. Here we did not undertake a full scan but instead focused on the qualitative features for some particularly motivated values of the scalar singlet mass. In the SMEFT the cutoff scale of the dimension six operator in the scalar potential is the only relevant parameter and we surveyed the whole range that facilitates a FOPT. 

For both scenarios we have found the intuitive expectation that stronger transitions produce faster walls holds.
However, the strength of the transition suitable for baryogenesis  is severely  limited with $\alpha\leq0.1$. For stronger transitions the friction of the plasma does not cease the acceleration of the wall before it reaches the Jouguet velocity. Above that velocity the heated plasma shell around the bubble disappears as our hydrodynamic solution becomes a detonation. This lowers the plasma friction significantly and such walls always reach very relativistic velocity $v_w\geq0.9$ for which we cannot compute the wall properties and assert the viability of baryogenesis. 
We find this behaviour in both models which suggests  this could be a generic trademark of any SM-like model which does not modify the SM plasma contents drastically. 

Our results also show that very strong transitions are not suitable for the semi-classical treatment as no solution for the moments can be found. In this case the assumptions of the fluid approximation are not satisfied as the wall becomes very thin and the WKB approximation is invalid. We also expect that very strong transitions correspond to large departures from thermodynamic equilibrium and a different formalism is needed in that case. In this paper we have focused only on the region of parameter space for which the wall properties can be computed. 
We have considered the presence of higher dimensional CP-violating operators and computed the BAU employing the upgraded fluid equations of \cite{Cline:2020jre}. While the scalar singlet model can easily yield the right amount of asymmetry, the EDM constraints on the CP-violating cutoff scale make SMEFT not capable of explaining the BAU.

Computation of the wall velocity is also crucial for the GW signals produced by the transition. We calculate the GW signals in both scenarios for transitions in which we can compute the properties of the wall. We find that only the strongest transitions of SMEFT fall within the sensitivity band of the LISA experiment while AEDGE operating at a slightly higher frequency will not be able to observe any of these signals. Both models, of course, support also stronger transitions which would be clearly visible in LISA and AEDGE, however, for these cases we find the plasma friction will not stop the wall acceleration before breakdown of the fluid approximation and its justified to simply assume $v_w\approx 1$. We provide a simple and quite accurate approximation for the wall velocity requiring only the strength of the transition and potential which takes this effect into account. From our results we infer that GW signals produced by simple SM extensions visible in future experiments are likely to only be produced in strong transitions with $v_w\approx 1 $. This does not mean that plasma can be neglected altogether in these results and bubble collision can always play an important role as this requires $\alpha \gg 1$ instead. However, observable signals produced by plasma related sources are likely to be produced mostly by detonations with highly relativistic wall velocities.

\section*{Acknowledgements}
The authors would like to thank Benoit Laurent for helpful correspondence regarding the modified fluid equations. This work was supported by the Polish National Science Center grant 2018/31/D/ST2/02048. ML was also supported by the Polish National Agency for Academic Exchange within Polish Returns Programme under agreement \\ PPN/PPO/2020/1/00013/U/00001.

%\appendix \label{app}
%%%%%%%----APPENDIX A---- %%%%%%%%%%
\newpage

\appendix \label{app}

\section{Finite temperature effective potential} \label{sec:appendix_singlet}
In this paper we adopt the customary procedure of calculating the effective potential in the Landau gauge where the contribution from the Goldstone bosons is independent from the massive gauge bosons and ghosts do not contribute.

Generically, the effective finite temperature potential is given by 
\begin{equation}
V_{\text{eff}}(\phi,T) = V_0(\phi) + V_{\text{CW}}(\phi) + V_{\text{T}}(\phi,T).
\end{equation}
where $V_0$ gives the tree-level contribution, $V_{\text{CW}}$ represents the Coleman-Weinberg potential  \cite{Weinberg:1973am} and $V_{\text{T}}$ the finite temperature contribution.  In the equation above we have written a generic field dependence but it should be understood that $\phi$ could denote the background field values $h$ and $s$ in the scalar singlet extension or simply the Higgs background vev in case of the SMEFT.  

In this paper we consider the effective potential calculated at one-loop order and we choose to follow the on-shell prescription \cite{Delaunay:2007wb, Curtin:2014jma} in which the one-loop contributions do not disturb the minimization conditions at tree-level. In this case, the Coleman-Weinberg contribution can be written as
\begin{equation}
V_{\text{CW}}(\phi) = \sum_i (-1)^{F_i}  \frac{d_i}{64 \pi^2} \left[ m_{i}^4(\phi) \left(  \log{\frac{m_{i}^2(\phi)}{m_{0i}^2}}  - \frac{3}{2}   \right) +2m_i^2(\phi) m_{0i}^2 \right],
\end{equation}
where the index $i$ runs over all particles contributing to the potential with $F_i=0$ ($1$) for bosons (fermions), $d_i$ is the number of degrees of freedom of the particle species while $m_i(\phi)$  is the field dependent mass of particle $i$ and $m_{0i}$ its value at the EW vacuum. As mentioned above, this form of the Coleman-Weinberg potential ensures that the zero temperature vacuum conditions are completely determined by the tree-level contribution. In other words
\begin{equation}
\frac{d V_{\text{CW}}(\phi)}{d \phi} \Big{|}_{\phi_0} = \frac{d^2 V_{\text{CW}}(\phi)}{d \phi^2} \Big{|}_{\phi_0} = 0, 
\end{equation}
where $\phi_0$ corresponds to the field values at the zero temperature EW vacuum, i.e., $\phi_0= (v,0)$ for scalar singlet and $\phi_0 = v$ in the SMEFT. 

The contribution to the Coleman-Weinberg potential from the Goldstone bosons requires special care since for 
small field values the squared mass parameter becomes negative and furthermore it vanishes at the EWSB minimum leading to infrared divergences of the effective potential. It has been shown in \cite{Elias-Miro:2014pca,Martin:2014bca} that proper resummation of the Goldstone boson contributions must be performed to avoid such divergences. We do not include the contributions from the Goldstone bosons as it has been shown \cite{Martin:2014bca} that the numerical impact of the resummation procedure as a function of the renormalization scale is very small.

The finite temperature piece is given by
\begin{equation}
V_T(\phi, T) = \frac{T^4}{2\pi^2}  \sum_i d_i J_{\mp} \left( \frac{m_i(\phi)}{T} \right),
\end{equation}
where the $J_{\mp}$ functions are defined as  
\begin{equation}
J_{\mp}(x)  = \pm \int^{\infty}_{0} dy y^2 \log\left( 1 \mp e^{-\sqrt{y^2 + x^2}} \right),
\end{equation}
and the upper (lower) sign is for bosons (fermions).

%%%%%%%----APPENDIX B---- %%%%%%%%%%

\section{Field dependent and thermal masses}\label{sec:appendix_SMEFT}

In this section we present the relevant formulas for the field dependent masses which are inputs for the one-loop and thermal contributions to the effective potential. In the models studied in this paper the thermal masses for the SM particles are given by 
\be
m_W^2  =\frac{g^2}{4}h^2, \quad m_Z^2=\frac{g^2+g'^2}{4}h^2, \quad  m_t^2=\frac{y_t^2}{2}h^2.
\ee
The field dependent masses for the scalar particles are modified with respect to the SM and are presented in the following two sub-appendices. 

\subsection{Scalar singlet extension}
At any field value the physical masses correspond to the eigenvalues of the Hessian matrix of the scalar potential, namely

\begin{equation}
m^2_{ij}(h,s)=\begin{pmatrix}
\frac{m_h^2}{2}\left(3\frac{h^2}{v^2} -1  \right) + s^2 \frac{\lambda_{hs}}{2}&   h s\lambda_{hs}    \\
  h s\lambda_{hs} & m_s^2 + \frac{h^2-v^2}{2}\lambda_{hs} + 3 \lambda_s s^2
\end{pmatrix}.
\end{equation}

We add the effect of thermal masses that correct the behavior at high temperatures. In our model they are written as \cite{Weinberg:1974hy, Arnold:1992rz}
\begin{equation}
 \Pi_h = \left( \frac{3g^2}{16} + \frac{g'^2}{16} + \frac{\lambda}{2} + \frac{y_t^2}{4} + \frac{\lambda_{hs}}{24}  \right) T^2,
\end{equation}
\begin{equation}
\Pi_s =\left(   \frac{\lambda_{hs}}{6} + \frac{\lambda_s}{4} \right)T^2,
\end{equation}
\begin{equation}
\Pi_{\text{Gauge}} = T^2 \text{diag}( \frac{11}{6} g^2 ,  \frac{11}{6}g'^2   ).
\end{equation}

A truncated full dressing implementation \cite{Curtin:2016urg} amounts to the replacement 
\begin{equation}
m^2_{ii}(h,s) \rightarrow m^2_{ij}(h,s,T) \equiv m^2_{ii}(h,s)  + \Pi_i
\end{equation}
 in the one-loop potential at finite temperature. This procedure is also called daisy resummation.

\subsection{SMEFT}

In this model the addition of the dimension six operator gives rise to a Higgs mass
\be
m^2_h  = -m^2+3\lambda h^2 +  \frac{15}{4}\frac{h^4}{\Lambda^2},
\ee
while the thermal masses are 
\begin{equation}
\begin{split}
\Pi_{h }(T)&=\frac{T^2}{4v^2}\left(m_h^2+2m_W^2+m_Z^2+2m_t^2\right)-\frac{3}{4} T^2 \frac{v^2}{\Lambda^2} \, ,
\\
\Pi_W(T)&=\frac{22}{3}\frac{m^2_W}{v^2}T^2. 
\end{split}
\end{equation}

\bibliographystyle{JHEP}
\bibliography{Baryogen}  
  
\end{document}